\documentclass[12pt,english]{article}
\usepackage[T1]{fontenc}
\usepackage[utf8]{inputenc}
\usepackage{geometry}
\usepackage{verbatim}
\usepackage{float}
\usepackage{amsmath}
\usepackage{amsthm}
\usepackage{amssymb}
\usepackage{stmaryrd}
\usepackage{graphicx}
\usepackage{setspace}
\usepackage{wasysym}
\usepackage{tikz-network}
\usepackage{tikz-cd}
\usepackage{tikz}

\usepackage[authoryear]{natbib}
\usepackage{comment}
\usepackage{color}
\definecolor{MyDarkBlue}{rgb}{0,0.08,0.45}
\usepackage[unicode=true,pdfusetitle, bookmarks=false,bookmarksnumbered=false,bookmarksopen=false, breaklinks=false,pdfborder={0 0 1},colorlinks=true, linkcolor = MyDarkBlue, citecolor = MyDarkBlue]{hyperref}
\usepackage{breakurl}
\usepackage{graphicx}
\usepackage{caption}
\usepackage{subcaption}
\usepackage{titling}
\usepackage{subcaption}
\makeatletter
\theoremstyle{definition}
\newtheorem{defn}{\protect\definitionname}
\theoremstyle{plain}
\newtheorem{lem}{\protect\lemmaname}
\theoremstyle{plain}
\newtheorem{cor}{\protect\corollaryname}
\theoremstyle{remark}
\newtheorem{rem}{\protect\remarkname}
\theoremstyle{plain}

\theoremstyle{plain}
\newtheorem{thm}{\protect\theoremname}
\theoremstyle{plain}
\newtheorem{prop}{\protect\propositionname}
\newtheorem*{prop*}{Proposition}
\theoremstyle{plain}
\newtheorem{claim}{\protect\claimname}

\newcommand{\lambdabar}{\mathchoice
  {\lambdabar@{.88ex}{1.10ex}{.12ex}{18}{.08em}}
  {\lambdabar@{.88ex}{1.10ex}{.12ex}{18}{.08em}}
  {\lambdabar@{.70ex}{.85ex}{.10ex}{20}{.14em}}
  {\lambdabar@{.70ex}{.70ex}{.09ex}{20}{.12em}}
}
\newcommand{\lambdabar@}[5]{%
  \mathord{\ooalign{%
    \raisebox{#1}{\hskip#5\rotatebox{#4}{\rule{#2}{#3}}}\cr
    $\lambda$\cr
  }}%
}

\usepackage[bottom]{footmisc}
\makeatother

\usepackage{babel}
\providecommand{\corollaryname}{Corollary}
\providecommand{\definitionname}{Definition}
\providecommand{\lemmaname}{Lemma}
\providecommand{\propositionname}{Proposition}
\providecommand{\remarkname}{Remark}
\providecommand{\theoremname}{Theorem}
\providecommand{\claimname}{Claim}

\providecommand{\E}{\mathbb{E}}
\providecommand{\Var}{\operatorname{Var}}
\providecommand{\Cov}{\operatorname{Cov}}
\providecommand{\1}{\mathbf{1}}

\begin{document}
\setlength{\droptitle}{-5em}
\title{Network Beliefs and Behavior with Peer Effects\thanks{ This paper supersedes a previous paper with the title ``Game Under Network Uncertainty.'' We would like
to thank Eric Bahel, Francis Bloch, Niloy Bose, Christophe Bravard,
Krishna Dasaratha, Bhaskar Dutta, Ben Golub, Sumit Joshi, Matthew
Kovach, Ahmed Saber Mahmud, Antoine Mandel, Mihai Manea, Luca Merlino,
Agnieszka Rusinowska, Alireza Tahbaz-Salehi, Rakesh Vohra, Leaat Yariv
as well as seminar participants of the 17th Annual Conference on Economic
Growth and Development ISI New Delhi, the Eighth Annual NSF Conference
on Network Science and Economics, Midwest Economic Theory, Game Theory
Lisbon Meetings 2023, CRETE 2023, Queen Mary University, Universite
de Franche-Comte, Besancon, Universite Paris 1 Sorbonne, Ashoka University
Annual Economics Conference, Bocconi University, and KAIST College of Business for their insightful
comments and suggestions. Sudipta
Sarangi would like to acknowledge the support of Collegium de Lyon and the French Institutes
of Advanced Study where a part of this research was carried out.}}
\author{Promit K. Chaudhuri\thanks{Department of Economics, Virginia Tech
}
\and Matthew O. Jackson \thanks{Department of Economics, Stanford University, and Santa Fe Institute
} \and Sudipta Sarangi\thanks{Department of Economics, Virginia Tech
}
\and Hector Tzavellas\thanks{Department of Economics, Virginia Tech}
}
\date{May 2026}
\maketitle
\begin{abstract}
\begin{singlespace}
Individuals often act without knowing the full structure of the social networks in which they are embedded. We develop a framework in which agents form beliefs about their networks and use these beliefs to anticipate peers’ actions. Because peers reason similarly, equilibrium is governed by Iterative Belief Centrality, nesting complete-information and degree-based models as special cases. When beliefs satisfy a natural monotonicity in connectedness, rational iteration amplifies behavioral differences: more-connected agents act more, less-connected agents act less, relative to homogeneous-belief benchmarks. Correlation in network positions further increases dispersion, yielding a tractable theory of peer driven network behavior.
\end{singlespace}

\textbf{JEL Classifications:} C72, D81, D85

\textbf{Keywords:} Incomplete Information, Network Games, Uncertainty, Social Networks, Peer Effects, Network Perceptions, Assortativity
\end{abstract}
\newpage{}

\section{Introduction}
\label{sec:intro}

People's knowledge of the networks in which they are embedded tends to be incomplete and varies substantially across individuals. Krackhardt's \citeyearpar{krackhardt1987cognitive} seminal work on Cognitive Social Structures demonstrates that individuals form subjective representations of who is connected to whom, and that these representations differ systematically across observers. 
Individuals report significantly different sets of ties and patterns of connectedness for the same network \citep{kumbasar1994systematic,dessi2016network}, and their perceptions of their own and others' centrality can differ substantially from the true network \citep{kilduff2008organizational,kovavrik2025perception}.

Such beliefs about networks have important behavioral consequences. Individuals who believe themselves to be influential are more likely to engage in leadership and visible behaviors \citep{flynn2006helping}. Perceived network centrality predicts assertiveness in advice networks \citep{brands2020perceived}, while differences between perceived and actual connectedness predict mental well-being \citep{fang2025differences}. In online environments, users' beliefs about prevalence and influence are shaped by locally observed connections, leading to phenomena such as the virality paradox and the majority illusion \citep{hodas2013friendship,lerman2016majority}. Perceived influence can also shape political participation and collective action. Individuals abstain from movements when they believe themselves peripheral, and can emerge as mobilizers if they overestimate their reach \citep{finkel1985reciprocal}.

In this paper we provide a model of how people's beliefs and network knowledge shape their behaviors, and how those beliefs depend in systematic ways on network position and how this can explain phenomena such as those mentioned above.
Importantly, heterogeneity in beliefs and discrepancies between peoples' perceptions and true network need not be a sign of irrationality. Individuals only observe a small part of the network and must infer the rest, and because their local positions differ, so do their beliefs. 

These observations highlight that behavior in networks depends on individuals beliefs about network structure, which can systematically deviate from the actual realized structure, and can be heterogeneous across individuals in ways that reflect their local position. 
Despite this growing body of evidence,the link between beliefs about network structure and behavior remains largely underexplored.

The lack of formal results reflects two dominant modeling approaches for network influenced behaviors.\footnote{See \cite{jackson2015games} for a survey on network
games.} The first assumes complete network information. In such models
\citep{ballester2006s,bramoulle2007public,bramoulle2014strategic},  Behavior is derived as a Nash equilibrium based
on the full realized network structure, rendering beliefs irrelevant for equilibrium outcomes. The second approach assumes that agents do not know the network and all have \emph{homogeneous beliefs} about its structure: while agents may differ in their degree, they share the same beliefs about how others are connected. While these \emph{degree-based} models (e.g., \cite{jackson2006diffusion,jackson2007diffusion,jackson2007relating,sundararajan2008local,galeotti2011complex})
incorporate heterogeneity through realized degree, they do not capture the fact that individuals hold distinct, position-dependent perspectives on the network, which may in turn shape behavior.\footnote{\cite{galeotti2010network} allow for degree-dependent beliefs under an affiliation assumption to derive monotonicity results, as we discuss below.}

In this paper, we show that people's limited information about the network has strong and systematic implications for their behaviors when payoffs are peer-interactive.  When agents infer the broader network from their local observations, they form heterogeneous perceptions that generate behaviors consistent with many of the empirical patterns described above.  We develop a model in which agents form beliefs based on their position in the network, leading to systematic differences in behavior relative to both complete-information settings and models with homogeneous incomplete information. We show that even fully rational individuals can end up with heterogeneous and/or biased beliefs about what the rest of the network looks like. This leads to more extreme differences in behaviors than those that are generated by differences in network position with full information, or where individuals all have homogeneous beliefs about the rest of the network.

We introduce a general information framework that nests the behavioral predictions stemming from the approaches mentioned above as special cases. Our model proceeds as follows. A network involving $n$ agents, modeled as an $n\times n$ adjacency matrix, is drawn from a commonly-known distribution. Agents learn some information about the network---for instance, their own connections---and make inferences about the rest of the network.\footnote{Knowledge of the network architecture need not be restricted to first order neighborhoods. For example, our model admits situations in which one agent only observes their direct neighbors while another agent can also see their neighbors' neighbors.} Agents then form beliefs about the network and about what other agents know. These beliefs determine interim expectations of the strategic complementarities (or substitutabilities) present in the network, and agents choose actions to maximize expected payoffs. To keep things tractable, we focus on payoffs based on the seminal linear-quadratic formulation of \citet{ballester2006s}.

Our first set of results characterizes equilibrium behavior under arbitrary priors over networks.\footnote{We first establish the existence and uniqueness of a Bayesian Nash equilibrium (henceforth, simply equilibrium), which is necessarily in pure strategies} We show that equilibrium actions can be written as an infinite expansion over \emph{expected} walks. The intuition is simple. Under complete information, agents compute realized walks, which correspond to strategic chains of influence, and equilibrium actions are proportional to Katz--Bonacich centrality \citep{ballester2006s}. Under incomplete information, agents compute expected chains of influence, where expectations reflect both uncertainty about the network, and more importantly, beliefs about how others perceive the network. In exerting actions, therefore, agents form higher order beliefs through nested expectations, generating a \emph{belief hierarchy} about neighbors' beliefs, neighbors'  neighbors beliefs, and so on. Equilibrium actions therefore reflect a belief-based notion of influence rather than realized influence. We refer to this object as \emph{Iterative Belief Centrality}. Under complete information, belief iteration collapses and Iterative Belief Centrality coincides with Katz--Bonacich centrality.

To illustrate the formation and behavioral consequences of belief hierarchies, we present a simple example. A network is generated under one of two possible densities, one high and one low, but the realized density is unknown to agents. Agents observe only their local connections and infer global connectivity from these observations. Low-degree agents update to place higher probability toward a sparse network, while high-degree agents update toward a dense one. Crucially, agents internalize that others are forming inferences from their own observations. High-degree agents expect others to believe the network is dense, while low-degree agents expect the opposite. These higher-order beliefs translate into systematic differences in equilibrium behavior. Under strategic complements, high-degree agents exert higher actions and low-degree agents exert lower actions. An analogous pattern holds under substitutes. Importantly, the gap between low- and high-degree behavior exceed the one implied by homogeneous-belief benchmarks. The example illustrates how heterogeneous local observations generate correlated belief hierarchies and amplify behavioral differences through Iterative Belief Centrality.

We proceed to show that the mechanism illustrated by the example extends beyond that specific environment. We provide general conditions under which heterogeneous incomplete information generates systematic differences in behavior. When higher realized connectivity shifts beliefs toward higher connectivity throughout the belief hierarchy and this ordering is preserved under conditioning, beliefs propagate monotonically through the network. Higher-degree agents expect stronger spillovers not only directly, but at every distance, because they believe others also face stronger spillovers. Under strategic complements, this \textit{monotone propagation of beliefs} yields monotone differences in equilibrium actions. The theorem thus generalizes the example by showing when heterogeneous local information leads to persistent and ordered differences in perceived influence and behavior.

To further clarify the role of beliefs, we study when two agents belief hierarchies coincide and thus exert the same equilibrium action. For two agents to behave identically, they must not only occupy locally similar positions, but also have similar information and beliefs about how the network could have formed and what others know. We show that this occurs when agents occupy \textit{isomorphic local neighborhoods} and hold \textit{isomorphic priors}---local neighborhoods and priors that do not distinguish between relabeled versions of the same structure. 

The importance of this result is two-fold. First, in the  environment where agents observe only their direct connections, we show that under isomorphic priors, agents with the same degree exert the same action. Degree therefore becomes a sufficient statistic for perceived influence and behavior, not because it captures realized influence, but because symmetry collapses belief hierarchies. This sharply contrasts with the complete-information benchmark, in which degree alone cannot generically summarize influence or behavior. Second, this result provides a foundation for degree-based models of incomplete information. In our framework, degree-based behavior is not imposed, rather, it emerges as a special case in which belief iteration collapses. Moreover, because we build priors over networks rather than degree distributions, the induced degree-based representation of equilibrium actions inherits a correlation structure. Thus, even when agents with the same degree choose the same action, agents of different degrees can hold heterogeneous beliefs about structure beyond their neighborhood, and these beliefs can be correlated. This allows us to \emph{quantify} the roles of degree correlation and belief heterogeneity in shaping equilibrium behavior---features that previous theories have yet to address.\footnote{While degree distributions may exogenously be endowed with a correlation structure, one does not need to be specified to define them. Existing models of incomplete information are build on degree-distribution priors, but all rely on the assumption of degree independence to make predictions. Our approach not only shows that behaviors based on degree-distributions are emergent from a more general information structure, but also provides a rational for where degree correlation can arise from and how it impacts behaviors.}

What allows us to quantify these effects is a tractable characterization of equilibrium actions in terms of a \emph{belief transition matrix}: a kernel of conditional degree distributions describing how agents of a given degree believe they are connected to others. This matrix summarizes how local observations map into perceived influence and how beliefs propagate through higher-order expectations. Using this representation, we show that heterogeneous beliefs can generate equilibrium action profiles that differ sharply from homogeneous benchmarks. For instance, actions need not be monotone in degree. More importantly, the transition matrix also determines the persistence of belief correlations through higher-order belief iteration. This in turn governs how belief heterogeneity aggregates, shaping both the level and dispersion of equilibrium actions. We capture this persistence using a family of \emph{path assortativity} measures and show that all such effects are bounded by the mixing rate of the belief transition matrix. Slow mixing allows belief differences to echo and amplify, increasing both average behavior and the variance of behaviors, while fast mixing causes beliefs and behavior to converge rapidly.

These results provide a new explanation for the empirical phenomena described above. Consider for instance the issue of  civic participation and political efficacy. The literature emphasizes that participation enhances citizens' sense of efficacy---their belief that actions influence outcomes. In our framework, this corresponds to environments in which individuals perceive high assortativity and slow mixing. Similar neighborhood expectations raise path assortativities and amplify actions and spillovers. Societies with slower mixing should therefore exhibit more persistent participation. The same mechanism helps explain digital paradoxes. Biased sampling exposes users to high-activity, high-degree agents, making the perceived kernel size-biased. Our results then reproduce the virality paradox: actions and beliefs are pulled toward what looks common locally, increasing average behavior and dispersion.

\subsection*{Related Literature}

A canonical example of biased perceptions in networks is the friendship paradox \citep{feld1991your}, which implies that local samples can be systematically non-representative of the broader network.  The most connected individuals in a network tend to be observed by more individuals.   If people do not account for this bias, it can  translate into systematic differences in behavior across people and from settings with complete information about the network \citep{jackson2019,jackson2020human}.
Our analysis here shows how heterogeneous behavior occurs even without any bias in beliefs.    Our uncertainty points to a different sort of effect on behaviors:  leading to increased heterogeneity, and as a function of position, rather than raising all behaviors.  This is a quite complementary effect.  

As mentioned above, most models of incomplete-information network games are degree-based. In these models, agents have homogeneous beliefs about the remainder of the network and differ only in their own degree. A central implication is that equilibrium behavior is tightly linked to degree. In games with either strategic complementarities or substitutabilities, actions are monotone in degree, and both the mean and variance of actions are determined by the corresponding moments of the degree distribution.

Our approach departs from this literature by allowing agents to update beliefs from partial observations of the network itself. This yields a characterization of equilibrium behavior under general information structures and priors over networks, rather than over degree distributions. While degree-based behavior emerges as a special case under isomorphic priors, the framework permits these degree-based beliefs to be network-dependent and correlated across agents. As a result, equilibrium behavior can differ qualitatively from standard degree-based predictions. Actions need not be monotone in degree, and aggregate outcomes depend not only on degree moments but also on the mixing rate of the belief transition matrix that governs how beliefs propagate.

Two previous studies on games of complements with heterogeneous beliefs and misperceptions are \citet{jackson2019}, who examines how projecting connectivity from local neighborhoods to the rest of the network influences behavior, and \citet{frick2022dispersed}, who study how beliefs about correlations of connectedness can lead to biased behavior. While these studies provide important examples of how perceptions influence behavior, we provide a more general theory of the relationship between perceptions and networked behavior. In these prior studies, misperceptions arise from mistaken extrapolation. Here we show that misperceptions can arise even under fully rational Bayesian updating, and that heterogeneity in network positions can amplify their behavioral consequences.

Closer to our model is \citet{breza2018seeing}, who study a quadratic game in which agents lack complete information regarding the network. In their model, however, agents' expectations regarding the rest of the network (beyond friends) are independent of their information, which precludes the belief heterogeneity we study. By contrast, we show how local connectivity provides information about the broader network and systematically influences behavior in ways that we can characterize.

\citet{golub2020expectations} study how higher-order beliefs propagate through a known network to shape stable conventions, showing that nested expectations are key drivers of equilibrium outcomes in coordination environments. Our notion of Iterative Belief Centrality is closely related to the equilibrium characterization in their setting. In both frameworks, equilibrium behavior arises from iterating beliefs about others, and others' beliefs about others, etc. through the network. The key distinction lies in the uncertainty. In \citet{golub2020expectations}, the network structure is common knowledge and the authors explore agents' higher-order beliefs about others' actions. By contrast, our agents can be uncertain about the network itself and infer structure from their limited local information. As a result, belief iteration in our model is built from perceptions of the network, generating heterogeneous belief hierarchies even under full rationality. Iterative Belief Centrality therefore extends the logic of belief iteration to environments in which beliefs are determined by a network that is itself uncertain; and this introduces systematic differences in agents' beliefs based on their positions, which are the focus of our investigation.

Other work introducing incomplete information into network games is less related. \citet*{lambert} show existence and generic uniqueness of equilibria in quadratic games with general information structures, but do not address how uncertainty about the network itself maps to behavior. \citet{de2015network} study a linear-quadratic game in which agents lack information regarding parameters other than the network (e.g., complementarity strength and returns), whereas we focus on incomplete information about the network and how it impacts beliefs differentially.

\citet{ruiz2021network} propose a model of how agents form mental representations of network structure and examine effects on equilibrium existence and welfare. Unlike their work, our focus is primarily on behavior: we relate network uncertainty to observable behavioral patterns under general belief structures and characterize how variation in perceptions translates into systematic differences in actions.

The rest of the paper is structured as follows. In Section \ref{sec:model} we provide our model. In Section \ref{sec:equi} we  characterize equilibrium actions and provide conditions for monotone belief propagation. In Section \ref{sec:sym} we analyze when belief hierarchies coincide and identify conditions for Iterative Belief Centrality to collapse to degree-based behavior. In Section \ref{sec:hete} we discuss the role of belief heterogeneity and correlation. Section \ref{sec:gen} extends our model to allow for arbitrary network information. In Section \ref{sec:con} we offer concluding remarks. The proofs of our main results are contained in the Appendix. Additional discussion, examples, and omitted proofs are contained in the Online Appendix.

\section{Model}
\label{sec:model}


We model incomplete information using the canonical \cite{harsanyi} approach: Nature draws a profile of preference types---here essentially a network---and agents observe parts of it, update beliefs, and choose actions. 
In particular, a useful novelty in our approach is that 
uncertainty over preference parameters corresponds with uncertainty over networks, allowing us to speak interchangeably about preference parameter uncertainty and network uncertainty.

\subsection{Networks}

There is a finite set of agents $N=\{1,2,...,n\}$.

A \emph{network} (or \emph{graph}) $\mathbf{g}$ is represented by an adjacency matrix
$\mathbf{g}=[g_{ij}]\in\mathbb{R}^{n\times n}_+$, where $g_{ij}>0$ if a link
from $i$ to $j$ exists, and $g_{ij}=0$ otherwise. 

Note that links may be directed and weighted.

We let $\mathbf{g}_{i}$ denote the $i^{th}$ row of $\mathbf g$, i.e.\
$\mathbf{g}_{i}=(g_{i1},g_{i2},....,g_{in})$.

We let $\mathcal{G}_{n}$ be the set of possible networks, and for
convenience take this to be a finite set.  This is without loss of
generality if links are $0,1$, but helps in avoiding measure theoretic
definitions in the case where we allow general weights. 

The \emph{neighborhood} of agent $i$ in network $\mathbf g$ is $N_i(\mathbf g)=\{j\in N:\text{ }g_{ij}>0\}$,
and the cardinality of this set is $i$'s \emph{degree}:
$d_i(\mathbf g)\equiv\vert N_i(\mathbf g)\vert$.

\subsection{Payoffs}

Each agent $i$ chooses an action $a_i\in\mathbb R_+$.
Following \citet{ballester2006s}, payoffs as a function of actions are 
\begin{equation}
u_i(a_i,\mathbf a_{-i},\mathbf g)
=
a_i-\frac{1}{2}c\,a_i^2+\lambda a_i\sum_{j\in N} g_{ij}a_j
\label{eq:game}
\end{equation}
with $c>0$. Actions are strategic complements if $\lambda>0$ and substitutes if $\lambda<0$.

Note that arbitrarily heterogeneous complementarities are allowed, since we admit $g_{ij}$ that is any nonnegative real number.  Thus, $\lambda$ is redundant, but still useful for doing comparative statics.

\subsection{Types and Information}

Following \cite{harsanyi}, each agent $i\in N$ has a finite set of ``types'' $T_i$ that capture their information about the network and other agents' types, etc.  Here a type relates to what an agent knows about the network, knows about what others know about the network, what others know about what they know about the network, etc.
We construct these type sets as follows.

For each network $\mathbf g\in\mathcal G_n$, let
\[
\mathcal{A}_{i}(\mathbf g)\subseteq N
\]
denote the set of agents whose connections are observed by agent $i$
when the realized network is $\mathbf g$. 

For instance, if $\mathcal{A}_i(\mathbf{g})=\{i\}$ for all $i$, then agents observe only their own links. 
If $\mathcal{A}_i(\mathbf{g})=N$ for all $i$, then agents observe the entire matrix, corresponding to complete information (i.e., the \cite{ballester2006s} setting).

For now, we assume that $i\in\mathcal{A}_{i}(\mathbf g)$
for all $\mathbf g\in\mathcal G_n$, so that each agent knows who they are connected to.

The collection  $\left\{ \mathcal{A}_{i}(\cdot)\right\} _{i\in N}$, defines the information structure and is common knowledge.

Agent $i$'s type induced by a network $\mathbf g$ is 
\[
t_i(\mathbf g):=\big(\mathbf g_j\big)_{j\in \mathcal A_i(\mathbf g)},
\]
which consists of the rows of $\mathbf g$ corresponding to the agents observed by $i$.
The type set of agent $i$ is
\begin{equation}
T_i:=\{t_i(\mathbf g):\mathbf g\in\mathcal G_n\},
\label{eq:types}
\end{equation}
and the type space of the game is $T\equiv\prod_{i\in N}T_i$.

The following figure illustrates an example of the information structure. Panel (a) shows the realized network $\mathbf{g}$, in which player 2 is assumed to observe only its neighbors $\mathcal{A}_2(\mathbf{g})=\{2\}$ while player 4 observes both its neighbors and their neighbors $\mathcal{A}_4(\mathbf{g})=\{2,4\}$.
\begin{figure}[h!]
    \centering

    \begin{subfigure}[b]{0.32\textwidth}
        \centering
        \begin{tikzpicture}[scale=0.9,line width=1pt]
            \Vertex[label=2,opacity=0.2]{B}
            \Vertex[x=2,label=4,opacity=0.2]{D}
            \Vertex[x=-2,y=1,label=1,opacity=0.2]{A}
            \Vertex[x=-2,y=-1,label=3,opacity=0.2]{C}
            \Edge(B)(D)
            \Edge(B)(C)
            \Edge(B)(A)
            \Edge(A)(C)
        \end{tikzpicture}
        \caption{Realized graph}
    \end{subfigure}
    \hfill
    \begin{subfigure}[b]{0.32\textwidth}
        \centering
        \begin{tikzpicture}[scale=0.9,line width=1pt]
            \Vertex[label=2,opacity=0.2]{B}
            \Vertex[x=2,label=4,opacity=0.2]{D}
            \Vertex[x=-2,y=1,label=1,opacity=0.2]{A}
            \Vertex[x=-2,y=-1,label=3,opacity=0.2]{C}
            \Edge(B)(D)
            \Edge(B)(C)
            \Edge(B)(A)
            \Edge[label=\large$?$,position=left,style={dashed}](A)(C)
            \Edge[label=\large$?$,position={above=2mm},style={dashed}](A)(D)
            \Edge[label=\large$?$,position={below=2mm},style={dashed}](C)(D)
        \end{tikzpicture}
        \caption{Player 2's type}
    \end{subfigure}
    \hfill
    \begin{subfigure}[b]{0.32\textwidth}
        \centering
        \begin{tikzpicture}[scale=0.9,line width=1pt]
            \Vertex[label=2,opacity=0.2]{B}
            \Vertex[x=2,label=4,opacity=0.2]{D}
            \Vertex[x=-2,y=1,label=1,opacity=0.2]{A}
            \Vertex[x=-2,y=-1,label=3,opacity=0.2]{C}
            \Edge(B)(D)
            \Edge(B)(C)
            \Edge(B)(A)
            \Edge[label=\large$?$,position=left,style={dashed}](A)(C)
        \end{tikzpicture}
        \caption{Player 4's type}
    \end{subfigure}

\end{figure}

Before proceeding, we note that this structure of types where agents know their own row of connections plus possibly the rows of some other agents, 
is \emph{with} loss of generality.   Nonetheless, as will become clear, it is not necessary for the results.  However, it helps simplify the exposition and many expressions. In Section \ref{sec:gen} we present a more general formulation that admits arbitrary network information and nests this row-based formulation.

\subsection{Beliefs and Compatibility}

In our game, players choosing actions with limited
information about the network means that they have incomplete incomplete
information about the preference parameters, $g_{ij}$, of other players.
We now describe how to represent this uncertainty, and
how we can guarantee that uncertainty over preferences mirrors uncertainty over networks.

Let $p\in\Delta(T)$ denote the probability distribution over the type space, where
$\Delta(T)$ denotes the set of probability distributions over $T$. Nature moves
first and draws a type profile $\mathbf t:=(t_1,\ldots,t_n)\in T$.

For any $t_i\in T_i$, define the set of networks consistent with $t_i$ by
\[
g(t_i):=\{\mathbf g\in\mathcal G_n:\ t_i=t_i(\mathbf g)\}.
\]
For any $\mathbf t\in T$, let
\[
g(\mathbf t):=\bigcap_{i\in N} g(t_i).
\]
be the networks consistent with the type tuple $\mathbf{t}$.  
Note that
given that $t_i$ includes at least $i$'s row, the network is completely tied down by the vector of types, and so $g(\mathbf t)$  is either a unique network or empty.\footnote{Suppose $\mathbf{g},\mathbf{g}^{\prime}\in g\left(\mathbf{t}\right)\neq\emptyset$, then $N_i\left(\mathbf{g}\right)=N_i\left(\mathbf{g}^{\prime}\right)$ for all $i\in N$, which implies $\mathbf{g}=\mathbf{g}^\prime$. Thus $g\left(\mathbf{t}\right)=\mathbf{g}$ or $g\left(\mathbf{t}\right)=\emptyset$.}


We say that $p\in\Delta(T)$ is \emph{compatible} if $p(\mathbf t)=0$ whenever $g(\mathbf t)=\emptyset$.  Note that this implies that $p(\mathbf t)>0$ implies $|g(\mathbf t)|=1$.
Let $\Delta_A(T)$ denote the set of all such priors.

\begin{lem}
\label{lem:admissibility}
The map $\psi:\Delta_A\left(T\right)\rightarrow\Delta\left(\mathcal{G}_n\right)$ defined by 
\[
\left(\psi p\right)\left(\mathbf{g}\right) = p\left(\boldsymbol{t}\right)
\]
such that $g(\mathbf t)=\{\mathbf g\}$, is a bijection between compatible preference priors $\Delta_A(T)$ and
network priors $\Delta(\mathcal G_n)$. Equivalently, if 
\(p\in \Delta_A(T)\) and \(g(\boldsymbol{t})=\{\textbf{g}\}\), then $
\hat p(g)=p(t)$ where $\psi\left(p\right) = \hat{p}$.
\end{lem}

The lemma shows that compatibility creates a one-to-one correspondence between preference uncertainty and network uncertainty. Because compatible type profiles uniquely identify a network, each realization of preferences parameters corresponds to one feasible network, and each network induces a unique compatible type profile. As a result, any compatible preference prior can be represented uniquely as a prior over networks, so Nature can be viewed as choosing a network, with agents observing partial information about it.

\subsection{Belief Updating}

Under compatibility, agents know that Nature draws a network $\mathbf g\in\mathcal G_n$
according to $\hat p$, which implies a corresponding type profile. An agent $i$ with
realized type $t_i$ updates beliefs according to Bayes' rule, which can then equivalently be thought of as updating over types or networks. 

\subsection{Equilibrium}

A strategy for agent $i$ is a function $a_i:T_i\to\mathbb R_+$.

An equilibrium is a strategy profile $a=(a_1,\ldots,a_n)$ such that for all $i$ and $t_i$,
\begin{equation}
a_i(t_i)\in\arg\max_{x\ge0}\text{ }
x-\frac{c}{2}x^2
+\lambda x \sum_{j\in N} g_{ij}^{t_i}\,
\mathbb E_p\!\left[\left. a_j(t_j)\ \right|\ t_i\right],
\label{eq:equilibrium2}
\end{equation}
where $g_{ij}^{t_i}$ denote the value of $g_{ij}$ when agent $i$ is of type $t_i$.

\subsection{Existence and Uniqueness of Equilibria}

An equilibrium is characterized by a fixed point of the system in \eqref{eq:equilibrium2}.
Let
\[
d^*
=
\max_{\mathbf g\in \mathcal G_n,\ i\in N}\sum_{j\in N}g_{ij}
\]
be the maximum possible row sum of feasible adjacency matrices. For example, if $\mathcal G_n$ are all $0,1$
networks and contains the complete simple graph, then $d^*=n-1$.   More generally, $d^*$ could be larger or smaller than $n-1$ depending on the number of connections agents tend to have and how large the weights on those connections are.  

\begin{lem}
\label{lem:exist}
There exists a unique and interior pure strategy equilibrium for
$|\lambda|\in\left[0,\frac{c}{d^*}\right)$.
\end{lem}

Note that in the case of complements, if for any possible row sum $\mathcal G_n$ also
contains a regular network corresponding to that row sum, then
$d^*=\max_{\mathbf g\in\mathcal G_n}\rho(\mathbf g)$, where $\rho(\mathbf g)$ is the
spectral radius of $\mathbf g$. In this case, the bound on the local complementarity
parameter $\lambda$ in Lemma~\ref{lem:exist} is the  analog of the complete-information
bound.

The linear-quadratic game played on networks exhibits direct and indirect complementarities
or substitutabilities (henceforth \textit{modularities}). 
The maximal modularity from direct links is
$\lambda c^{-1} d^*$. Iterating then yields $(1-\lambda c^{-1}d^{*})^{-1}$ as the maximal
modularity from the network.

\section{Equilibrium Structure and Analysis}
\label{sec:equi}

In this section, we characterize equilibrium behavior. 
Unlike the complete information environment, agents do not necessarily behave based on their realized influence in the network, but on their \emph{perceived influence} formed from their local information and the belief updating it induces.

\subsection{Iterative Belief Centrality}

Recall that $a_i(t_i)$ denotes the equilibrium action of agent $i$ whose realized
type is $t_i\in T_i$. 

Let $\kappa=\lambda c^{-1}$. The first-order condition associated with
\eqref{eq:equilibrium2} implies that
\begin{equation}
a_i(t_i)
= c^{-1}
+\kappa
\sum_{j\in N}
g_{ij}^{t_i}\,
\mathbb E_p\!\left[\left. a_j \right| t_i \right],
\label{eq:equilibriumFOC}
\end{equation}
Iterating on \eqref{eq:equilibriumFOC} yields the following characterization.

For any $s\in\mathbb N_+$, let $j_1,j_2,\ldots,j_s$ denote an arbitrary sequence of agents.

\begin{thm}
\label{thm:character}
If $|\lambda|\in\left[0,\frac{c}{d^*}\right)$, then for any realized type profile
$\mathbf t\in T$, equilibrium actions satisfy
\begin{equation}
a_i(t_i)
= c^{-1}\left(1+
\sum_{s=1}^{\infty}
\kappa^{s}
\beta_{i,t_i}^{(s)}\right)
\quad
\forall\, i\in N,\ \forall\, t_i\in T_i,
\label{eq:BNE-action}
\end{equation}
where
\begin{equation}
\beta_{i,t_i}^{(s)}
=
\sum_{j_1,\ldots,j_s=1}^{n}
\;
\sum_{\{t_{j_k}\in T_{j_k}:1\le k\le s\}}
g_{ij_1}^{t_i}
g_{j_1j_2}^{t_{j_1}}
\cdots
g_{j_{s-1}j_s}^{t_{j_{s-1}}}
\,
p(t_{j_s}\mid t_{j_{s-1}})
\cdots
p(t_{j_1}\mid t_i).
\label{eq:BNE-walk}
\end{equation}
\end{thm}

\medskip

The expression in \eqref{eq:BNE-action} is a standard multiplier expansion formula.
Each term $\kappa^s \beta_{i,t_i}^{(s)}$ captures the contribution of indirect
effects operating at distance $s$, so that equilibrium actions are obtained by
aggregating the direct effect and all higher-order spillovers. In contrast to
standard settings, these higher-order effects are not determined by the realized
network alone, but also by how agents evaluate the strength of interactions across
successive layers of beliefs about the network.

To make this distinction precise, we compare to the complete information
Nash equilibrium of \cite{ballester2006s}:
\[
a_i^{c}(\mathbf g)
=c^{-1} \left(1+
\sum_{s=1}^{\infty}
\kappa^s
\left[
\sum_{j_1,\ldots,j_s=1}^{n}
g_{ij_1}
g_{j_1j_2}
\cdots
g_{j_{s-1}j_s}
\right]\right).
\]
Each term in brackets represents a walk of length $s$
originating from agent $i$. Each walk captures the strategic complementarity or substitutability (\textit{modularity}) generated by a particular sequence of links connecting $i$ to $j_s$. Taking all such walks into account, this shows that under complete information, agents exert action in proportion to how strategically influential they are in the network. This corresponds to their Katz--Bonacich centrality, which measures agents \emph{structural} influence.

Under incomplete information, agents do not observe the full network and only have beliefs about their influence. Nonetheless, agents are aware that they are embedded in a network and internalize the existence of walks of arbitrary length. Since these
walks determine the magnitude of strategic spillovers, agents form expectations about their strength. In Theorem~\ref{thm:character}, these are captured by terms $\kappa^s\beta_{i,t_i}^{(s)}$ which measure the expected modularity from all walks of length $s$. These  expectations are taken with respect to both agents' beliefs about both the structure of the network and others' beliefs. Specifically, agents internalize the uncertainty about how others perceive the network and their strategic position within it generating higher-order belief effects. Equilibrium behavior therefore reflects \emph{perceived influence formed through an iterative hierarchy of beliefs}. We refer to this object as \emph{Iterative Belief Centrality}.%
\footnote{Of course, the usual discussion applies:  it is clear that in practice people do not perform such calculations in making behavioral decisions.  This includes the complete information benchmark, which is already exponentially complex. Just as with any  equilibrium analysis, this provides an ``as if'' modeling device, and also identifies stable points from which no individual would want to deviate given the information available to them.   Any other vector of actions would be unstable.} 

\medskip

Before proceeding, several remarks are in order.

\begin{rem}
If $\hat p(\mathbf g)=1$ and $\hat p(\mathbf g')=0$ for all
$\mathbf g'\neq \mathbf g$, then
$a_i(t_i)=a_i^{c}(\mathbf g)$ for all $i$.
\end{rem}

Complete information corresponds to a degenerate belief environment.  Belief iteration collapses, and perceived influence coincides with actual influence.

\begin{rem} $\beta_{i,t_{i}}^{(s)}\neq \mathbb{E}\!\left(\sum_{j_{1},j_{2},..,j_{s}=1}^{n} g_{ij_{1}}^{t_{i}}g_{j_{1}j_{2}}^{t_{j_{1}}}\cdots g_{j_{s-1}j_{s}}^{t_{j_{s-1}}}\right)$. \end{rem} 

\begin{rem} $\beta_{i,t_{i}}^{(s)}\neq \mathbb{E}\!\left(\left. \sum_{j_{1},j_{2},..,j_{s}=1}^{n} g_{ij_{1}}^{t_{i}}g_{j_{1}j_{2}}^{t_{j_{1}}}\cdots g_{j_{s-1}j_{s}}^{t_{j_{s-1}}} \right|\;t_i\right)$. \end{rem} 

 The second remark emphasizes that equilibrium is \emph{interim}: beliefs are updated from local observations, so expected influence does not equal its ex ante expectation. The third remark states that merely internalizing ones own type when forming expectation of walks does not capture the higher-order belief structure in \eqref{eq:BNE-walk}; Iterative Belief Centrality weights walks by the likelihood of encountering types that themselves form different expectations downstream.\footnote{A stark illustration arises in an example of core--periphery priors, where some agents can infer the full network while others cannot; even those who learn the realized network do not behave as they would have under complete information because they internalize the fact that others remain uncertain. Details are presented in the Online Appendix.}

\begin{rem}
Equilibrium actions can be non-monotonic in agents' degrees.
\end{rem}

This mirrors the complete-information version of the game. While equilibria may fail to be monotonic in degrees, there are many natural applications in 
which behavior is monotonic in degree. 
We identify a general condition for this to occur in
Section~\ref{sec:generalized_example}.

Before that, we illustrate why we care about
monotonicity, and some of the powerful implications of our model in an example.

\subsection{An Example}
\label{sec:example}

Consider an example in which the probability structure is such that
the network is randomly formed according to an Erdos-Renyi random graph: an undirected unweighted network
in which each pair of agents is connected independently with some probability $p$.  However, 
$p$ can take on either a high or low value.  

Once the $p$ is chosen by nature and the network is realized, agents see their own connections before choosing an action, but do not know the rest of the network or the $p$.  
The key aspect of our model is that agents make inferences about $p$ from what they see in the network.

In particular,  suppose
that Nature first chooses a probability $p\in\{p_{H},\,p_{L}\}$ with
$\mathbb{P}(p=p_{H})=\alpha\in[0,1]$ and $p_{H}>p_{L}$. Following
the realization of $p$, a network is produced according to an Erdos-Reyni
(ER) network formation process with its chosen $p$. Once the network
is realized, agents observe their direct connections and nothing more; i.e., $\mathcal{A}_{i}\left(\mathbf{g}\right)=\{i\}$.

A key insight about equilibrium behavior is the following. 
An agent with a
higher realized degree places greater posterior weight on $p_H$ and therefore expects
others to have higher degrees, while an agent with a lower realized degree places less weight on $p_H$ and more weight on $p_L$. 
Importantly, this logic iterates: a higher-degree agent
expects neighbors to also have updated toward $p_H$, and expects those neighbors to expect higher degrees among their own neighbors, and so on. 

This mechanism can be understood as a \emph{belief propagation} process, whereby an agent's locally observed degree shapes their expectations about the beliefs held by other agents further away. In this example, belief propagation is \emph{monotone}: agents with higher  realized degrees systematically assign higher beliefs about the degrees of others throughout the network.

Because actions are increasing in expected neighbors' actions under strategic complements, this monotone propagation of beliefs implies monotone differences in actions. Higher-degree agents choose higher actions because they expect
stronger spillovers, both directly and through the entire network and hierarchy of beliefs. 
Similarly, low-degree agents reason in the opposite direction. 

\begin{figure}[H]
\begin{centering}
\begin{minipage}[t]{0.45\columnwidth}
\begin{center}
\includegraphics[scale=0.6]{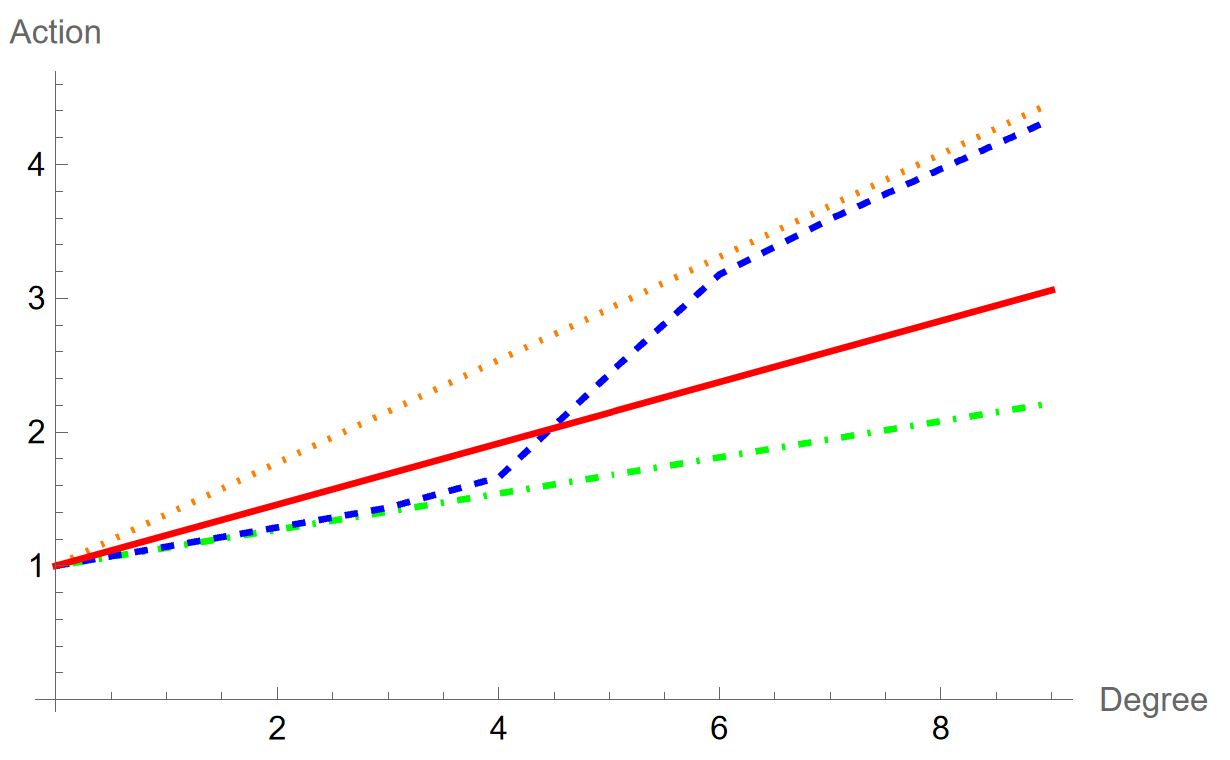} 
\par\end{center}%
\end{minipage}\hfill{}%
\begin{minipage}[t]{0.45\columnwidth}%
\begin{center}
\includegraphics[scale=0.6]{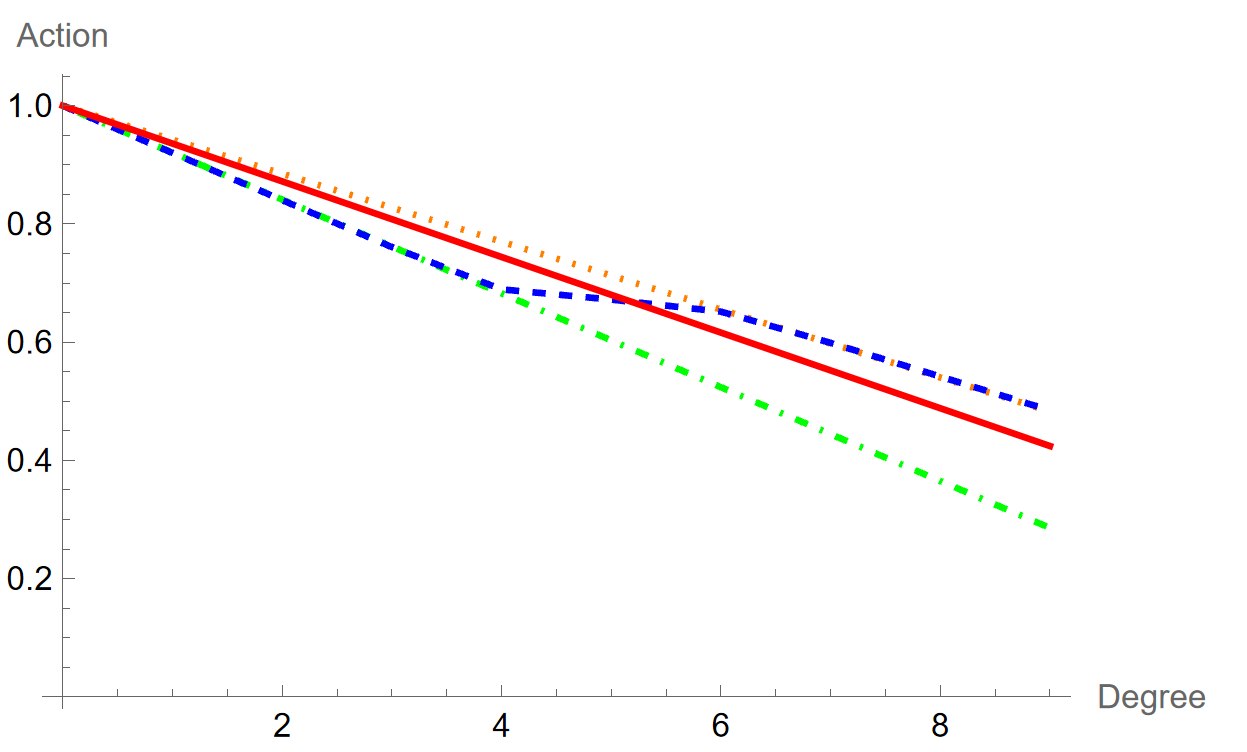} 
\par\end{center}%
\end{minipage}
\par\end{centering}
\caption{\label{fig1}Equilibrium actions under $\alpha=0$ (green dot-dashed), $\alpha=1$ (red dotted), $\alpha=0.3$ (blue dashed), and based on the average degree distribution (red solid). The left panel shows the case of strategic complements, and the right panel shows the case of substitutes.}
\end{figure}

Figure~\ref{fig1} illustrates equilibrium actions for different values of $\alpha$ and compares them to a degree-based benchmark in which agents only exert actions based on the unconditional degree distribution.\footnote{We revisit this example in Section \ref{sec:prof} where we provide a full derivation of equilibrium actions.}

The incomplete information equilibrium behavior is sandwiched between the extremes in which agents know with certainty whether the network is  dense or sparse. 
However, an additional key feature here is that equilibrium actions are {\sl not} linear in degree. 
This contrasts with the benchmark in which agents do not update about the network based on their degree but are still uncertain (the middle solid red line).  
Agents' also update about the structure of the network and hence about the incentives of their neighbors.   
Under strategic complements, low-degree agents choose lower actions and high-degree agents choose higher actions relative to the `average' degree-based benchmark. This nonlinearity reflects the fact that belief updating works at multiple levels and feeds back.  The example provides a simple illustration of how Iterative Belief Centrality results in amplified behavioral differences relative to degree-based benchmarks.

This mechanism is reminiscent of phenomena such as the friendship paradox \citep{feld1991your,jackson2019}, where beliefs about others' connections are biased by network structure. Here, however, uncertainty does not merely bias beliefs in one
direction; instead, it amplifies behavior in opposite directions for low- and high-degree agents because these agents form systematically different beliefs about the rest of the network.

This example contrasts with  
models of incomplete-information network games 
in which agents know the parameters governing network formation, just not the realized network. Such models presume that all agents hold the same belief regarding the density of the network, and therefore form \emph{homogeneous} beliefs hierarchies about how others are likely to be connected beyond their local neighborhoods. Here, however, we admit more general uncertainty. As the example makes clear, the \emph{heterogeneous}  hierarchies it generates leads to systematic and intuitive differences in behavior.

\subsection{Monotone Belief-Propagation}
\label{sec:generalized_example}

Next we generalize the mechanism illustrated in the previous
example to a wide and natural class of beliefs. 
We show that as long as agents' beliefs about the rest of the network update in a way that is monotone in their own degree, then belief-propagation is also monotonic which then leads agents with higher degrees to expect higher actions across the entire network, and agents with lower degrees to expect the opposite.

Throughout this section we focus on the case of first-order information, so that
$\mathcal A_i(\mathbf g)=\{i\}$ for all $i$ and all $\mathbf g\in\mathcal G_n$. 

Compatibility then allows us to write $N_i(\mathbf g)=N_i(t_i)$ and $d(\mathbf{g})=d_i(t_i)$.

\medskip
\noindent\textbf{Degree Order.}

Since types correspond to rows of the network, each type $t_i\in T_i$ induces a
degree $d(t_i)$. This allows us to order row types according to the magnitude of
their induced degree:
\[
t_i\preceq_d t_i' \quad\Longleftrightarrow\quad d(t_i)\le d(t_i').
\]
Similarly, for vectors of types $r=(t^{(1)},\ldots,t^{(m)})$ with $t^{(\ell)}\in T_{k_\ell}$,
define the coordinatewise degree order
\[
r\preceq_d r' \quad\Longleftrightarrow\quad
t^{(\ell)}\preceq_d t'^{(\ell)} \text{ for all } \ell.
\]

A function $\Phi:\prod_{\ell=1}^m T_{k_\ell}\to\mathbb R$ is
\emph{degree-monotone} if $r\preceq_d r'$ implies $\Phi(r)\le \Phi(r')$.
Intuitively, a degree-monotone function assigns higher values when each
component corresponds to a more connected position, so that increases in
degree, holding other components fixed, translate into higher outcomes.

We now define a natural condition on beliefs that generalizes the inference pattern of the example above.

\noindent\textbf{Row-Based Dominance}
\label{ass:RBD}

We say that a setting satisfies \emph{Row-Based Dominance} if 
for every agent $i$, any two row types $t_i,t_i'\in T_i$ with $d(t_i')>d(t_i)$, and for every
coordinatewise degree-nondecreasing function
\[\Phi:T_{j_1}\times\cdots\times T_{j_m}\to\mathbb R\]
defined on the rows of any finite
collection of other agents $\{j_1,\ldots,j_m\}$, it follows that
\begin{equation}
\mathbb E\!\left[\Phi\big(t_{j_1},\ldots,t_{j_m}\big)\,\middle|\, t_i'\right]
\;\ge\;
\mathbb E\!\left[\Phi\big(t_{j_1},\ldots,t_{j_m}\big)\,\middle|\, t_i\right].
 \label{eq:RBDeq}   
\end{equation}

Row-Based Dominance requires that an agent's own higher realized connectivity systematically shifts that agent's beliefs toward higher connectivity throughout the network. It is a natural
monotone likelihood ordering on beliefs so that whenever an agent observes a row type that induces a higher degree, their posterior expectations of any degree-increasing function of others' row types are weakly higher. Since agents exert action based on expected strategic spillovers along walks of arbitrary length, Row-Based Dominance implies that higher-degree realizations systematically shift these belief-based walks upward. Specifically, higher-degree agents expect stronger spillovers from neighbors, expect those neighbors to face stronger spillovers themselves, and so on. This gives the following result.

\begin{thm}
\label{thm:iterated_multiplier}
Suppose that $\lambda>0$, $\mathcal A_i(\mathbf g)=\{i\}$, and the setting satisfies Row-Based Dominance.
Consider any agent $i$ and two row types $t_i,t_i'\in T_i$ with $d(t_i')>d(t_i)$, $m\ge 1$,
and a labeled path $j_0=i,j_1,\ldots,j_m$. 
If for each $\ell=0,\ldots,m-1$, $p\!\left(j_{\ell+1}\in N_{j_\ell}(t_{j_\ell}) \mid t_i\right)>0$, then the equilibrium action expected at the $m$-step node is weakly higher under the higher-degree type:
\[
\mathbb E\!\left[a_{j_m}(t_{j_m})\,\middle|\, t_i'\right]
\;\ge\;
\mathbb E\!\left[a_{j_m}(t_{j_m})\,\middle|\, t_i\right].
\]
\end{thm}

Under strategic complements, monotone propagation of beliefs implies more extreme expected actions throughout the network. Agents with higher realized degrees expect stronger complementarities at every distance and hence higher actions, while lower-degree agents expect weaker complementarities and lower actions.\footnote{We note that under strategic substitutes belief propagation is still active, but is not monotone.}

Row-Based Dominance is closely related to the affiliation condition in \citet{galeotti2010network}, where agents' payoff-relevant types are degrees and higher realized degree makes higher neighbor degrees more likely. Row-Based Dominance plays a similar role, but for beliefs about network structure rather than degree alone. Since agents are uncertain about the network, types encode local connectivity, and Row-Based Dominance requires that higher realized connectivity shifts beliefs upward throughout the belief hierarchy, ensuring monotone belief propagation. In Section~\ref{sec:hete}, we show that under a special class of priors, Row-Based Dominance reduces to a degree-affiliation condition, recovering the assumption in \cite{galeotti2010network} as a special case.

In addition to establishing that higher realized degree leads agents to expect higher actions, Row-Based Dominance also allows for a comparison between equilibrium behavior under heterogeneous beliefs and  homogeneous-belief benchmarks used in degree-based models of incomplete information. To formalize this comparison, recall that in degree-based models, conditional on being linked, agents of different degrees hold the same beliefs about their neighbors' behavior. In such environments, realized degree affects behavior only through the number of links, not through beliefs about how influence propagates beyond the local neighborhood.

Let $a^{\star}$ denote the equilibrium action profile under a compatible prior satisfying Row-Based Dominance. For any row type $t_i$ with $d(t_i)>0$, define the expected action of a randomly selected neighbor by
\[
m(t_i)
\;:=\;
\mathbb E\!\left[a_j^{\star}(t_j)\,\middle|\, t_i,\ g_{ij}^{t_i}=1\right],
\]
and let $m(d):=\mathbb E[m(t_i)\mid d(t_i)=d]$ denote its degree-conditioned counterpart. Under homogeneous beliefs, this expectation is constant across degrees; denote the common value by $m^{H}$.

We now present a corollary that shows that the intuition from the our motivating example in Section \ref{sec:example} is quite general and holds for any network formation process in which more connected agents expect that other agents are more connected too.

\begin{cor}
\label{cor:cutoff_homogeneous}
Suppose that $\lambda>0$, $\mathcal A_i(\mathbf g)=\{i\}$, and the setting satisfies Row-Based Dominance.
If there exist degrees $d_1<d_2$ such that $m(d_1) \;<\; m^{H} \;<\; m(d_2)$, then there exists a cutoff degree $d^{\ast}$ such that
\[
\mathbb E\!\left[a_i^{\star}(t_i)\mid d(t_i)=d\right]
<
c^{-1}+\kappa d\, m^{H}
\quad\text{for all } d<d^{\ast},
\]
and
\[
\mathbb E\!\left[a_i^{\star}(t_i)\mid d(t_i)=d\right]
>
c^{-1}+\kappa d\, m^{H}
\quad\text{for all } d>d^{\ast}.
\]
\end{cor}

The corollary illustrates the power of the iterated belief structure and its implications.
Under homogeneous beliefs, agents of all degrees share the same expectation about their neighbors' behavior, so equilibrium actions scale linearly with degree. Under Row-Based Dominance, by contrast, higher degree agents have beliefs that are shifted upwards, not only on their neighbors' degrees, but on their neighbors' neighbors, etc., and all of those agents' beliefs.   As a result, the expected action of a neighbor becomes increasing in degree. When this degree-conditioned expectation crosses the homogeneous-belief benchmark, equilibrium actions must cross it as well. Lower-degree agents exert lower actions than those predicted by homogeneous-beliefs, while higher-degree agents exert more, with a cutoff degree separating the two regimes.

These results also have important implications for applications and policy. For instance, the corollary implies that informational interventions can affect behavior even without changing the network itself. Disclosing aggregate network statistics such as the degree distribution can compress differences in agents' perceptions of others' connectivity and thereby reduce the dispersion in behavior generated by belief amplification. 
Conversely, when agents rely only on local observations, heterogeneous perceptions naturally arise, leading individuals in
similar structural positions to behave quite differently depending on how they
infer the broader network.

This connects directly to the motivation of the paper. A large body of empirical evidence shows that individuals hold systematically different perceptions of the same network and that these perceptions shape behavior in settings ranging from organizational decision-making to online environments and collective action. Our results provide a theoretical mechanism for these patterns: differences in local observations generate differences in beliefs, which are then propagated through higher-order expectations and amplified in equilibrium behavior. Consequently, two environments with identical underlying networks can exhibit very different outcomes depending on how agents perceive their position and the
structure around them. This highlights that perceived influence, rather than structural position alone, is a key determinant of behavior in networked settings.

\section{Symmetries and the Collapse of Iterative Belief Centrality}
\label{sec:sym}

In Section \ref{sec:equi} we showed that equilibrium actions can be interpreted as
\emph{Iterative Belief Centrality}. Agents choose actions based on perceived influence
formed through an iterative hierarchy of beliefs about others' beliefs. We now ask a natural follow up question: \emph{when do these hierarchies coincide, leading agents to choose the same action?}

The underlying logic is straightforward.
Equilibrium behavior is determined by how agents perceive their position in the network,
and this perception is entirely shaped by what they observe and how they interpret it.
If two agents face the same informational situation, both in terms of what they see and how
they map these observations into beliefs about the rest of the network and others' beliefs, 
then they solve the same decision problem and behave identically. 

To formalize this idea, it is helpful to separate three interdependent factors that determine whether two
agents will behave identically: (i) the \emph{depth} to which they observe the network,
(ii) the \emph{local structure} they observe up to that depth, and (iii) the \emph{beliefs}
they form about the further unobserved network and about what other agents know and believe.

\medskip

\noindent\textbf{(i) Symmetry of observational depth.}
If one agent observes only their direct neighbors while another observes neighbors'
neighbors, then the basis on which belief hierarchies are formed are ex ante different,
and Iterative Belief Centrality need not coincide. This symmetry, therefore, requires that agents
have access to the same informational depth. To formalize this, let $N^m_i(\mathbf{g})$
denote the set of agents at distance no more than $m$ from $i$, with
$N^0_i(\mathbf{g})=\{i\}$. The information structure in which all agents observe the
network up to radius $m+1$ around themselves (and no more) is
\[
\mathcal{A}_{i}(\mathbf{g})=N^m_i(\mathbf{g}).
\]

\begin{figure}[H]
\begin{centering}
\begin{minipage}[t]{0.45\columnwidth}%
\begin{center}
\includegraphics[scale=0.95]{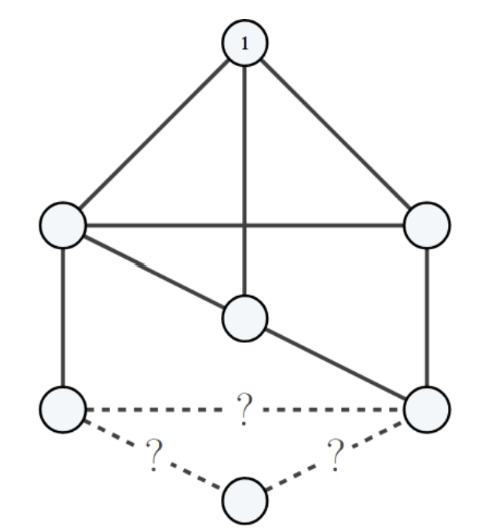}
\par\end{center}%
\end{minipage}\hfill{}%
\begin{minipage}[t]{0.45\columnwidth}%
\begin{center}
\includegraphics[scale=0.95]{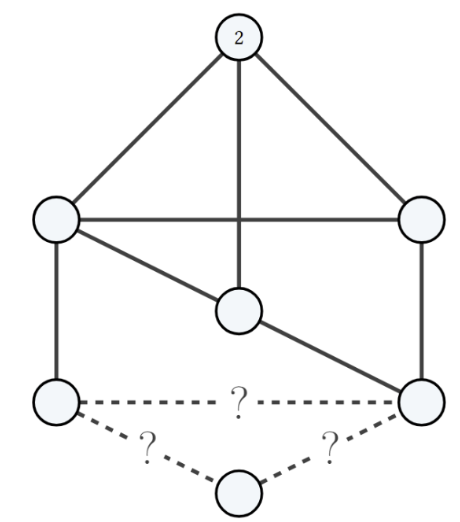}
\par\end{center}%
\end{minipage}
\par\end{centering}
\caption{Two agents $1$ (left) and $2$ (right) are isomorphic under second-order
information $\mathcal{A}_{i}(\mathbf{g})=N_i^1(\mathbf{g})$. Hence, the agents observe
their neighbors and each neighbor's neighbors.}
\label{fig:isomorphic}
\end{figure}

\noindent\textbf{(ii) Symmetry of observed local structure.}
Even when agents observe the network to the same depth, their realized observations may
differ. To capture when two agents face symmetric local environments, we focus on the
structure of the subgraphs they observe. Figure \ref{fig:isomorphic} illustrates the
relevant notion. Suppose that $\mathcal{A}_{i}(\mathbf{g})=N^1_i(\mathbf{g})$, so agents
observe their neighbors and their neighbors' neighbors. In the figure, agent 1 on the
left and agent 2 on the right are each directly connected to three neighbors. Moreover, within the range
of their observations, the observed pattern of connections among those neighbors is identical.
Although the agents fundamentally occupy different positions in the global network, their observed
local neighborhoods differ only in the labels of those nodes within it, i.e., their identities. This motivates our notion
of symmetry in local observation: two agents are symmetric if the subgraphs they observe
are \emph{isomorphic}.

Since compatibility implies that each realized type corresponds to a unique network, we
formalize this isomorphism in terms of the networks induced by types. We say that types
$t_{i}\in T_{i}$ and $t_{j}\in T_{j}$ are \textit{isomorphic} (or $t_{i}\simeq t_{j}$) if for any $\mathbf{g}\in g\left(t_i\right)$ and $\mathbf{g}^\prime\in g\left(t_j\right)$
\begin{itemize}
\item there exists $m\in \{0,\ldots, n-1\}$ such that
$\mathcal{A}_{i}(\mathbf{g})=N_i^m(\mathbf{g})$ and
$\mathcal{A}_{j}(\mathbf{g}')=N_j^m(\mathbf{g}')$; and
\item there is a bijection $f:N\to N$ with $f(i)=j$ such that
$f\!\left(N_k(\mathbf{g})\right)= N_{f(k)}(\mathbf{g}')$ for every
$k\in N^m_i(\mathbf{g})$.
\end{itemize}

\noindent\textbf{(iii) Symmetry of beliefs about the unobserved network.}
Isomorphism of observed neighborhoods ensures that two agents receive the same
\emph{local} information in terms of the pattern of connections they observe. However, Iterative Belief Centrality also depends on how they
extrapolate from what they see and how they think others extrapolate. Thus, two agents
will behave identically only if their posterior beliefs about the rest of the network,
as well as their beliefs about what others know, are also the same.

The next theorem provides sufficient conditions guaranteeing that isomorphic types
generate identical belief hierarchies and hence identical equilibrium behavior. Although
the statement is naturally expressed using isomorphisms, the economic content is simple:
agents who see the same local structure and have no ex ante reason to distinguish between
relabeled versions of the unobserved network form the same hierarchy of beliefs, and
therefore, have the same perceived influence.

\begin{thm}
\label{thm:same-action}
Suppose $\mathcal{A}_{i}(\mathbf{g})=N^m_i(\mathbf{g})$ and
$\mathcal{G}_{n}=\mathcal{G}_{n}^U$ is the set of all unweighted and undirected networks.
Let $\left\{ \left[\mathbf{g}^{1}\right],\left[\mathbf{g}^{2}\right],\ldots,
\left[\mathbf{g}^{q}\right]\right\} $ be the partition of $\mathcal{G}_{n}$ into its
isomorphism classes. If a prior $p\in\Delta_{A}(T)$ satisfies
$\hat{p}(\mathbf{g})=\hat{p}(\mathbf{g}^{\prime})$ whenever
$\mathbf{g},\mathbf{g}^{\prime}\in\left[\mathbf{g}^{k}\right]$, then if
$t_{i}\simeq t_{j}$ it follows that $a_{i}(t_{i})=a_{j}(t_{j})$.
\end{thm}

Isomorphic types ensure that agents are locally identical, while isomorphic priors guarantee that posterior beliefs and all higher-order beliefs are invariant under relabeling of unobserved nodes. The uniform distribution, $\hat{p}(\mathbf{g})=\frac{1}{|\mathcal{G}_{n}^{U}|}$ for each
$\mathbf{g}\in \mathcal{G}_{n}^{U}$, as well as Erd\H{o}s--R\'enyi random network formation,
are special cases of such isomorphic priors, but the condition is much more general. What
matters is that if some network is possible under the prior, then relabelings of that
network are equally likely. Otherwise, two agents with identical local observations could
rationally assign different probabilities to what the broader network might look like,
and their belief hierarchies, and hence Iterative Belief Centralities, could differ.

Besides providing sufficient conditions that lead to agent to take the same action, this theorem is the bridge between our general framework and degree-based models of incomplete information.  In the next subsection, we formally show how the symmetries identified here deliver a tractable degree-based representation of equilibrium behavior under first-order information thus establishing a microfoundation for degree-based behaviors.

\subsection{Degree Collapse}
\label{sec:symdeg}

We now study the consequences of Theorem \ref{thm:same-action} for the game in which
agents observe only their own direct connections and are endowed with isomorphic priors.
This environment corresponds to a case in which belief hierarchies collapse. Specifically,
agents cannot distinguish among networks beyond their local neighborhoods, and symmetry
implies that agents who see the same degree form the same hierarchy of beliefs about
others' behavior. As a result, Iterative Belief Centrality reduces to a function of
degree.

\begin{prop}
\label{prop:Degree-br}
In the first-order information environment under isomorphic priors, the system of best
responses characterizing equilibrium actions is given by
\begin{equation}
\boldsymbol{a}=c^{-1}\cdot\mathbf{1}+\kappa \boldsymbol{M}_{d}\boldsymbol{P}\boldsymbol{a},
\label{eq:D-FOC}
\end{equation}
where $\boldsymbol{a}=(a_{1},..,a_{n-1})$ tracks the equilibrium as a function of each possible degree
$1$ to $n-1$,\footnote{Note that the equilibrium actions of agents who have degree 0 are
always equal to $c^{-1}$.}
$\boldsymbol{M}_{d}$ is a diagonal matrix whose $u^{th}$ diagonal element is equal to
$d_{u}=u\in\{1,2,..,n-1\}$, and $\boldsymbol{P}$ is a row-stochastic probability kernel with
\[
P_{uv}=\sum_{t_{j}\in T_{j}}p(t_{j}\mid t_{i}^{u})\cdot\mathbb{I}\left\{
\sum_{r\in N}g_{jr}^{t_{j}}=v\right\}
\]
for any $i,j\in N$ with type $t_{i}^{u}\in T_{i}$ such that $g_{ij}^{t_{i}^{u}}=1$ and
$\sum_{r\in N}g_{ir}^{t_{i}^{u}}=u$.
\end{prop}

The result follows from the fact that when $\mathcal{A}_{i}(\mathbf{g})=\{i\}$ for all
$\mathbf{g}$ and $i$, an agent's observed neighborhood is a star network with that agent at the center. Two such observed
neighborhoods are isomorphic if and only if the degrees of their respective central nodes
are the same. Theorem \ref{thm:same-action} therefore implies that degree becomes
sufficient for characterizing equilibrium behavior. Under first-order information and
relabeling-symmetric priors, agents with the same degree have isomorphic types and thus
form identical belief hierarchies and choose the same action. Equation
(\ref{eq:D-FOC}) provides a tractable representation of those actions.

It is important to emphasize, however, that even though this representation is indexed
only by degree, agents are not merely reacting to degree mechanically. They are still
computing expected chains of influence in the sense of Theorem \ref{thm:character}. The
difference is that symmetry reduces these expected chains and the beliefs that give rise to
them to depend only on degree. In this sense, Proposition \ref{prop:Degree-br} identifies
an environment in which Iterative Belief Centrality \emph{collapses} to a simple
degree-based statistic.

The significance of this collapse lies in the contrast with the complete-information version of
the game, as well as with degree-based models of incomplete information. 

First, recall that under complete information, agents exert actions in proportion to their \emph{structural} inluence in the realized
network. This, however, cannot be summarized by degree alone.\footnote{Generically, the total number of walks an agent has in the network cannot be summarized by that agent's degree alone. Other statistics, such as the degree sequence and the number of cycles are also needed to account for these walks. Degrees can only summarize total walks in trivial network topologies such as regular networks, i.e., networks in which all agents have the same degree.} By contrast, Proposition
\ref{prop:Degree-br} shows that under incomplete information, there are instances in which \emph{perceived} influence can be fully summarized by degree. This is not because degree captures realized structural influence, but because symmetry collapses higher-order belief iteration.

Second, Proposition
\ref{prop:Degree-br} clarifies the relationship between our model and degree-based models of incomplete
information. In such models, degree is \emph{assumed} to be the relevant state variable. In our model,\emph{ degree-based
behavior arises endogenously} from a more general information structure. Once priors
assign equal probability to relabelings of compatible networks, and agents observe only
their local neighborhoods, the belief hierarchies that define Iterative Belief Centrality
coincide across agents of the same degree.

Finally, in addition to endogenizing degree-based behaviors, Proposition \ref{prop:Degree-br} provides a tractable framework for quantifying how belief heterogeneity and correlation translate to behavioral outcomes. These are dimensions to behavior that existing models have yet to quantify as they typically rely on the assumption that beliefs are homogeneous across degree types for tractability. Because our priors are defined over networks rather than directly over degree distributions, the induced belief kernel $\boldsymbol{P}$ can embed rich patterns of
belief heterogeneity and correlation across degree types. While agents of the same degree are identical, agents of different degrees can hold different beliefs about the degrees and the behavior of nodes beyond their neighborhoods, and these beliefs can be correlated. 
Equation (\ref{eq:D-FOC}) allows us to study how this belief heterogeneity directly translates to heterogeneous behavior, a topic which we explore next.

\section{Heterogeneity, Correlation, and Collective Outcomes}
\label{sec:hete}

We now explore how \emph{heterogeneity} and \emph{correlation} in beliefs shape equilibrium behavior. In Section \ref{sec:sym} we identified conditions under which agents' belief hierarchies coincide, so that Iterative Belief Centrality collapses to a degree-based representation. Here we maintain same environment---agents observe only their local neighborhoods and priors are isomorphic. 

\medskip

We start by expressing equilibrium behavior implied by equation (\ref{eq:D-FOC}). Denote by $i_{k}$ an agent who lies $k$ links away from an agent whose degree is equal to $i$, and let $\mu_{i}$ represent that agent's expectation regarding its direct neighbors' degrees:
\[
\mu_{i}:=\mathbb{E}_{i}[d_{i_{1}}]=(\boldsymbol{P}\boldsymbol{d})_{i}=\sum_{q}P_{iq}d_{q},
\]
where $\boldsymbol{d}=(1,..,n-1)$. Then, equilibrium actions can be expressed as
\[
a(\boldsymbol{P})=c^{-1}\sum_{s\ge0}\kappa^{s}\boldsymbol{g}^{(s)},\quad\text{where}\quad \boldsymbol{g}^{(0)}= \boldsymbol{1},\quad \boldsymbol{g}^{(1)}=\boldsymbol{d},\quad \boldsymbol{g}^{(s+1)}=\boldsymbol{M}_{d}\boldsymbol{P}\boldsymbol{g}^{(s)}.
\]
Equivalently,
\[
a_{i}=c^{-1}\left(1+\kappa d_{i}+\kappa^{2}d_{i}\mu_{i}+\kappa^{3}d_{i}\mathbb{E}_{i}\left[d_{i_{1}}\mu_{i_{1}}\right]+\kappa^{4}d_{i}\mathbb{E}_{i}\left[d_{i_{1}}\mathbb{E}_{i_{1}}\left[d_{i_{2}}\mu_{i_{2}}\right]\right]+...\right)
\]

This expansion makes the connection to Iterative Belief Centrality and its connection to the belief-propagation mechanism transparent. The baseline term $d_i$ captures the direct effect of local connections, while $\mu_i$ captures a first-order belief about neighbor influence. The nested expectations then describe how perceived influence propagates beyond the first neighborhood through the belief hierarchy.

More importantly, the expansion shows that once belief hierarchies collapse to degree, belief heterogeneity also admits a simple form. While under general priors, row-based identity-dependent beliefs may lead to agents perceive the network very differently, under degree collapse, belief heterogeneity also collapses. Belief differences are summarized by $\mu_i$, and the kernel $\boldsymbol{P}$ governs how these differences propagate through higher-order beliefs. In this sense, $\mu_i$ acts as a first order belief input that can differ across degree types, and  $\boldsymbol{P}$ then determines how these inputs are transmitted as beliefs iterate.

This allows for a clear classification of equilibrium behaviors into two regimes. The \textit{homogeneous} regime is one in which $\mu_{i}=\mu_{j}$ for all $i\neq j$, and the \textit{heterogeneous} regime in which these expectations differ. Under homogeneity, beliefs about others are symmetric and wash out. Specifically, all nested expectations collapse to the same constant and each agent's behavior becomes a scaled version of degree. Under heterogeneity, initial differences in $\mu_i$ act like belief shocks that are transmitted through the network as beliefs iterate. The structure of $\boldsymbol{P}$ determines how  such belief differences are reflected in equilibrium behavior.

The following lemma will be useful in formalizing and quantifying these ideas.

\begin{lem}
\label{lem:revers}
Let $\delta:\{0,1,\ldots,n-1\}\to[0,1]$ such that for any $u\in \{0,1,\ldots,n-1\}$,
\[
\delta_{u}=\sum_{t_{i}\in T_{i}}p(t_{i})\cdot\mathbb{I}\left\{ \sum_{r=1}^{n}g_{ir}^{t_{i}}=u\right\}.
\]
Moreover, let $d$ be a random variable on $\{0,1,\ldots,n-1\}$ with $\mathbb{P}(d=u)=\delta_u$ and define
\[
\pi_{u}=\frac{u\cdot\rho_{u}}{\mathbb{E}_{\delta}\left[d\right]}.
\]
Then, (i) $\pi^{T}:=(\pi_{1},..,\pi_{n-1})$ is a stationary distribution of $\boldsymbol{P}$, (ii) $\boldsymbol{P}$ is reversible with respect to $\pi$, i.e., $\pi_{u}P_{uv}=\pi_{v}P_{vu}$, and (iii) $e\sim e_{uv}:=\pi_{u}P_{uv}$ corresponds to the ex-ante edge distribution.
\end{lem}

Here, $\delta$ corresponds to the ex-ante degree distribution implied by the prior. That is, $\delta_u$ is the probability (according to the prior distribution over networks) that a randomly selected node will have degree $u$. Similarly, $\pi$ corresponds to the ex-ante \textit{edge-based} degree distribution, where $\pi_u$ denotes the probability that a node of degree $u$ lies on either end of a randomly selected edge.

The lemma shows that the belief kernel $\boldsymbol{P}$ appearing in (\ref{eq:D-FOC}) is a reversible Markov chain with stationary distribution $\pi$. This allows us to interpret $\boldsymbol{P}$ as a \textit{belief transition matrix} as it summarizes how beliefs transition across degree types as agents form expectations about their direct and indirect neighbors.

\subsection{Homogeneous Benchmark}

We begin with the homogeneous-beliefs benchmark, which corresponds to a case in which higher-order belief iteration collapses to its simplest form. Under homogeneous beliefs, all degree types have the same first-order perception of neighbor influence, and this common perception is preserved at every subsequent belief step.

\begin{prop}
Let $\bar{d}=\mathbb{E}_{\pi}[d]$. Then, $\mu_{i}=\mu_{j}$ for all $i\neq j$ iff $a_{i}=c^{-1}\left(1+\frac{\kappa d_{i}}{1-\kappa\bar{d}}\right)$.
\label{prop:hom}
\end{prop}

The proposition shows that under homogeneous beliefs, behavior is completely captured by the mean edge-sampled degree $\bar{d}$. The result follows from the fact that $\boldsymbol{P}$ is stationary with respect to $\pi$ and the fact that under homogeneous beliefs, $\mu_i=\bar{d}$.\footnote{Note that the action characterization in Proposition \ref{prop:hom} is quantitatively identical to \cite{jackson2019} (Lemma 3) and qualitatively similar to \cite{galeotti2010network} (Proposition 1). They both consider degree-based models and their characterizations rely on the assumption of degree independence.  While degree independence implies belief homogeneity, the reverse in not always true. Our characterization, therefore, follows from  \emph{weaker} assumptions.}

Homogeneity in first-order beliefs means that all agents share the same perception about how influential their neighbors are. This in turn creates a global perception that the network is homogeneous and isotropic in terms of the externalities it produces. That is, from any agent's perspective, all walks of a given length yield the same expected action modularity regardless of which degree types appear along these walks. Thus, Iterative Belief Centrality reduces to a simple scaling of degree.

Denote by $a^H$ the equilibrium action profile under homogeneous beliefs. Using Proposition \ref{prop:hom}, we can summarize equilibrium predictions stemming from the homogeneous benchmark as follows.

\begin{cor}
Let $\sigma^{2}=Var_{\pi}(d)$. Then, under belief homogeneity:
\begin{enumerate}
\item The map $i\mapsto a_{i}^{H}$ is monotonic.
\item $\mathbb{E}_{\pi}[a^H]=c^{-1}\left(1+\tfrac{\kappa\bar{d}}{1-\kappa\bar{d}}\right)$
\item $Var_{\pi}(a^H)=c^{-2}\Big(\frac{\kappa\sigma}{1-\kappa\bar{d}}\Big)^{2}$
\end{enumerate}
\label{cor:hom}
\end{cor}

\subsection{Deviations from the Homogeneous Benchmark}
\label{sec:prof}

To illustrate how heterogeneous beliefs,
even after belief hierarchies collapse to degree, generate equilibrium behavior
that departs sharply from the homogeneous benchmark characterized above, we revisit the example introduced in Section \ref{sec:example}.

First we note that under Erd\H{o}s--R\'enyi network generation, links are formed independently of nodal identities. This implies that for any $\alpha\in[0,1]$, the resulting distribution over networks will assign equal probability to all graph relabelings, yielding an instance of isomorphic priors. This implies that instead of (\ref{eq:BNE-action}), we can characterize equilibrium actions using the simplified degree-collapsed system (\ref{eq:D-FOC}), with belief transition matrix
\begin{equation}
P_{uv}
= \binom{n-2}{v-1}
\frac{
\alpha p_{H}^{u+v-1}(1-p_H)^{2(n-1)-u-v}
+ (1-\alpha)p_{L}^{u+v-1}(1-p_L)^{2(n-1)-u-v}
}{
\alpha p_{H}^{\,u}(1-p_H)^{n-1-u}
+ (1-\alpha)p_{L}^{\,u}(1-p_L)^{n-1-u}
}.
\label{eq:d-prob}
\end{equation}
The associated edge-based stationary degree distribution is
\begin{equation}
\pi_v
= \binom{n-2}{v-1}\frac{\alpha p_H^{\,v}(1-p_H)^{n-1-v}
+ (1-\alpha)p_L^{\,v}(1-p_L)^{n-1-v}}{\alpha p_H + \left(1-\alpha \right)p_L}.
\label{eq:sta}
\end{equation}

Observe that when $\alpha\in\{0,1\}$, the kernel (\ref{eq:d-prob}) is independent of $u$ and coincides with (\ref{eq:sta}), generating homogeneous beliefs.\footnote{Here, degrees are also independent because links are formed independently.}
Applying Proposition~\ref{prop:hom} with $\bar d=1+(n-2)p$ for $p\in \{p_H,p_L\}$ yields the upper and lower envelopes in Figure \ref{fig1}.

For $\alpha\in(0,1)$, agents of different degrees assign different likelihoods to neighbors' degrees. This belief heterogeneity produces the discontinuous action profiles shown in the figure. 

The figure also reports a counterfactual in which agents beliefs are uniformly restricted to the edge-based degree distribution (\ref{eq:sta}). That is, agents exert action based on a belief transition matrix whose rows are all set to equal to (\ref{eq:sta}). This is again an instance of homogeneous beliefs. The corresponding actions therefore follow Proposition \ref{prop:hom} with $\bar d=1+(n-2)\bar p$, where $\bar p=\frac{\alpha p_H^2+(1-\alpha)p_L^2}{\alpha p_H+(1-\alpha)p_L}$.

The example illustrates that once beliefs are heterogeneous, Iterative Belief Centrality no longer collapses to a simple scaling of degree giving rise to equilibrium behaviors that depart from the homogeneous benchmark both quantitatively and qualitatively. 

First, as shown in the right panel of Figure \ref{fig1}, the action profile may fail to be monotone in degree. This follows because local differences in perceived influence alter the strength of best responses and can dominate the direct effect of degree. Under homogeneous beliefs, $\mu_i=\mu_j$ and such reversals are impossible. With heterogeneous beliefs, however, differences in perceived neighbor influence interact with degree to shape actions in nontrivial ways.

Second, relative to the homogeneous benchmark in which agents beliefs are uniformly restricted to $\pi$, some degree types exert more and others less. This cutoff behavior arises from the same belief-propagation mechanism identified in Section~\ref{sec:generalized_example}. When priors are isomorphic and belief hierarchies collapse to degree, Row-Based Dominance admits a simple characterization in terms of the belief transition matrix. Specifically, Row-Based Dominance collapses to the condition that higher realized degree induces stochastically higher beliefs about neighbors' degrees: for any increasing function $\varphi$,\footnote{This is
precisely an affiliation-type condition of the form imposed in
\citet{galeotti2010network}. Importantly, in our framework this condition is not imposed at the level of degree beliefs, but emerges endogenously from Row-Based Dominance applied to beliefs over network structure under isomorphic priors.}
\[
u'>u
\quad\Longrightarrow\quad
\sum_{v}P_{u'v}\varphi(v)\ \ge\ \sum_{v}P_{uv}\varphi(v).
\]
In the example, observing a higher degree increases the posterior probability that the network was generated in the high-density state. Since the neighbor-degree distribution is increasing in the density parameter, the kernel \eqref{eq:d-prob} satisfies this dominance condition. The cutoff behavior in Figure \ref{fig1} therefore reflects the same belief-propagation mechanism as in the general Row-Based Dominance results, expressed within the degree-based representation.

Third, beyond reshaping individual action profiles, heterogeneous beliefs also alter aggregate outcomes. Under homogeneous beliefs, the mean and variance of actions are pinned down entirely by the mean and variance of the degree distribution. Under heterogeneous beliefs, this is no longer the case. The example shows that
belief heterogeneity can increase both the average level of action and its dispersion across agents, even when the underlying degree distribution is held fixed. These effects arise because belief differences are correlated across links and persist through higher-order expectations.

The next subsection formalizes these observations by introducing belief-based assortativity measures and showing how the mixing properties of the belief transition matrix govern the mean and variance of equilibrium actions.

\subsection{Collective Behavior and Perceived Assortativity}

A key to characterizing aggregate behavior is the extent to which heterogeneous beliefs are \emph{correlated} across links and how persistent these correlations are as agents form beliefs about the beliefs of others downstream. To capture this, we introduce a family of measures that track how much information about degrees survives along successive ``belief steps'' under  $\boldsymbol{P}$.

Let $x(i):=\frac{d_{i}-\bar{d}}{\sigma}$ and $\left(X_{t}\right)_{t\geq0}$
be the Markov chain started at stationarity, $X_{0}\sim\pi$, with
transition kernel $\boldsymbol{P}$. Then, for any path $i_{0},..,i_{k-1}$ we
have 
\[
P\left(X_{0}=i_{0},...,X_{k-1}=i_{k-1}\right)=\pi_{i_{0}}P_{i_{0}i_{1}}...P_{i_{k-2}i_{k-1}}.
\]
For any binary tuple $b=(b_{0},...,b_{k-1})\in\{0,1\}^{k}$ define the \textit{path moment} 
\[
r^{b}=\mathbb{E}_{\pi_{i_{0}}P_{i_{0}i_{1}}...P_{i_{k-2}i_{k-1}}}\left[\prod_{t=0}^{k-1}x\left(X_{t}\right)^{b_{t}}\right].
\]

These path moments capture \textit{belief-based assortativity coefficients}. To see why, consider path $i_{0}\sim i_{1}$ with corresponding distribution $\pi_{i_{0}}P_{i_{0}i_{1}}$. Then, for the tuple $b=(1,1)$ we have:
\begin{align*}
r^{(1,1)} & =\frac{\sum_{i_{0},\,i_{1}}\pi_{i_{0}}P_{i_{0}i_{1}}d_{i_{0}}d_{i_{1}}-\bar{d}^{2}}{\sigma^{2}},
\end{align*}
where the equality follows from expanding the product in $r^b$ and then using the fact that $\boldsymbol{P}$ is a reversible chain.%
\footnote{In particular, $\mathbb{E}_{\pi P_{i_{0}i_{1}}}[d_{i_0}]=\sum_{i_{0},\,i_{1}}\pi_{i_{0}}P_{i_{0}i_{1}}d_{i_0}=\bar{d}$
and $\mathbb{E}_{\pi P_{i_{0}i_{1}}}[d_{i_{1}}]=\sum_{i_{0},\,i_{1}}\pi_{i_{0}}P_{i_{0}i_{1}}d_{i_{1}}=\sum_{i_{0},\,i_{1}}\pi_{i_{1}}P_{i_{1}i_{0}}d_{i_{1}}=\bar{d}$.}
Since $\pi_{i_{0}}P_{i_{0}i_{1}}$ corresponds to the edge probability $e_{i_{0}i_{1}}$,%
\footnote{The probability that a randomly selected edge connects a node of degree $i_0$ to one of degree $i_1$.}
this shows that $r^{(1,1)}$ is identical to the assortativity measure defined in \cite{newman2003mixing}, which measures the extent to which nodes of similar degrees are connected via a single edge in a given network. Here, however, $r^{(1,1)}$ measures the extent to which individuals \emph{believe} they are directly connected to nodes similar to them. Since the prior (and hence interim beliefs) are common knowledge, this assortativity measure captures a society-wide belief about whether similar individuals tend to be connected.

The family $\{r^{b}\}$ generalizes edge-based degree assortativity to higher-order variants. For instance, $r^{(1,0,..,0,1)}$ captures correlation between the degrees of two randomly selected nodes at long ranges, while $r^{(1,1,..,1,1)}$ captures correlation persistence across sequences of adjacent nodes. The cardinality $|b|=\sum_{t}b_{t}$ determines how many vertex degrees enter the correlation, while the number of successive $0$'s between $1$'s fixes the distances between them.

In our framework, behavior depends on how degree shapes beliefs about others. When beliefs are heterogeneous, these expectations differ across degree types. The coefficients $r^{b}$ provide a tool for gauging how such heterogeneity is correlated across links and how it propagates across successive belief steps. They measure how much information about a starting degree survives after one, two, or more belief steps under $\boldsymbol{P}$. If high-degree individuals expect to encounter other high-degree individuals, and those expectations persist along chains of beliefs, assortativity is high. If beliefs mix quickly, perceptions converge toward the global average and assortativity is low.

Comparing this to the homogeneous-beliefs benchmark, it is easy to verify that in that case $r^{b}=0$ for any binary pattern $b$. Homogeneous beliefs are already completely mixed at the stationary edge-based distribution, so there is no heterogeneity to propagate.

Let $1=\rho_{1}\geq\rho_{2}\geq...\geq\rho_{n-1}$ denote the spectrum
of $\boldsymbol{P}$. Denote by $\rho_{*}=\max\{|\rho_{2}|,\,|\rho_{n-1}|\}$, $\omega=\frac{d_{max}-\bar{d}}{\sigma}$,
and $\xi=\bar{d}+\sigma\omega\rho_{*}$. Let $a^{H}$ denote the equilibrium
action profile under homogeneous beliefs with the same stationary
distribution $\pi$. 

\begin{prop}
The difference between the average and variance in actions in the heterogeneous and homogeneous cases are bounded by the spectrum of the network:
\[
\left|\mathbb{E}_{\pi}\left(a\right)-\mathbb{E}_{\pi}\left(a^{H}\right)\right|\leq\frac{\left|\kappa\right|^{2}\sigma^{2}c^{-1}}{\left(1-\left|\kappa\right|\xi\right)^{3}}\rho_{*}
\]
and 
\[
\left|Var_{\pi}\left(a\right)-Var_{\pi}\left(a^{H}\right)\right|\leq\frac{6\left|\kappa\right|^{2}\sigma^{2}c^{-2}}{\left(1-\left|\kappa\right|\xi\right)^{6}}\rho_{*}
\]
\label{prop:mv} 
\end{prop}

The second-largest eigenvalue $\rho_{*}$ acts as a global cap on higher-order assortativity coefficient. When $\rho_{*}$ is small, the chain mixes rapidly, higher-order assortativity decays quickly, and agents' perceptions ``forget'' their starting point. In this case, even if there is local heterogeneity in beliefs, it does not propagate far through the belief hierarchy, and both the mean and variance of actions remain close to what we would expect under homogeneous beliefs.

Conversely, when $\rho_{*}$ is close to one, the chain mixes slowly, higher-order assortativity persists, and local differences in perceived influence continue to affect behavior across many belief steps. This persistence magnifies heterogeneity in actions and can shift the average away from the homogeneous benchmark.

The key insight is that agents sample/anticipate the network through their neighbors. When perceived assortativity is high and mixing is slow, neighbor sampling tends to reflect agents' own types back at them. High-degree agents expect highly connected neighbors, while low-degree agents expect weakly connected ones. Because these expectations persist along belief paths, perceived influence becomes increasingly type-dependent, amplifying differences in equilibrium behavior. By contrast, when mixing is fast, successive belief steps quickly resemble draws from the stationary distribution $\pi$, so local perceptions converge toward the global distribution and actions remain close to the homogeneous benchmark.

The role of the spectral parameter $\rho_*$ can be further illustrated in a simple configuration model with only two possible degrees $\mathcal{D}=\{d,d'\}$. Suppose the belief transition matrix over degree types is
\[
\boldsymbol{P}=
\begin{pmatrix}
p & 1-p\\
1-q & q
\end{pmatrix},
\]
with $p\neq q$. The mixing rate of the chain is $\rho_* =|p+q-1|$. 

Suppose $p+q>1$, so that agents assign higher probability to being connected to neighbors of the same degree type. The difference between the heterogeneous-beliefs equilibrium and the homogeneous benchmark can then be written as
\begin{equation}
\mathbb{E}_{\pi}(a)-\mathbb{E}_{\pi}(a^{H})
= C_{1}(\rho_*)\,\rho_* 
\qquad
Var_{\pi}(a)-Var_{\pi}(a^{H})
= C_{2}(\rho_*)\,\rho_*,
\label{eq:config}
\end{equation}
where $C_{1}(\rho_*)$ and $C_{2}(\rho_*)$ are strictly positive and increasing in $\rho_*$.\footnote{The details of this derivation are presented in the Online Appendix.} The intuition is transparent in this two-degree-type case. Here $\rho_* = p+q-1$, which measures the extent to which agents believe that neighbors have degrees similar to their own. When $p$ and $q$ are large, a type-$d$ agent expects to encounter type-$d$ neighbors with high probability (and similarly for type $d'$), so each belief step tends to preserve the current degree type and the chain converges slowly. When $p$ and $q$ are smaller, agents do not expect neighbors to resemble themselves, so beliefs quickly revert toward the stationary edge-based degree distribution $\pi$. Thus, in this example, $\rho_*$ governs both how strongly agents expect neighbors to be similar and how persistent those perceptions are along belief paths.

The example also suggests that environments with persistent assortativity are particularly relevant for understanding the effects of heterogeneous beliefs. In many network formation models, agents are more likely to connect with others who are similar to themselves, generating positive correlations in degree across links (see, for instance, \cite{jackson2007meeting}). In such environments, the degrees encountered along interaction paths are systematically positively correlated, so heterogeneity in perceived influence tends to reinforce itself rather than cancel out across belief paths. While the bounds derived in Proposition \ref{prop:mv} are general, they allow for both positive and negative correlations and are thus potentially conservative. When correlations are persistently positive, however, belief heterogeneity propagates more strongly through the belief hierarchy. The next result formalizes this idea by imposing a path-wise positive assortativity condition that captures environments in which postive degree correlations persist across belief paths.

Let $Z\sim\pi$, and conditional on $Z$ let $(X_t)_{t\ge 0}$ and $(Y_t)_{t\ge 0}$ be independent Markov chains with transition kernel $\boldsymbol{P}$
and common start $X_0=Y_0=Z$.
For binary patterns $b\in\{0,1\}^k$ and $c\in\{0,1\}^\ell$. Define the \emph{two-arm path moments}
\[
r^{b,c}:=\mathbb{E}_{\pi_{i_{0}}P_{i_{0}i_{1}}...P_{i_{k-2}i_{k-1}}P_{i_{0}j_{1}}\ldots P_{j_{l-2}j_{l-1}}}\!\left[\prod_{t=0}^{k-1} x(X_t)^{b_t}\prod_{s=0}^{\ell-1} x(Y_s)^{c_s}\right].
\]

The two-arm path moments capture how degree information propagates across \emph{multiple} belief paths originating from the same node. While the single-arm moments $r^b$ measure how information about a starting node's degree persists along one path, the two-arm moments $r^{b,c}$ measure how correlated degrees remain across different paths branching from that node. 

We say that $\boldsymbol{P}$ is \emph{Path-wise Positive Assortative} if:
(i) $r^b\ge 0$ for every binary pattern $b$ with $|b|\ge 2$, and
(ii) $r^{b,c}\ge r^b r^c$ for every pair of binary patterns $(b,c)$ of arbitrary lengths.

\begin{thm}
Suppose $\boldsymbol{P}$ has a positive spectrum, satisfies Row-Based Dominance and is Path-wise Positive Assortative. If $\lambda>0$, then there exists a $\gamma\in(0,1]$ such that
\[
\mathbb{E}_{\pi}\!\left[a(P)\right]-\mathbb{E}_{\pi}\!\left[a^{H}\right]
\;>\;
\frac{\kappa^{2}\sigma^{2}c^{-1}}{\left(1-\kappa\xi\right)^{3}}\;\gamma\,\rho_{*},
\]
and 
\[
\operatorname{Var}_{\pi}\!\left(a(P)\right)-\operatorname{Var}_{\pi}\!\left(a^{H}\right)
\;>\;\,\frac{\kappa^{4}\sigma^{4}c^{-2}}{\left(1-\kappa\xi\right)^{4}}\;\gamma^{2}\rho_{*}^{2}.
\]
\label{thm:tighter}
\end{thm}

Under path-wise positive assortativity degrees encountered along belief paths remain systematically positively correlated with the starting node's degree. As a result, belief heterogeneity compunds along belief hierarchies rather than averaging out. High-degree agents expect influential neighbors, who in turn expect influential neighbors, and so on, while low-degree agents expect weakly connected neighborhoods throughout the belief hierarchy. This then leads to both higher average activity and dispersion relative to the homogeneous benchmark with the magnitude governed by $\rho_{*}$. When $\rho_{*}$ is small, belief paths mix quickly and heterogeneous perceptions have limited aggregate effects. When $\rho_{*}$ is large, correlations persist across many belief steps, magnifying the impact of heterogeneous perceptions on collective outcomes.

Taken together, these results show that the collective implications of heterogeneous beliefs depend not only on the distribution of local observations but also on how perceptions propagate across the network. The key determinant is the persistence of belief correlations along interaction paths.

\section{Incorporating Arbitrary Network Information}
\label{sec:gen}

We have focused on an information structure in which agents' private information is row-based; i.e., agents know the identities of their direct
connections, and perhaps more.
This precludes settings in which people know how many future interactions they may have, but don't know with whom.  For instance, when learning a language, one knows that one will have many future conversations, but does not know the identities of everyone of the counterparties.   When choosing programs to use, a researcher might not know who all their future coauthors will be.   People often choose an action without fully knowing the network(s) in which their future interactions will take place.   

Despite the fact that we have presumed that people know the identities of their counterparties, this is not necessary for our main results and was more for notational convenience.   To make this point concrete,
we provide a more general
incomplete-information framework that nests our approach, degree-based approaches, and much more general information structures. 

Sticking to the Harsanyi formalism, let each agent $i$ have
a finite type set $T_i$. Let $T=T_1\times\cdots\times T_n$, and suppose that uncertainty is described
by a joint distribution
\[
P \in \Delta(\mathcal G_n \times T).
\]
Nature draws $(\mathbf g,\mathbf t)\sim P$, each agent observes $t_i$, and updates beliefs
about $(\mathbf g,\mathbf t_{-i})$ by Bayes' rule. 

In this formulation, types can encode any
information agents have about the network. This can include identities of neighbors, degree, noisy signals
about density, partial views of the graph, or even non-structural signals correlated with
network position. Our analysis of equilibrium behavior continues to apply whenever actions
are chosen based on interim expectations formed from these beliefs.
Effectively, one simply substitutes these type-based beliefs for the beliefs we have used in various results.

Both row-based and degree-based models fit within this general representation. Our baseline model corresponds to the restriction that types reveal agents' links, so that
\[
P(\mathbf g,\mathbf t)>0,\; P(\mathbf g',\mathbf t')>0,\; t_i=t_i'
\quad\Longrightarrow\quad
\mathbf g_i=\mathbf g_i'.
\]
This guarantees that $t_i$ pins down $i$'s realized neighborhood. 

Degree-based models correspond to the case in which agents' types reveal only degree. Let $D:=\{0,1,\ldots,n-1\}$ and define $T_i^{D}:=D$ for each $i$, with the interpretation
that $t_i=d_i(\mathbf g)$ is $i$'s realized degree. Let $\hat P\in\Delta(\mathcal G_n)$
be any distribution over networks, and define a joint distribution
$P^{D}\in\Delta(\mathcal G_n\times T^{D})$. Degree-based priors  then correspond to
\[
P^{D}(\mathbf g,\mathbf t)
:=\hat P(\mathbf g)\cdot \mathbb I\{t_i=d_i(\mathbf g)\ \text{for all } i\}.
\]
Homogeneous degree-based models are recovered as the special case in which
\[
P^{D}(d_j \mid d_i)=P^{D}(d_j)
\quad \text{for all } i.
\]

This general formulation makes clear that row-based and degree-based models differ only in the restrictions placed on how types relate with network realizations. 
These are not merely alternative modeling choices, but can also be connected endogenously. In Section \ref{sec:sym}, we show that under symmetry conditions on priors and information, starting from a row-based formulation, belief hierarchies can collapse so that equilibrium behavior depends only on degree. In this sense,
degree-based behavior need not be imposed ex ante through the information structure; rather, it can also emerge endogenously from richer identity-based information once symmetry eliminates
informational distinctions beyond degree.

The forces identified in the paper to extend beyond row-based types. Whenever agents' private information induces systematic differences in beliefs about others' beliefs and these differences are preserved under conditioning, iterative belief formation  generates analogous patterns of belief propagation, heterogeneity, and amplification. Thus, while the paper develops results in a tractable and transparent setting, the underlying mechanisms are not tied to rows and apply more broadly to games on networks with general forms of incomplete information.

\section{Concluding Remarks}
\label{sec:con}

We have developed a framework for studying games in networks under incomplete information about network structure. Focusing on the workhorse linear--quadratic game, we characterize equilibrium behavior when agents have limited information about the architecture of the network and form beliefs from local observations showing that equilibrium actions can be interpreted as \emph{Iterative Belief Centrality}. Agents exert action based on their perceived influence formed through an iterative hierarchy of beliefs about others' beliefs. This perspective unifies two leading benchmarks in the literature. Under complete information, belief iteration collapses and behavior is driven by Katz--Bonacich centrality; under symmetric incomplete information, belief iteration collapses in a different way and behavior is summarized by degree. In intermediate environments, belief iteration   generates systematic departures from both extremes.

We have shown that local network sampling can rationally lead to biased and heterogeneous perceptions of the network, and these perceptions can translate into extreme differences in behavior, including non-monotonicities and cutoff behavior relative to degree-based benchmarks. We have also shown that the magnitude and persistence of these behavioral differences are governed by how heterogeneous beliefs are correlated across links. These correlations are captured by the belief transition matrix and quantified through perceived assortativity and its higher-order variants. The mixing properties of the belief transition matrix provide a global index of how rapidly local belief differences  propagate and amplify through iterative belief steps.

Finally, the approach we take is broadly portable. The Harsanyi framework we apply is compatible with other network games featuring strategic spillovers among agents, and the logic of Iterative Belief Centrality extends beyond the linear--quadratic environment. This opens the door to incomplete-information variants of a wide range of network applications, including policy intervention and targeting \citep{galeotti2020targeting}, and environments with endogenous network formation \citep{konig,banerjee2024changes}. In such settings, an important direction for future work is to study how changes in network structure affect belief hierarchies, and to characterize welfare and policy implications of interventions that change how networks are perceived.

\section*{Appendix A. Proofs of Theorems}

\setlength\parindent{0pt}
\setlength{\parskip}{8pt}



\subsection*{Proof of Theorem \ref{thm:character}}

The best response of agent $i$ who is of type $t_{i}$ is given
by

\begin{align*}
a_{i}\left(t_{i}\right) & =c^{-1}+\kappa\sum_{j_{1}\in N}g_{ij_{1}}^{t_{i}}\sum_{t_{j_{1}}\in T_{j_{1}}}p\left(\left.t_{j_{1}}\right|t_{i}\right)a_{j_{1}}\left(t_{j_{1}}\right)
\end{align*}
Similarly, the best response of $j_{1}$ of type $t_{j_{1}}$ can
be written as 
\[
a_{j_{1}}\left(t_{j_{1}}\right)=c^{-1}+\kappa\sum_{j_{2}\in N}g_{j_{1}j_{2}}^{t_{j_{1}}}\sum_{t_{j_{2}}\in T_{j_{2}}}p\left(\left.t_{j_{2}}\right|t_{j_{1}}\right)a_{j_{2}}\left(t_{j_{2}}\right)
\]
And hence the action of agent $i$ of type $t_{i}$ is 
\[
a_{i}\left(t_{i}\right)=c^{-1}\left(1+\kappa\sum_{j_{1}\in N}g_{ij_{1}}^{t_{i}}+\kappa^{2}\sum_{j_{1}\in N}\sum_{j_{2}\in N}\sum_{t_{j_{1}}\in T_{j_{1}}}\sum_{t_{j_{2}}\in T_{j_{2}}}g_{ij_{1}}^{t_{i}}g_{j_{1}j_{2}}^{t_{j_{1}}}p\left(\left.t_{j_{1}}\right|t_{i}\right)p\left(\left.t_{j_{2}}\right|t_{j_{1}}\right)a_{j_{2}}\left(t_{j_{2}}\right)\right)
\]
Substituting for the best response of agent $j_{2}$ of type $t_{j_{2}}$,
and continuing in this way gives the result.

\subsection*{Proof of Theorem~\ref{thm:iterated_multiplier}}

We prove the theorem by backward induction along the labeled path
$j_0=i,j_1,\ldots,j_m$. 

We first show that for each agent $k$, the equilibrium strategy
$\sigma_k^\star:T_k\to\mathbb R$ under Row Based Dominance is degree-monotone.

Consider the linear--quadratic best-response operator:
\begin{equation}
(B\sigma)_k(t_k)
:=
c^{-1}+\kappa\,\mathbb E\!\left[\sum_{r} g_{kr}\,\sigma_r(t_r)\,\middle|\, t_k\right].
\label{eq:BR-operator-proof}
\end{equation}
Under first-order information, the coefficients $g_{kr}$ are known conditional on $t_k$,
so the expectation is taken only over neighbors' row types.

Define a profile $\sigma$ to be degree-monotone if each $\sigma_r$ is degree-monotone.
Suppose $\sigma$ is degree-monotone and fix $t_k,t_k'\in T_k$ with $d(t_k')>d(t_k)$.
Define the functional of neighbor rows
\[
\Phi_\sigma(R_k)
:=
\sum_{r\in N_k(t_k)} g_{kr}\,\sigma_r(t_r).
\]
Because each $\sigma_r$ is degree-monotone, $\Phi_\sigma$ is coordinatewise
degree-increasing. Applying \eqref{eq:RBDeq} with $x=k$ and $\Phi=\Phi_\sigma$
yields
\[
(B\sigma)_k(t_k')\ \ge\ (B\sigma)_k(t_k).
\]
Thus, whenever $\sigma$ is degree-monotone, so is $B\sigma$.

Starting from the constant profile $\sigma^{(0)}_r(\cdot)\equiv c^{-1}$, define
$\sigma^{(\ell+1)}:=B\sigma^{(\ell)}$. By Lemma~\ref{lem:exist}, $B$ is a contraction, so
$\sigma^{(\ell)}$ converges pointwise to the unique equilibrium $\sigma^\star$.
Degree-monotonicity is preserved under pointwise limits, hence
\begin{equation}
d(t_k')>d(t_k)\ \Longrightarrow\ \sigma_k^\star(t_k')\ge \sigma_k^\star(t_k)
\quad\text{for all }k.
\label{eq:sigma-star-degmono}
\end{equation}

Fix the path $j_0=i,j_1,\ldots,j_m$ and define continuation values
$V_m(t_{j_m}):=\sigma_{j_m}^\star(t_{j_m})$ and, for $\ell=m-1,\ldots,0$,
\[
V_\ell(t_{j_\ell})
:=
\mathbb E\!\left[
V_{\ell+1}(t_{j_{\ell+1}})
\ \middle|\ t_{j_\ell},\,\mathcal H_\ell
\right],
\]
where $\mathcal H_\ell$ denotes the realized history along the path up to node $j_\ell$.
The positive-probability condition in the theorem ensures these conditional expectations
are well defined under both posteriors induced by $t_i$ and $t_i'$.

We prove by backward induction that $V_\ell$ is degree-monotone for all $\ell$.

The base case $\ell=m$ follows from \eqref{eq:sigma-star-degmono}. For the inductive step,
fix $\ell\in\{0,\ldots,m-1\}$ and suppose $V_{\ell+1}$ is degree-monotone. Consider
agent $j_\ell$ and two row types $t,t'\in T_{j_\ell}$ with $d(t')>d(t)$.
Let $R_{j_\ell}$ denote the random vector of neighbor row types of $j_\ell$ under the
interim belief induced by conditioning on $t$ (and analogously under $t'$), and define the
functional
\[
\Phi_{\ell}(R_{j_\ell})
:=
\mathbb E\!\left[
V_{\ell+1}(t_{j_{\ell+1}})
\ \middle|\ t_{j_\ell},\,\mathcal H_\ell,\,R_{j_\ell}
\right].
\]
By construction, $\Phi_\ell$ is a coordinatewise degree-increasing functional of
$R_{j_\ell}$ because $V_{\ell+1}$ is degree-monotone and the next-step node
$j_{\ell+1}$ is selected along the path via a neighbor relation.
Applying \eqref{eq:RBDeq} to agent $x=j_\ell$ with $\Phi=\Phi_\ell$
implies $V_\ell(t')\ge V_\ell(t)$ whenever $d(t')>d(t)$. Hence $V_\ell$ is degree-monotone.

Lastly, applying degree-monotonicity of $V_0$ to $t_i,t_i'$ yields
\[
\mathbb E\!\left[a_{j_m}(t_{j_m})\,\middle|\, t_i'\right]
\;\ge\;
\mathbb E\!\left[a_{j_m}(t_{j_m})\,\middle|\, t_i\right],
\]
which completes the proof.

\subsection*{Proof of Theorem \ref{thm:same-action}}


We start with the following definition:
\begin{defn}
For any graph $\mathbf{g}\in\mathcal{G}_{n}$ and a bijection $f:N\longrightarrow N$,
we define by $\mathbf{g}^{f}\in\mathcal{G}_{n}$ the graph for
which $g_{f\left(i\right)f\left(j\right)}^{f}=g_{ij}$. 
\end{defn}
From now on, we will assume that there is $m^{th}$ order information
prevailing in the game, where $m\geq1$. Moreover, assume that $t_{i}$
and $t_{q}$ are isomorphic, so that there exists a bijection $f:N\longrightarrow N$
with $f\left(i\right)=q$ such that $f\left(N_{k}\left(\mathbf{g}\right)\right)=N_{f\left(k\right)}\left(\mathbf{g}^{\prime}\right)$
for every $k\in N_{i}^{m}\left(\mathbf{g}\right)$, where $\mathbf{g}\in g\left(t_i\right)$ and $\mathbf{g}^\prime\in g\left(t_q\right)$. For the bijection
$f$ and for any agent $l$'s type $t_{l}=t_{l}\left(\mathbf{g}\right)$,
the image of $t_{l}$ under $f$ is given by $f\left(t_{l}\right)= t_{f\left(l\right)}\left(\mathbf{g}^f\right)$.

The rest of the proof involves showing that the walk characterization
of the equilibrium (\ref{eq:BNE-action})-(\ref{eq:BNE-walk}) is
invariant under local isomorphism. In particular, we show that both the walks, and their corresponding probabilities under symmetric priors,are invariant to the isomorphism. 
We consider two cases, depending
on the length of walks relative to the prevailing order of information.

\noindent{{\textbf{Case 1:} $s\leq m$}}

Since agents are aware of the identities of their $m^{th}$ order
neighbors, then for all $s\leq m$, and for any enumeration $\left(j_{1},j_{2},\ldots,j_{s}\right)\in N^{s}$,
agents are certain about the value $g_{ij_{1}}^{t_{i}}g_{j_{1}j_{2}}^{t_{j_{1}}}...g_{j_{s-1}j_{s}}^{t_{j_{s-1}}}$,
i.e., whether they are $0$ or $1$. Hence, for $s\leq m$ we have
\begin{align*}
\beta_{i,t_{i}}^{(s)} & =\sum_{j_{1},j_{2},..,j_{s}=1}^{n}\text{ }g_{ij_{1}}^{t_{i}}g_{j_{1}j_{2}}^{t_{i}}...g_{j_{s-1}j_{s}}^{t_{i}}\sum_{\left\{ t_{j_{k}}\in T_{j_{k}}:1\leq k\leq s\right\} }p(t_{j_{s-1}}|t_{j_{s-2}})p(t_{j_{s-2}}|t_{j_{s-3}})...p(t_{j_{1}}|t_{i})\\
 & =\sum_{j_{1},j_{2},..,j_{s}=1}^{n}\text{ }g_{ij_{1}}^{t_{i}}g_{j_{1}j_{2}}^{t_{i}}...g_{j_{s-1}j_{s}}^{t_{i}}
\end{align*}
where the second equality is due to the fact that $\sum_{t_{j_{k}}\in T_{j_{k}}}p(t_{j_{k-1}}|t_{j_{k-2}})=1$
for all $2\leq k\leq s$ and $\sum_{t_{j_{1}}\in T_{j_{1}}}p(t_{j_{1}}|t_{i})=1$.
Similarly, for agent $q=f\left(i\right)$ who is of type $t_{q}$
we can write $\beta_{q,t_{q}}^{(s)}=\sum_{f\left(j_{1}\right),..,f\left(j_{s}\right)=1}^{n}\text{ }g_{f\left(i\right)f\left(j_{1}\right)}^{\prime t_{q}}...g_{f\left(j_{s-1}\right)f\left(j_{s}\right)}^{\prime}$. 
Next, we use the fact that $t_{i}$ and $t_{q}$ is isomorphic with
$f\left(.\right)$ being a bijection, as well as the fact that $s\leq m$
to write 
\begin{align*}
\beta_{i,t_{i}}^{(s)} & =\sum_{j_{1},j_{2},..,j_{s}=1}^{n}\text{ }g_{ij_{1}}^{t_{i}}g_{j_{1}j_{2}}...g_{j_{s-1}j_{s}}=\sum_{j_{1}\in N_{i}\left(\mathbf{g}\right)}\sum_{j_{2}\in N_{j_{1}}\left(\mathbf{g}\right)}\ldots\sum_{j_{s}\in N_{j_{s-1}}\left(\mathbf{g}\right)}\;1\\
 & =\sum_{f\left(j_{1}\right)\in N_{q}\left(\mathbf{g}^{\prime}\right)}\ldots\sum_{f\left(j_{s}\right)\in N_{f\left(j_{s-1}\right)}\left(\mathbf{g}^{\prime}\right)}\;1=\beta_{q,t_{q}}^{(s)}
\end{align*}
Therefore, $\beta_{i,t_{i}}^{(s)}=\beta_{q,t_{q}}^{(s)}$ for all
$s\leq m$.

\noindent{{\textbf{Case 2:} $s\geq m+1$}}

Agent $j_{1}$ being of type $t_{j_{1}}$ for any enumeration $\left(j_{2},j_{3},\ldots,j_{m}\right)\in N^{m-1}$
is certain about the value $g_{j_{1}j_{2}}^{t_{j_{1}}}...g_{j_{m}j_{m+1}}^{t_{j_{m}}}$,
whether its $0$ or $1$. Note that whenever $g_{j_{1}j_{2}}^{t_{j_{1}}}...g_{j_{m}j_{m+1}}^{t_{j_{m}}}=1$,
we have that $j_{m}\in t_{j_{1}}$. Hence, $g_{j_{m}j_{m+1}}^{t_{j_{m}}}=g_{j_{m}j_{m+1}}^{t_{j_{1}}}$.
This generalizes to any $j_{k}$. Therefore, for $s=m+l$ we have
\[
\beta_{i,t_{i}}^{(m+l)}=\sum_{j_{1},j_{2},..,j_{m+l}=1}^{n}\text{\ensuremath{\sum_{\left\{ t_{j_{k}}\in T_{j_{k}}:1\leq k\leq l\right\} }} \ensuremath{g_{ij_{1}}^{t_{i}}}...\ensuremath{g_{j_{m-1}j_{m}}^{t_{i}}}}g_{j_{m}j_{m+1}}^{t_{j_{1}}}\ldots g_{j_{m+l-1}j_{m+l}}^{t_{j_{l}}}p(t_{j_{l}}|t_{j_{l-1}})\ldots p(t_{j_{1}}|t_{i}).
\]
Next, define the set $\mathcal{T}_{t_{i}}\subseteq\vartimes_{k=1}^{l}T_{j_{k}}$
as follows 
\[
\mathcal{T}_{t_{i}}:=\left\{ \left(t_{j_{1}},\ldots,t_{j_{l}}\right)\in\vartimes_{k=1}^{l}T_{j_{k}}:j_{m+s}\in N_{j_{m+s-1}}\left(g\left(t_{j_{s}}\right)\right)\:\text{ for all }1\leq s\leq l\,\wedge\,g\left(t_{i}\right)\bigcap_{k=1}^{l}g\left(t_{j_{k}}\right)\neq\phi\right\} 
\]
where the first part of the condition, $j_{m+s}\in N_{j_{m+s-1}}\left(g\left(t_{j_{s}}\right)\right)\:\text{ for all }1\leq s\leq l$,
ensures that under the type $\left(t_{j_{1}},\ldots,t_{j_{l}}\right)$,
there exists a walk $j_{m}\sim j_{m+1}\sim\ldots\sim j_{m+l}$, i.e.,
$g_{j_{m}j_{m+1}}^{t_{j_{1}}}\ldots g_{j_{m+l-1}j_{m+l}}^{t_{j_{l}}}=1$.
The second part of the condition, $g\left(t_{i}\right)\bigcap_{k=1}^{l}g\left(t_{j_{k}}\right)\neq\phi$,
ensures that the type $\left(t_{j_{1}},\ldots,t_{j_{l}}\right)$ is
actually feasible, that is, $P(t_{j_{l}}|t_{j_{l-1}})\ldots P(t_{j_{1}}|t_{i})>0$.
These allows us to rewrite $\beta_{i,t_{i}}^{(m+l)}$ as follows 
\[
\beta_{i,t_{i}}^{(m+l)}=\sum_{j_{1}\in N_{i}\left(\mathbf{g}\right)}\ldots\sum_{j_{m}\in N_{j_{m-1}}\left(\mathbf{g}\right)}\text{\ensuremath{\sum_{\left(t_{j_{1}},\ldots,t_{j_{l}}\right)\in\mathcal{T}_{t_{i}}}} }p(t_{j_{l}}|t_{j_{l-1}})\ldots p(t_{j_{1}}|t_{i})
\]

\begin{claim}
\label{claim2} \textit{\label{SBNN-L5}For any $l\geq1$, $\left(t_{j_{1}},\ldots,t_{j_{l}}\right)\in\mathcal{T}_{t_{i}}$
iff $\left(f\left(t_{j_{1}}\right),\ldots,f\left(t_{j_{l}}\right)\right)\in\mathcal{T}_{t_{q}}$.} 
\end{claim}

Claim \ref{SBNN-L5} shows that for every feasible type-tuple $\left(t_{j_{1}},\ldots,t_{j_{l}}\right)$
giving rise to the walk $j_{m}\sim j_{m+1}\sim\ldots\sim j_{m+l}$,
the type-tuple $\left(f\left(t_{j_{1}}\right),\ldots,f\left(t_{j_{l}}\right)\right)$
is also feasible and will give rise to the walk $f\left(j_{m}\right)\sim f\left(j_{m+1}\right)\sim\ldots\sim f\left(j_{m+l}\right)$,
and vice versa. Therefore, we can write 

\begin{equation*}
    \beta_{q,t_{q}}^{(m+l)}=\sum_{j_{1}\in N_{i}\left(\mathbf{g}\right)}\ldots\sum_{j_{m}\in N_{j_{m-1}}\left(\mathbf{g}\right)}\text{\ensuremath{\sum_{\left(t_{j_{1}},\ldots,t_{j_{l}}\right)\in\mathcal{T}_{t_{i}}}}\ensuremath{p\left(\left.f\left(t_{j_{l}}\right)\right|f\left(t_{j_{l-1}}\right)\right)\ldots}\ensuremath{p\left(\left.f\left(t_{j_{1}}\right)\right|t_{q}\right)} }
\end{equation*}
where the equality is due to the fact that $t_{i}$ and $t_{q}$
are isomorphic where $\mathcal{A}_{i}\left(\mathbf{g}\right)=N_{i}^{m}\left(\mathbf{g}\right)$
and $\mathcal{A}_{q}\left(\mathbf{g}^{\prime}\right)=N_{q}^{m}\left(\mathbf{g}^{\prime}\right)$,
and also by invoking the Claim \ref{SBNN-L5}.

Claim \ref{claim3} guarantees that under symmetric beliefs, the conditional beliefs of two agents about their unlabeled
extended neighborhood is the same if they are of isomorphic types. 

\begin{claim}
\label{claim3} \textit{For any }$p\in\Delta_{A}\left(T\right)$ satisfies
$\hat{p}\left(\mathbf{g}\right)=\hat{p}\left(\mathbf{g}^{\prime}\right)\text{ whenever }\mathbf{g},\mathbf{g}^{\prime}\in\left[\mathbf{g}^{k}\right]$\textit{
for some $k$, we have that $p\left(\left.t_{j}\right|t_{i}\right)=p\left(\left.t_{s}\right|t_{q}\right)$
whenever $t_{s} = f\left(t_{j}\right)$ and $f\left(t_{i}\right) = t_{q}$ for some bijection $f:N\rightarrow N$. } 
\end{claim}

The Claim implies that \textit{$p\left(\left.f\left(t_{j_{s}}\right)\right|f\left(t_{j_{s-1}}\right)\right)=p\left(\left.t_{j_{s}}\right|t_{j_{s-1}}\right)$
}for $1\leq s\leq l$ and \textit{$p\left(\left.f\left(t_{j_{1}}\right)\right|t_{q}\right)=p\left(\left.t_{j_{1}}\right|t_{i}\right)$
}. This is because $f\left(t_{i}\right)=t_q$. Hence, for any $l\geq1$,
\begin{align*}
\beta_{q,t_{q}}^{(m+l)} & =\sum_{j_{1}\in N_{i}\left(\mathbf{g}\right)}\ldots\sum_{j_{m}\in N_{j_{m-1}}\left(\mathbf{g}\right)}\text{\ensuremath{\sum_{\left(t_{j_{1}},\ldots,t_{j_{l}}\right)\in\mathcal{T}_{t_{i}}}} }p(t_{j_{l}}|t_{j_{l-1}})\ldots p(t_{j_{1}}|t_{i})=\beta_{i,t_{i}}^{(m+l)}
\end{align*}
Combining Cases 1 and 2 gives 
\[
a_{i}^{*}\left(t_{i}\left(\mathbf{g}\right)\right)=c^{-1}\sum_{s=0}^{\infty}\kappa^{s}\beta_{i,t_{i}}^{(s)}=c^{-1}\sum_{s=0}^{\infty}\kappa^{s}\beta_{q,t_{q}}^{(s)}=a_{q}^{*}\left(t_{q}\left(\mathbf{g}^{\prime}\right)\right)
\]

\subsection*{Proof of Theorem \ref{thm:tighter}}

Let $\{v_k\}_{k=1}^m$ be an orthonormal eigenbasis with $Pv_k=\rho_k v_k$.
Since $\langle x,\mathbf 1\rangle_\pi=0$, write $x=\sum_{k\ge 2}\alpha_k v_k$ with $\sum_{k\ge 2}\alpha_k^2=\|x\|_{2,\pi}^2=1$.
Then
\[
r^{(1,1)}=\langle x,Px\rangle_\pi=\sum_{k\ge 2}\rho_k\alpha_k^2\ \ge\ \rho_2\alpha_2^2=\rho_*\gamma,
\]
where $\alpha_k=\langle x,v_k\rangle_\pi$ and where the inequality follows by the postive spectrum assumption. Note that by construction $\|x\|_{2,\pi}^2=\sum_i\pi_i x_i^2=1$ and $\|v_2\|_{2,\pi}=1$. Hence Cauchy--Schwarz gives
$|\langle x,v_2\rangle_\pi|\le 1$, and therefore $\gamma=\langle x,v_2\rangle_\pi^2\leq 1$.

It remains to show that $\gamma>0$. Under RBD, the eigenvector $v_2$ may be chosen to be increasing (see Lemma 22.17 in \cite{levin2017markov}). Since $\langle v_2,\mathbf{1}\rangle_\pi=0$, it cannot be constant. Therefore there exist $u<u'$ such that
$
v_{2,u}<v_{2,u'}.
$
Using $\langle x,\mathbf 1\rangle_\pi=0$ and $\langle v_2,\mathbf 1\rangle_\pi=0$, we may write
\[
\langle x,v_2\rangle_\pi
=
\langle x,v_2\rangle_\pi
-
\langle x,\mathbf 1\rangle_\pi\langle v_2,\mathbf 1\rangle_\pi
=
\frac12\sum_{s,q=1}^{n-1}\pi_s\pi_q\,(x_s-x_q)(v_{2,s}-v_{2,q}).
\]
Because $x$ is strictly increasing, and $v_2$ is increasing, every summand in the sum is nonnegative. In addition, for the pair $(s,q)=(u',u)$, both differences are strictly positive, so the corresponding term is strictly positive. Therefore $\langle x,v_2\rangle_\pi>0,$ which implies $\gamma\in(0,1].$

\textbf{Mean lower bound.} As in the proof of Proposition \ref{prop:mv}, we have
\[
\mathbb{E}_\pi[a(P)]-\mathbb{E}_\pi[a^H]
=
\sum_{s\ge 2}\lambda^s
\sum_{\substack{b\in\{0,1\}^s\\ |b|>1}}
\sigma^{|b|}\bar d^{\,s-|b|}\,r^b.
\]
Under Path-wise Positive Assortativity, every term in the sum is positive. Hence we may retain only the subfamily $|b|=2$ and then further retain only
the adjacent two-mark pattern $(1,1,0,\dots,0)$, whose moment equals $r^{(1,1)}$ by reversibility. This gives
\[
\mathbb{E}_\pi[a(P)]-\mathbb{E}_\pi[a^H]
\ \ge\
c^{-1}\sum_{s\ge 2}\kappa^s\,\sigma^2\bar d^{\,s-2}\,r^{(1,1)}
=
\frac{\kappa^2\sigma^2c^{-1}}{1-\kappa\bar d}\,r^{(1,1)}.
\]
Using the bound for $r^{(1,1)}$, the fact that $\bar d\le \xi$  yields, and $1-\kappa\xi\in(0,1]$ yields
\[
\mathbb{E}_\pi[a(P)]-\mathbb{E}_\pi[a^H]
\ \ge\
\frac{\kappa^2\sigma^2c^{-1}}{1-\kappa\xi}\,\gamma\rho_*
=
\frac{\kappa^2\sigma^2c^{-1}}{(1-\kappa\xi)^3}\,(1-\kappa\xi)^2\,\gamma\rho_*\ge\
\frac{\kappa^2\sigma^2c^{-1}}{(1-\kappa\xi)^3}\,\gamma\rho_*.
\]

\textbf{Variance lower bound.} As in the proof of Proposition \ref{prop:mv}, we have
\[
\operatorname{Var}_\pi(a(P))
=
c^{-2}\sum_{s,\ell\ge 1}\lambda^{s+\ell}
\Big(\mathbb{E}_\pi[g^{(s)}g^{(\ell)}]-\mathbb{E}_\pi[g^{(s)}]\mathbb{E}_\pi[g^{(\ell)}]\Big),
\]
and each covariance block can be written (by the two-arm representation) as a finite sum with nonnegative coefficients multiplying
terms of the form $r^{b,c}-r^b r^c$. Under Path-wise Positive Assortativity, $r^{b,c}-r^b r^c\ge 0$ for all $(b,c)$, hence every block in the expansion is nonnegative.
Therefore we may retain only the $(s,\ell)=(2,2)$ contribution and within it only the pattern $b=c=(1,1)$, which yields
\[
\operatorname{Var}_\pi(a(P))-\operatorname{Var}_\pi(a^H)\ \ge\ c^{-2}\kappa^4\sigma^4\,(r^{(1,1)})^2.
\]
Using the fact that $(r^{(1,1)})^2\ge \gamma^2\rho_*^2$, and $\kappa\xi<1$,
we may write
\[
\operatorname{Var}_\pi(a(P))-\operatorname{Var}_\pi(a^H)
\ \ge\,\frac{\kappa^4\sigma^4c^{-2}}{(1-\kappa\xi)^6}\,(1-\kappa\xi)^2\,\gamma^2\rho_*^2
\].

\clearpage
\renewcommand{\thepage}{OA.\arabic{page}}
\setcounter{page}{1}
\setcounter{footnote}{0}
\setcounter{lem}{3}
\setcounter{claim}{2}
\setlength{\droptitle}{0pt}
\begin{center}
{\Large\bfseries Online Appendix to ``Network Beliefs and Behavior with Peer Effects''\par}
\vspace{0.75em}
{Promit K. Chaudhuri, Matthew O. Jackson, Sudipta Sarangi, and Hector Tzavellas\par}
\vspace{0.5em}
{May 2026\par}
\end{center}
\begin{abstract}
\begin{singlespace}
This Online Appendix is structured as follows. In Appendix B we present omitted proofs and derivations. Appendix C extends our analysis of Iterative Belief Centrality to setting where individuals hold exogenously imposed biased perspectives on the network. Appendix D presents additional examples. \\
\end{singlespace}

\end{abstract}
\newpage{}

\renewcommand{\theequation}{O.\arabic{equation}}
\providecommand{\theHequation}{O.\arabic{equation}}
\renewcommand{\theHequation}{O.\arabic{equation}}
\setcounter{equation}{0}
\setlength\parindent{0pt}
\setlength{\parskip}{8pt}

\section*{Appendix B. Omitted Proofs and Derivations}

\subsection*{Proof of Lemma~\ref{lem:admissibility}}

Let $p\in \Delta_A(T)$ and $\mathbf t\in T$ satisfy $p(\mathbf t)>0$. By compatibility,
$g(\mathbf t)\neq \emptyset$ (Consistency) and $|g(\mathbf t)|=1$
(Identification). Hence there exists a unique $\mathbf g\in \mathcal G_n$
such that $g(\mathbf t)=\{\mathbf g\}$.

Fix $\mathbf g\in \mathcal G_n$ and define a type profile
$\mathbf t(\mathbf g)=(t_1(\mathbf g),\ldots,t_n(\mathbf g))\in T$ by
\[
t_i(\mathbf g):=\big(\mathbf g_j\big)_{j\in \mathcal A_i(\mathbf g)}
\quad\text{for each } i\in N.
\]
By construction, $\mathbf g\in g(\mathbf t(\mathbf g))$, so
$g(\mathbf t(\mathbf g))\neq \emptyset$. Moreover, if $\mathbf g'\in g(\mathbf t(\mathbf g))$,
then $t_i(\mathbf g')=t_i(\mathbf g)$ for all $i\in N$, hence $\mathbf t(\mathbf g')=\mathbf t(\mathbf g)$.
In particular, $\mathbf g$ and $\mathbf g'$ generate the same type profile.

Suppose $g(\mathbf t(\mathbf g))=\{\mathbf g\}$ fails. Then there exists $\mathbf g'\neq \mathbf g$
such that $\mathbf g'\in g(\mathbf t(\mathbf g))$. Consider the type profile $\mathbf t(\mathbf g)$.
If $p(\mathbf t(\mathbf g))>0$, compatibility (Identification) would imply
$|g(\mathbf t(\mathbf g))|=1$, a contradiction. Hence $p(\mathbf t(\mathbf g))=0$
whenever $|g(\mathbf t(\mathbf g))|>1$. Therefore, for every $\mathbf g\in\mathcal G_n$
there exists a unique $\mathbf t\in T$ on the support of $p$ such that $g(\mathbf t)=\{\mathbf g\}$.
(Equivalently, the map $\mathbf g\mapsto \mathbf t(\mathbf g)$ restricted to the support of $p$
is injective.)

Fix $p\in\Delta_A(T)$ and define $\psi\left(p\right)=\hat p$ by
\[
\hat p\left(\mathbf g\right) = p \left(\mathbf{t}\left(\mathbf{g}\right)\right),\qquad\text{for all } \mathbf{g}\in\mathcal{G}_n
\]
This is well defined since $\mathbf{t}\left(\mathbf{g}\right)\in T$ is unique for each $\mathbf{g}\in\mathcal{G}_n$. Moreover, $\hat p(\mathbf g)\ge 0$ for all $\mathbf g$ and
\[
\sum_{\mathbf g\in\mathcal G_n}\hat p(\mathbf g)
=
\sum_{\mathbf g\in\mathcal G_n}\sum_{\mathbf t:\ g(\mathbf t)=\{\mathbf g\}} p(\mathbf t)
=
\sum_{\mathbf t\in T} p(\mathbf t)
=
1,
\]
so $\hat p\in\Delta(\mathcal G_n)$.

To prove injectivity, suppose that $\psi(p)=\psi(p')$. Let $\mathbf{t}\in T$ be arbitrary.

If $g(\mathbf{t})=\emptyset$, then by compatibility of $p$ and $p'$, we must have
\[
p(\mathbf{t})=p'(\mathbf{t})=0.
\]

If $g(\mathbf{t})=\{\mathbf{g}\}$ for some $\mathbf{g}\in \mathcal{G}_n$, then the uniqueness of the type profile induced by $\mathbf{g}$ implies that $\mathbf{t}=\mathbf{t}\left(\mathbf{g}\right)$. Therefore,
\[
p(\mathbf{t})=p(\mathbf{t}(\mathbf{g}))=(\psi (p))(\mathbf{g})=(\psi(p'))(\mathbf{g})=p'(\mathbf{t}(\mathbf{g}))=p'(\mathbf{t}).
\]
Thus, in either case, $p(\mathbf{t})=p'(\mathbf{t})$. Since this holds for every $\mathbf{t}\in T$, it follows that $p=p'$. Hence, $\psi$ is injective.

To prove surjectivity, take any $\hat p\in \Delta(\mathcal{G}_n)$. Define a probability distribution $p\in \Delta_A(T)$ by setting
\[
p(\mathbf{t}(\mathbf{g})):=\hat p(\mathbf{g})\qquad \text{for each } \mathbf{g}\in \mathcal{G}_n,
\]

Because each network $\mathbf{g}\in \mathcal{G}_n$ induces a unique type profile $t(g)$, the above definition is well defined. Also, $p$ is compatible by construction. Finally, for every $\mathbf{g}\in \mathcal{G}_n$,
\[
\psi(p)(\mathbf{g})=p(\mathbf{t}(\mathbf{g}))=\hat p(\mathbf{g}).
\]
Therefore, $\psi(p)=\hat p$, establishing surjectivity. And hence $\psi$ is a bijection.

\subsection*{Proof of Lemma \ref{lem:exist}}

The best response of each player is given by
\[
a_{i}\left(t_{i}\right)=\begin{cases}
c^{-1}+\kappa\sum_{j\in N}g_{ij}^{t_{i}}\sum_{\mathbf{t}_{j}\in T_{j}}p\left(\left.t_{j}\right|t_{i}\right)a_{j}\left(t_{j}\right) & \text{ if }c^{-1}+\kappa\sum_{j\in N}g_{ij}^{t_{i}}\sum_{\mathbf{t}_{j}\in T_{j}}p\left(\left.t_{j}\right|t_{i}\right)a_{j}\left(t_{j}\right)\geq0\\
0 & \text{ otherwise}
\end{cases}
\]
Observe that, if $\lambda>0$, then $\kappa>0$ and thus $c^{-1}+\kappa\sum_{j\in N}g_{ij}^{t_{i}}\sum_{t_{j}\in T_{j}}p\left(\left.t_{j}\right|t_{i}\right)a_{j}\left(t_{j}\right)\geq0$
holds true always. Hence, for $\lambda>0$ the best response is given
by $c^{-1}+\kappa\sum_{j\in N}g_{ij}^{t_{i}}\sum_{\mathbf{t}_{j}\in T_{j}}p\left(\left.t_{j}\right|t_{i}\right)a_{j}\left(t_{j}\right)$
and for any equilibrium that exists it is always interior. Using standard contraction argument it can be shown that equilibrium exists and unique when $0<\kappa<1/d_{max}$. 

We prove it for the case when $\lambda<0$.

This lemma has two parts. In the first part we prove uniqueness
along with existence of equilibrium and in the second part we prove
interiority of the equilibrium. 

\noindent{\textbf{Part I: Existence and Uniequness}}

We have $\left|T_{i}\right|=\gamma_{i}$ and $\left|T\right|=\gamma=\sum_{i\in N}\gamma_{i}$.
For each $i\in N$ define a function
\[
\varphi_i:T_{i}\longrightarrow\left\{ 1,\ldots,\gamma_i\right\} 
\]
and $\sum_{i\in N}\gamma_i = \gamma$. Then consider $K=\left[0,c^{-1}\right]^{\gamma}$ and $F:\left[0,c^{-1}\right]^{\gamma}\longrightarrow\mathbb{R}^{\gamma}$
such that 
\[
F\left(\mathbf{a}\right)=\left(F_{1}\left(\mathbf{a}\right),F_{2}\left(\mathbf{a}\right),\ldots,F_{n}\left(\mathbf{a}\right)\right)
\]
with $F_{i}\left(\mathbf{a}\right):\left[0,c^{-1}\right]^{\gamma}\longrightarrow\mathbb{R}^{\gamma_{i}}$
and $F_{i}\left(\mathbf{a}\right)=\left(F_{i,1}\left(\mathbf{a}\right),F_{i,2}\left(\mathbf{a}\right),\ldots,F_{i,\gamma_{i}}\left(\mathbf{a}\right)\right)$
where 
\[
F_{i,s}\left(\mathbf{a}\right)=a_{i}\left(t_{i}\right)-\kappa\sum_{j\in N}g_{ij}^{t_{i}}\sum_{t_{j}\in T_{j}}p\left(\left.t_{j}\right|t_{i}\right)a_{j}\left(t_{j}\right)-c^{-1}
\]
for some $t_{i}\in T_{i}$ such that $\varphi_i\left(t_{i}\right)=s$. 
\begin{defn}
A vector $\mathbf{x}^{*}\in\mathbb{R}^{n}$ solves the \textbf{Variational
Inequality $VI\left(K,F\right)$ }with set $K\subseteq\mathbb{R}^{n}$
and an operator $F:K\rightarrow\mathbb{R}^{n}$ if and only if
\[
\left(\mathbf{x}-\mathbf{x}^{*}\right)^{T}F\left(\mathbf{x}^{*}\right)\geq0\qquad\qquad\forall\mathbf{x\in}K
\]
\end{defn}

From \cite{zenou2024sign}, since the game is well behaved,
we have for $\mathbf{a}^{*}=\left(\mathbf{a^{*}}_{i}\right)_{i\in N}$ such that:
\[
a_{i}^{*}\left(t_{i}\right)\in\arg\max_{x\in\left[0,c^{-1}\right]}u_{i}\left(x;\mathbf{a}_{i}^{-t_{i}},\mathbf{a}_{-i}\right)\quad\forall t_{i}\in T_{i},i\in N\iff\left(\mathbf{a}-\mathbf{a}^{*}\right)^{T}F\left(\mathbf{a}^{*}\right)\geq0\qquad\forall\mathbf{a\in}K
\]

Let
\[
\nabla{}_{\mathbf{a}}F\left(\mathbf{a}\right)=\left(\begin{array}{cccc}
A_{11} & \ldots & \ldots & A_{1\gamma}\\
\vdots & \ldots & \ldots & \vdots\\
\vdots & \ldots & \ldots & \vdots\\
A_{\gamma1} & \ldots & \ldots & A_{\gamma\gamma}
\end{array}\right)
\]
where
\begin{itemize}
\item For $q\in\left\{ 1,\ldots,\gamma\right\} $ such that $\varphi\left(t_{i}\right)=\sum_{j<q}\left|\gamma_{j}\right|+q$
for some $i\in N$, $t_{i}\in T_{i}$, $A_{qq}=\frac{\partial F_{i,q}\left(\mathbf{a}\right)}{\partial a_{i}\left(t_{i}\right)}=1$.
\item For $q,s\in\left\{ 1,\ldots,\gamma\right\} $ such that $\varphi\left(t_{i}\right)=\sum_{j<q}\left|\gamma_{j}\right|+q$
and $\varphi\left(t_{i}^{\prime}\right)=\sum_{j<q}\left|\gamma_{j}\right|+s$
for some $i\in N$, $t_{i},t_{i}^{\prime}\in T_{i}$,
$A_{qs}=\frac{\partial F_{i,q}\left(\mathbf{a}\right)}{\partial a_{i}\left(t_{i}^{\prime}\right)}=0$. 
\item For $q,s\in\left\{ 1,\ldots,\gamma\right\} $ such that $\varphi\left(t_{i}\right)=\sum_{j<q}\left|\gamma_{j}\right|+q$
and $\varphi\left(t_{j}\right)=\sum_{j<q}\left|\gamma_{j}\right|+s$
for some $i,j\in N$, $t_{i}\in T_{i}$ and $t_{j}\in T_{j}$,
$A_{qs}=\frac{\partial F_{i,q}\left(\mathbf{a}\right)}{\partial a_{j}\left(t_{j}\right)}=-\kappa g_{ij}^{t_{i}}p\left(\left.t_{j}\right|t_{i}\right)$.
\end{itemize}
Also define 
\[
\varUpsilon=\left(\begin{array}{cccc}
\zeta_{11} & -\zeta_{12} & \ldots & -\zeta_{1\gamma}\\
-\zeta_{21} & \zeta_{22} & \ldots & -\zeta_{2\gamma}\\
\vdots & \ldots & \ldots & \vdots\\
-\zeta_{\gamma1} & \ldots & \ldots & \zeta_{\gamma\gamma}
\end{array}\right)
\]
where
\begin{itemize}
\item For $q\in\left\{ 1,\ldots,\gamma\right\} $ such that $\varphi\left(t_{i}\right)=\sum_{j<q}\left|\gamma_{j}\right|+q$
for some $i\in N$, $t_{i}\in T_{i}$, $\zeta_{qq}=A_{qq}=1$.
\item For $q,s\in\left\{ 1,\ldots,\gamma\right\} $ such that $\varphi\left(t_{i}\right)=\sum_{j<q}\left|\gamma_{j}\right|+q$
and $\varphi\left(t_{i}^{\prime}\right)=\sum_{j<q}\left|\gamma_{j}\right|+s$
for some $i\in N$, $t_{i},t_{i}^{\prime}\in T_{i}$,
$\zeta_{qs}=A_{qs}=0$. 
\item For $q,s\in\left\{ 1,\ldots,\gamma\right\} $ such that $\varphi\left(t_{i}\right)=\sum_{j<q}\left|\gamma_{j}\right|+q$
and $\varphi\left(t_{j}\right)=\sum_{j<q}\left|\gamma_{j}\right|+s$
for some $i,j\in N$, $t_{i}\in T_{i}$ and $t_{j}\in T_{j}$,
$\zeta_{qs}=A_{qs}=-\kappa g_{ij}^{t_{i}}p\left(\left.t_{j}\right|t_{i}\right)$.
\end{itemize}
\begin{prop*}
(\cite{facchinei201012}) If $\varUpsilon$is a P-Matrix (i.e. determinants
of all the principal minors are positive) then the $VI\left(K,F\right)$
admits a unique solution in $\left[0,c^{-1}\right]^{\gamma}$.
\end{prop*}

The following proposition gives a condition for a matrix to be a P-Matrix.
\begin{prop*}
(\cite{mckenzie1960matrices}) If a matrix with strictly dominant diagonals has
positive diagonal entries then it is a P-Matrix. 
\end{prop*}

Let $d_{max}=\max\left\{ \sum_{j\in N}g_{ij}^{t_{i}}:t_{i}\in T_{i}\wedge i\in N\right\} $
then for any $i\in N$ and $t_{i}\in T_{i}$, we have $\left|\kappa\right|d_{i}^{t_{i}}<\left|\kappa\right|d_{max}$.
And $\left|\kappa\right|d_{max}<1\Rightarrow\left|\kappa\right|d_{i}^{t_{i}}<1$
for all $i\in N,t_{i}\in T_{i}$. Hence, 
\[
\left|\kappa\right|\sum_{j\in N}\sum_{t_{j}\in T_{k}}g_{ij}^{t_{i}}p\left(\left.t_{j}\right|t_{i}\right)=\left|\kappa\right|d_{i}^{t_{i}}<1,\quad\forall i\in N,t_{i}\in T_{i}
\]
or equivalently
\[
\sum_{s\neq q}\left|A_{qs}\right|<\left|A_{qq}\right|\qquad\forall q\in\left\{ 1,\ldots,\gamma\right\}
\]
Thus, $\varUpsilon$ is a diagonally dominant matrix with positive
diagonal elements and hence it is a P-Matrix. And thus, $VI\left(K,F\right)$
admits a unique solution in $\left[0,c^{-1}\right]^{\gamma}$.

\noindent{\textbf{Part II: Interiority}}

Suppose for some $t_i \in T$, $a^*\left(t_i\right)=0$. Then from the best responses, we have that 
\[
c^{-1}+\kappa\sum_{j\in N}g_{ij}^{t_{i}}\sum_{\mathbf{t}_{j}\in T_{j}}p\left(\left.t_{j}\right|t_{i}\right)a_{j}\left(t_{j}\right) \leq 0 \Rightarrow \sum_{j\in N}g_{ij}^{t_{i}}\sum_{\mathbf{t}_{j}\in T_{j}}p\left(\left.t_{j}\right|t_{i}\right)a_{j}\left(t_{j}\right)\geq \frac{c^{-1}}{\left|\kappa\right|}
\]
But, we have that $a_j\left(t_j\right) \leq c^{-1}$ for all $t_j\in T$, and hence 
\[
\sum_{j\in N}g_{ij}^{t_{i}}\sum_{\mathbf{t}_{j}\in T_{j}}p\left(\left.t_{j}\right|t_{i}\right)a_{j}\left(t_{j}\right) \leq c^{-1} \sum_{j\in N}g_{ij}^{t_{i}}\sum_{\mathbf{t}_{j}\in T_{j}}p\left(\left.t_{j}\right|t_{i}\right) = c^{-1}d_i^{t_i}\leq c^{-1}d_{max}
\]
Thus comparing, the two bounds on $\sum_{j\in N}g_{ij}^{t_{i}}\sum_{\mathbf{t}_{j}\in T_{j}}p\left(\left.t_{j}\right|t_{i}\right)a_{j}\left(t_{j}\right)$, we get
\[
c^{-1}d_{max} \geq \frac{c^{-1}}{\left|\kappa\right|} \Rightarrow d_{max} \geq \frac{1}{\left|\kappa\right|}
\]
But, this is a contradiction, since $\left|\kappa\right|d_{max}\leq 1$. Therefore, $a^*\left(t_i\right)> 0$ for all $t_i \in T$. 

\subsection*{Proof of Corollary \ref{cor:cutoff_homogeneous}}

Fix $d\in\{0,1,\ldots,n-1\}$ and write
\[
\bar a(d):=\mathbb E\!\left[a_i^{\star}(t_i)\mid d(t_i)=d\right].
\]
Using the first-order condition
\[
a_i^{\star}(t_i)
=
c^{-1}+\kappa\sum_{j:g_{ij}^{t_i}=1}
\mathbb E\!\left[a_j^{\star}(t_j)\mid t_i\right],
\]
and conditioning on the event $d(t_i)=d$, we obtain
\begin{equation}
\bar a(d)=c^{-1}+\kappa\,d\,m(d),
\label{eq:adegproof}
\end{equation}
where
\[
m(d):=\mathbb E[m(t_i)\mid d(t_i)=d]
\qquad\text{and}\qquad
m(t_i):=\mathbb E\!\left[a_j^{\star}(t_j)\mid t_i,\ g_{ij}^{t_i}=1\right].
\]
Under homogeneous beliefs, equilibrium actions satisfy
$a^{H}(d)=c^{-1}+\kappa d\,m^{H}$.

We first show that $m(d)$ is weakly increasing in $d$ under Row-Based Dominance. Fix an agent $i$ and two row types $t_i,t_i'\in T_i$ with $d(t_i')>d(t_i)$.  
Apply Theorem~\ref{thm:iterated_multiplier} to the one-step path
$j_0=i,j_1=j$.  
The theorem implies that, for any neighbor $j$ of $i$,
\[
\mathbb E\!\left[a_j^{\star}(t_j)\,\middle|\, t_i'\right]
\;\ge\;
\mathbb E\!\left[a_j^{\star}(t_j)\,\middle|\, t_i\right].
\]
Because $m(t_i)$ is itself an expectation of $a_j^{\star}(t_j)$ taken over neighbors of $i$,
this inequality implies
\[
m(t_i')\ge m(t_i)
\qquad\text{whenever } d(t_i')>d(t_i).
\]
Averaging over row types within each degree class yields that
$m(d)$ is weakly increasing in $d$.

Next, let:
\[
\Delta(d):=\bar a(d)-\big(c^{-1}+\kappa d\,m^{H}\big).
\]
Using \eqref{eq:adegproof}, we obtain
\[
\Delta(d)=\kappa d\big(m(d)-m^{H}\big).
\]
Since $\lambda>0 \Rightarrow \kappa >0$ and $d\ge 0$, the sign of $\Delta(d)$ coincides with the sign of
$m(d)-m^{H}$. Because $m(\cdot)$ is weakly increasing and the assumption of the corollary provides
degrees $d_1<d_2$ such that
\[
m(d_1)<m^{H}<m(d_2),
\]
there exists a cutoff degree $d^{\ast}$ such that
$m(d)\le m^{H}$ for $d<d^{\ast}$ and $m(d)\ge m^{H}$ for $d>d^{\ast}$. Therefore,
\[
\bar a(d)<c^{-1}+\kappa d\,m^{H}\quad\text{for all }d<d^{\ast},
\qquad
\bar a(d)>c^{-1}+\kappa d\,m^{H}\quad\text{for all }d>d^{\ast},
\]
which establishes the desired cutoff comparison.

\subsection*{Proof of Claim \ref{claim2}}

Suppose $\left(t_{j_{1}},\ldots,t_{j_{l}}\right)\in\mathcal{T}_{t_{i}}$,
then $g\left(t_{i}\right)\bigcap_{k=1}^{l}g\left(t_{j_{k}}\right)\neq\phi$.
Let $\tilde{\mathbf{g}}\in g\left(t_{i}\right)\bigcap_{k=1}^{l}g\left(t_{j_{k}}\right)$,
then 
\[
\mathcal{A}_{i}\left(\mathbf{g}\right)=N_{i}^{m}\left(\mathbf{g}\right)=N_{i}^{m}\left(\tilde{\mathbf{g}}\right)
\]
Since for agent $i$, the types that are induced by the graphs $\mathbf{g}$
and $\tilde{\mathbf{g}}$ are same, they have the same neighborhood
up to $\left(m-1\right)^{th}$ order under $\mathbf{g}$ as well as
$\tilde{\mathbf{g}}$. Moreover, the order of connections is also
preserved, 
\[
\tilde{g}_{js}=g_{js},\qquad\forall j\in N_{i}^{m}\left(\mathbf{g}\right)
\]

Then, by the construction of $\tilde{\mathbf{g}}^{f}$, we have that,
\[
\tilde{g}_{f\left(j\right)f\left(s\right)}^{f}=\tilde{g}_{js}=g_{js}=g_{f\left(j\right)f\left(s\right)}^{\prime},\qquad\forall j\in N_{i}^{m}\left(\mathbf{g}\right)
\]
where the third equality is due to the fact that $t_{i}$ and $t_{q}$
are isomorphic, i.e., $\mathbf{g}$ and $\mathbf{g}^{\prime}$ have
the same local structure with respect to $i$ and $q=f\left(i\right)$,
where $\mathcal{A}_{i}\left(\mathbf{g}\right)=N_{i}^{m}\left(\mathbf{g}\right)$
and $\mathcal{A}_{q}\left(\mathbf{g}\right)=N_{q}^{m}\left(\mathbf{g}^{\prime}\right)$.

Next, using the fact that $\mathbf{g},\tilde{\mathbf{g}}$ have the
same local structure from $i$'s point of view, we get, 
\[
N_{q}^{m}\left(\tilde{\mathbf{g}}^{f}\right)=\left\{ f\left(j_{k}\right):j_{k}\in N_{i}^{m}\left(\tilde{\mathbf{g}}\right)\right\} =\left\{ f\left(j_{k}\right):j_{k}\in N_{i}^{m}\left(\mathbf{g}\right)\right\} =N_{q}^{m}\left(\mathbf{g}^{\prime}\right)
\]
Thus, 
\[
\tilde{\mathbf{g}}_{f\left(j\right)}^{f}=\mathbf{g}_{f\left(j\right)}^{\prime},\qquad\forall j\in N_{i}^{m}\left(\tilde{\mathbf{g}}\right)
\]
which results in the construction of $\mathcal{A}_{q}\left(\mathbf{g}^{\prime}\right)$
from $\tilde{\mathbf{g}}^{f}$, i.e. 
\[
\mathcal{A}_{q}\left(\mathbf{g}^{\prime}\right):=N_{q}^{m}\left(\mathbf{g}^{\prime}\right)=N_{q}^{m}\left(\tilde{\mathbf{g}}^{f}\right)
\]
Hence, we have $\tilde{\mathbf{g}}^{f}\in g\left(t_{q}\right)$. Similarly
it can be proved that , $\tilde{\mathbf{g}}^{f}\in g\left(t_{q}\right)\Rightarrow\tilde{\mathbf{g}}\in g\left(t_{i}\right)$.
It then follows from Definition 1 that $\tilde{\mathbf{g}}\in\bigcap_{k=1}^{l}g\left(t_{j_{k}}\right)$
iff $\tilde{\mathbf{g}}^{f}\in\bigcap_{k=1}^{l}g\left(f\left(t_{j_{k}}\right)\right)$.
Hence, $g\left(t_{i}\right)\bigcap_{k=1}^{l}g\left(t_{j_{k}}\right)\neq\phi$
iff $g\left(t_{q}\right)\bigcap_{k=1}^{l}g\left(f\left(t_{j_{k}}\right)\right)\neq\phi$.

Again, say $\tilde{\mathbf{g}}\in g\left(t_{i}\right)\bigcap_{k=1}^{l}g\left(t_{j_{k}}\right)$
and $\left(t_{j_{1}},\ldots,t_{j_{l}}\right)\in\mathcal{T}_{t_{i}}$
implies $j_{m+s}\in N_{j_{m+s-1}}\left(\tilde{\mathbf{g}}_{j_{m+s-1}}\right)$
$\text{for all }1\leq s\leq l$. From the definition of $\tilde{\mathbf{g}}^{f}$,
we have that $f\left(j_{m+s}\right)\in N_{f\left(j_{m+s-1}\right)}\left(\tilde{\mathbf{g}}^{f}\right)$
$\text{for all }1\leq s\leq l$. Similarly it can be show that $f\left(j_{m+s}\right)\in N_{f\left(j_{m+s-1}\right)}\left(\tilde{\mathbf{g}}^{f}\right)$
implies $j_{m+s}\in N_{j_{m+s-1}}\left(\tilde{\mathbf{g}}\right)$.

$\blacksquare$

\subsection*{Proof of Claim \ref{claim3}}

Define a relation $\mathcal{R}$ on $\mathcal{G}_{n}^{U}$ such that,
$\mathbf{g}_{u}\mathcal{R}\mathbf{g}_{v}$ iff $\mathbf{g}_{u}$ is
isomorphic to $\mathbf{g}_{v}$. It follows that $\mathcal{R}$ is
an equivalence relation on $\mathcal{G}_{n}^{U}$. Let $\left\{ \left[\mathbf{g}^{1}\right],\left[\mathbf{g}^{2}\right],\ldots,\left[\mathbf{g}^{q}\right]\right\} $
be the set of all equivalence classes defined on $\mathcal{G}_{n}$.
Define a prior distribution $p\in\Delta_{A}\left(T\right)$ such that
for some $k\in\left\{ 1,\ldots,q\right\} ,$we have $\hat{p}\left(\mathbf{g}\right)=\hat{p}\left(\mathbf{g}^{\prime}\right)\text{ whenever }\mathbf{g},\mathbf{g}^{\prime}\in\left[\mathbf{g}^{k}\right]$.

Let $\mathbf{g}\in g\left(t_{i}\right)\cap\left[\mathbf{g}^{l}\right]$
and $f:N\longrightarrow N$ be a bijection with $f\left(i\right)=q$
and $i\sim j$ if and only if $f\left(i\right)\sim f\left(j\right)$.
Let $\mathbf{g}^{f}$ be such that $g_{f\left(i\right)f\left(j\right)}^{f}=g_{ij}$.
It follows from Definition 1 that $\mathbf{g}$ and $\mathbf{g}^{f}$
are isomorphic. Hence, $\mathbf{g}^{f}\in\left[\mathbf{g}^{l}\right]$. Consider the type pairs $\left(t_{i},t_{j}\right)\in T_{i}^{\left(m\right)}\times T_{j}^{\left(m\right)}$
and $\left(t_{q},t_{s}\right)\in T_{q}^{\left(m\right)}\times T_{s}^{\left(m\right)}$such
that $t_{q}=f\left(t_{i}\right)$ and $t_{s}=f\left(t_{j}\right)$. We show that $\left|\left\{ \mathbf{g}\in\mathcal{G}_{n}^{U}:\mathbf{g}\in g\left(t_{i}\right)\cap\left[\mathbf{g}^{l}\right]\right\} \right|=\left|\left\{ \mathbf{g}\in\mathcal{G}_{n}^{U}:\mathbf{g}\in g\left(t_{q}\right)\cap\left[\mathbf{g}^{l}\right]\right\} \right|$
for all $l\in\left\{ 1,2,\ldots,m\right\} $.

We have $j\in N_{i}^{m}\left(\mathbf{g}\right)\Rightarrow f\left(j\right)\in N_{q}^{m}\left(\mathbf{g}^{f}\right)$
and vice versa. Hence, $\mathbf{g}^{f}\in g\left(t_{q}\right)$. As
a result, for every $\mathbf{g}\in g\left(t_{i}\right)\cap\left[\mathbf{g}^{l}\right]$
there exists a $\mathbf{g}^{f}\in g\left(t_{i}\right)\cap\left[\mathbf{g}^{l}\right]$.
Similarly we can show that for every $\mathbf{g}\in g\left(t_{q}\right)\cap\left[\mathbf{g}^{l}\right]$
there exists a $\mathbf{g}^{f}\in g\left(t_{i}\right)\cap\left[\mathbf{g}^{l}\right]$.
Thus, $\left|\left\{ \mathbf{g}\in\mathcal{G}_{n}:\mathbf{g}\in g\left(t_{i}\right)\cap\left[\mathbf{g}^{l}\right]\right\} \right|=\left|\left\{ \mathbf{g}\in\mathcal{G}_{n}:\mathbf{g}\in g\left(t_{q}\right)\cap\left[\mathbf{g}^{l}\right]\right\} \right|$.

A similar argument shows that 
\[
\left|\left\{ \mathbf{g}\in\mathcal{G}_{n}:\mathbf{g}\in g\left(t_{i}\right)\cap g\left(t_{j}\right)\cap\left[\mathbf{g}^{l}\right]\right\} \right|=\left|\left\{ \mathbf{g}\in\mathcal{G}_{n}:\mathbf{g}\in g\left(t_{q}\right)\cap g\left(t_{s}\right)\cap\left[\mathbf{g}^{l}\right]\right\} \right|
\]
so that 
\[
p\left(t_{i}\right)=\sum_{\mathbf{g}\in\mathcal{G}_{n}}\hat{p}\left(\mathbf{g}\right)\cdot\mathbb{I}\left\{ \mathbf{g}\in g\left(t_{i}\right)\right\} =\sum_{l=1}^{q}\sum_{\mathbf{g}\in\left[\mathbf{g}^{l}\right]}\hat{p}\left(\mathbf{g}\right)\cdot\mathbb{I}\left\{ \mathbf{g}\in g\left(t_{i}\right)\right\} .
\]

Lastly, since $\hat{p}\left(\mathbf{g}\right)=p_{l}$ for all $\mathbf{g}\in\left[\mathbf{g}^{l}\right]$
(i.e., equal probability is placed on all the graphs in the same equivalence
class), it follows that 
\begin{align*}
p\left(t_{i}\right) & =\sum_{l=1}^{q}p_{l}\sum_{\mathbf{g}\in\left[\mathbf{g}^{l}\right]}\mathbb{I}\left\{ \mathbf{g}\in g\left(t_{i}\right)\right\} =\sum_{l=1}^{m}p_{l}\cdot\left|\left\{ \mathbf{g}\in\mathcal{G}_{n}:\mathbf{g}\in g\left(t_{i}\right)\cap\left[\mathbf{g}^{l}\right]\right\} \right|\\
 & =\sum_{l=1}^{m}p_{l}\cdot\left|\left\{ \mathbf{g}\in\mathcal{G}_{n}:\mathbf{g}\in g\left(t_{q}\right)\cap\left[\mathbf{g}^{l}\right]\right\} \right|=p\left(t_{q}\right).
\end{align*}
Similarly, we can show $p\left(t_{i},t_{j}\right)=p\left(t_{q},t_{s}\right)$,
and thus, $p\left(\left.t_{j}\right|t_{i}\right)=p\left(\left.t_{s}\right|t_{q}\right)$.

\subsection*{Proof of Proposition \ref{prop:Degree-br}}

In the first order information, from the definition
of isomorphic types, $t_{i}\simeq t_{s}$ iff $\sum_{r\in N}g_{ir}^{t_{i}}=\sum_{r\in N}g_{sr}^{t_{s}}$,
for any $t_{i}\in T_{i}$ and $t_{s}\in T_{s}$. Then from Theorem
\ref{thm:same-action}, for all $u\in\left\{ 1,\ldots,n-1\right\} $
\[
a_{i}\left(t_{i}^{u}\right)=a_{s}\left(t_{s}^{u}\right)=a_{u}
\]
for any $i,s\in N$ with types $t_{i}^{u}\in T_{i}$, $t_{s}^{u}\in T_{s}$
such that $\sum_{r\in N}g_{ir}^{t_{i}^{u}}=\sum_{r\in N}g_{sr}^{t_{s}^{u}}=u$.
Then 
\begin{align*}
a_{i}\left(t_{i}^{u}\right) & =c^{-1}+\kappa\sum_{r\in N}g_{ir}^{t_{i}^{u}}\sum_{t_{r}\in T_{r}}p\left(\left.t_{r}\right|t_{i}^{u}\right)a_{r}\left(t_{r}\right)\\
 & =c^{-1}+\kappa\sum_{r\in N}g_{ir}^{t_{i}^{u}}\sum_{v=1}^{n-1}\sum_{t_{r}\in T_{r}}p\left(\left.t_{r}\right|t_{i}^{u}\right)\cdot\mathbb{I}\left\{ \sum_{s\in N}g_{rs}^{t_{r}}=v\right\} \cdot a_{v}
\end{align*}
Thus, for any $j\in N$ such that $g_{ij}^{t_{i}^{u}}=1$ we will
have that 
\[
a_{i}\left(t_{i}^{u}\right)=c^{-1}+\kappa u\sum_{v=1}^{n-1}\sum_{t_{j}\in T_{j}}p\left(\left.t_{j}\right|t_{i}^{u}\right)\cdot\mathbb{I}\left\{ \sum_{r\in N}g_{jr}^{t_{j}}=v\right\} \cdot a_{v}
\]
And since for any $t_{i}^{u}\in\cup_{q\in N}T_{q}$ such that $\sum_{r\in N}g_{ir}^{t_{i}^{u}}=u$,
we have that $a_{i}\left(t_{i}^{u}\right)=a_{u}$ and 
\[
a_{u}=c^{-1}+\kappa u\sum_{v=1}^{n-1}\sum_{t_{j}\in T_{j}}p\left(\left.t_{j}\right|t_{i}^{u}\right)\cdot\mathbb{I}\left\{ \sum_{r\in N}g_{jr}^{t_{j}}=v\right\} \cdot a_{v}
\]

\subsection*{Proof of Lemma \ref{lem:revers}}

Under symmetric priors, for any arbitrary $t_{i}^{u},t_{j}^{u}\in\cup_{q\in N}T_{q}$
such that $\sum_{r\in N}g_{ir}^{t_{i}^{u}}=\sum_{r\in N}g_{jr}^{t_{j}^{u}}=u$,
we have that $p\left(t_{i}^{u}\right)=p\left(t_{j}^{u}\right)=p_{u}$.
As a result,
$$
\rho_{u} =\sum_{t_{i}\in T_{i}}p\left(t_{i}\right)\cdot\mathbb{I}\left\{ \sum_{r=1}^{n}g_{ir}^{t_{i}}=u\right\}={n-1 \choose u}p_{u}
$$
Thus, 
$$
\pi_{u}  =\frac{u\cdot\rho_{u}}{\delta}=\frac{1}{\delta}\cdot\left(n-1\right){n-2 \choose u-1}p_{u}
$$
Moreover, for any arbitrary $t_{i}^{u}\in\cup_{q\in N}T_{q}$ such that
$\sum_{r\in N}g_{ir}^{t_{i}^{u}}=u$ and $g_{ij}^{t_{i}^{u}}=1$ we
have,
\begin{align*}
P_{uv} & =\sum_{t_{j}\in T_{j}}p\left(\left.t_{j}\right|t_{i}^{u}\right)\cdot\mathbb{I}\left\{ \sum_{r\in N}g_{jr}^{t_{j}}=v\right\}=\frac{1}{p\left(t_{i}^{u}\right)}\cdot\sum_{t_{j}\in T_{j}}p\left(t_{j},t_{i}^{u}\right)\cdot\mathbb{I}\left\{ \sum_{r\in N}g_{jr}^{t_{j}}=v\right\}\\
 & =\frac{1}{p_{u}}\cdot\sum_{\mathbf{g}\in\mathcal{G}_{n}}\hat{p}\left(\mathbf{g}\right)\cdot\mathbb{I}\left\{ \mathbf{g}_{i}=t_{i}^{u}\wedge\sum_{r\in N}g_{jr}=v\wedge g_{ij}=1\right\} 
\end{align*}

\noindent{\textit{Statement} $(i)$:}

We have$\left[\pi^{T}\boldsymbol{P}\right]_{v}=\sum_{u=1}^{n-1}\pi_{u}P_{uv}$.
Moreover,
\[
\pi_{u}P_{uv}=\left(\frac{n-1}{\delta}\right){n-2 \choose u-1}\sum_{\mathbf{g}\in\mathcal{G}_{n}}\hat{p}\left(\mathbf{g}\right)\cdot\mathbb{I}\left\{ \mathbf{g}_{i}=t_{i}^{u}\wedge\sum_{r\in N}g_{jr}=v\wedge g_{ij}=1\right\} 
\]
As a consequence of symmetric priors we have that for any two types
$t_{i}^{u},\tilde{t}_{i}^{u}\in T_{i}$ such that $\sum_{r=1}^{n}g_{ir}^{t_{i}^{u}}=\sum_{r=1}^{n}g_{ir}^{\tilde{t}_{i}^{u}}=u$
and $g_{ij}^{t_{i}^{u}}=1=g_{iq}^{\tilde{t}_{i}^{u}}$, it will be
the case that 
\[
\sum_{\mathbf{g}\in\mathcal{G}_{n}}\hat{p}\left(\mathbf{g}\right)\cdot\mathbb{I}\left\{ \mathbf{g}_{i}=t_{i}^{u}\wedge\sum_{r\in N}g_{jr}=v\wedge g_{ij}=1\right\} =\sum_{\mathbf{g}\in\mathcal{G}_{n}}\hat{p}\left(\mathbf{g}\right)\cdot\mathbb{I}\left\{ \mathbf{g}_{i}=\tilde{t}_{i}^{u}\wedge\sum_{r\in N}g_{qr}=v\wedge g_{iq}=1\right\} 
\]
Thus,
\begin{align*}
 & \sum_{t_{i}\in T_{i}}\sum_{\mathbf{g}\in\mathcal{G}_{n}}\hat{p}\left(\mathbf{g}\right)\cdot\mathbb{I}\left\{ \mathbf{g}_{i}=t_{i}\wedge\sum_{r\in N}g_{jr}=v\wedge g_{ij}=1\wedge\sum_{r\in N}g_{ir}=u\right\} \\
 & ={n-2 \choose u-1}\sum_{\mathbf{g}\in\mathcal{G}_{n}}\hat{p}\left(\mathbf{g}\right)\cdot\mathbb{I}\left\{ \mathbf{g}_{i}=t_{i}^{u}\wedge\sum_{r\in N}g_{jr}=v\wedge g_{ij}=1\right\} 
\end{align*}
which holds for an arbitrary type $t_{i}^{u}\in T_{i}$ for which $\sum_{r=1}^{n}g_{ir}^{t_{i}^{u}}=u$
and $g_{ij}^{t_{i}^{u}}=1$. Then, we can write
\[
\pi_{u}P_{uv}=\left(\frac{n-1}{\delta}\right)\sum_{t_{i}\in T_{i}}\sum_{\mathbf{g}\in\mathcal{G}_{n}}\hat{p}\left(\mathbf{g}\right)\cdot\mathbb{I}\left\{ \mathbf{g}_{i}=t_{i}\wedge\sum_{r\in N}g_{jr}=v\wedge g_{ij}=1\wedge\sum_{r\in N}g_{ir}=u\right\} 
\]
Moreover,
\begin{align*}
\sum_{u=1}^{n-1}\pi_{u}P_{uv} & =\left(\frac{n-1}{\delta}\right)\sum_{u=1}^{n-1}\sum_{t_{i}\in T_{i}}\sum_{\mathbf{g}\in\mathcal{G}_{n}}\hat{p}\left(\mathbf{g}\right)\cdot\mathbb{I}\left\{ \mathbf{g}_{i}=t_{i}\wedge\sum_{r\in N}g_{jr}=v\wedge g_{ij}=1\wedge\sum_{r\in N}g_{ir}=u\right\} \\
 & =\left(\frac{n-1}{\delta}\right)\sum_{t_{i}\in T_{i}}\sum_{\mathbf{g}\in\mathcal{G}_{n}}\hat{p}\left(\mathbf{g}\right)\cdot\mathbb{I}\left\{ \mathbf{g}_{i}=t_{i}\wedge\sum_{r\in N}g_{jr}=v\wedge g_{ij}=1\right\} \\
 & =\left(\frac{n-1}{\delta}\right)\sum_{\mathbf{g}\in\mathcal{G}_{n}}\hat{p}\left(\mathbf{g}\right)\cdot\mathbb{I}\left\{ \sum_{r\in N}g_{jr}=v\wedge g_{ij}=1\right\} \\
 & =\left(\frac{n-1}{\delta}\right){n-2 \choose v-1}p_{v}\\
 & =\pi_{v}
\end{align*}
Hence, $\left[\pi^{T}\boldsymbol{P}\right]_{v}=\pi_{v}$ and $\pi$
is a stationary distribution of $\boldsymbol{P}.$

\noindent{\textit{Statement} $(ii)$:}

For any $u,v\in\left\{ 1,\ldots,n-1\right\} $
\begin{align*}
\pi_{u}P_{uv} & =\left(\frac{n-1}{\delta}\right)\sum_{t_{i}\in T_{i}}\sum_{\mathbf{g}\in\mathcal{G}_{n}}\hat{p}\left(\mathbf{g}\right)\cdot\mathbb{I}\left\{ \mathbf{g}_{i}=t_{i}\wedge\sum_{r\in N}g_{jr}=v\wedge g_{ij}=1\wedge\sum_{r\in N}g_{ir}=u\right\} \\
 & =\left(\frac{n-1}{\delta}\right)\sum_{\mathbf{g}\in\mathcal{G}_{n}}\hat{p}\left(\mathbf{g}\right)\cdot\mathbb{I}\left\{ \sum_{r\in N}g_{jr}=v\wedge g_{ij}=1\wedge\sum_{r\in N}g_{ir}=u\right\} 
\end{align*}
And,
\begin{align*}
\pi_{v}P_{vu} & =\left(\frac{n-1}{\delta}\right)\sum_{t_{i}\in T_{i}}\sum_{\mathbf{g}\in\mathcal{G}_{n}}\hat{p}\left(\mathbf{g}\right)\cdot\mathbb{I}\left\{ \mathbf{g}_{i}=t_{i}\wedge\sum_{r\in N}g_{jr}=u\wedge g_{ij}=1\wedge\sum_{r\in N}g_{ir}=v\right\} \\
 & =\left(\frac{n-1}{\delta}\right)\sum_{\mathbf{g}\in\mathcal{G}_{n}}\hat{p}\left(\mathbf{g}\right)\cdot\mathbb{I}\left\{ \sum_{r\in N}g_{jr}=v\wedge g_{ij}=1\wedge\sum_{r\in N}g_{ir}=u\right\} 
\end{align*}
Thus, $\pi_{u}P_{uv}=\pi_{v}P_{vu}$, for any $u,v\in\left\{ 1,\ldots,n-1\right\} $
and hence, $\boldsymbol{P}$ is reversible with respect to $\pi.$

\noindent{\textit{Statement} $(iii)$:}

From the expression of $\pi_{u}P_{uv}$ we see that $\pi_{u}P_{uv}$
gives the probability that a degree $u$ agent is connected to a degree
$v$ agent. Thus, $e\sim e_{ij}:=\pi_{i}P_{ij}$ corresponds to the
ex-ante edge distribution.

\subsection*{Proof of Proposition \ref{prop:hom}}

Suppose $\mu_{i}=\mu_{j}=\mu$ for all $i\neq j$. Then, from the
definition $\mu_{i}=\left(\boldsymbol{P}\boldsymbol{d}\right)_{i}$
and $\mathbb{E}_{\pi}[d]=\pi^{T}\boldsymbol{d}=\bar{d}$ we can write
that 
\[
\boldsymbol{P}\boldsymbol{d}=\mu\boldsymbol{1}\Rightarrow\pi^{T}\boldsymbol{P}\boldsymbol{d}=\mu\cdot\pi^{T}\boldsymbol{1}\Rightarrow\pi^{T}\boldsymbol{d}=\mu\Rightarrow\bar{d}=\mu
\]
where the second equality is due to the fact that $\pi^{T}$ is the
stationary distribution of $\boldsymbol{P}$ and $\pi^{T}\boldsymbol{1}=1$.
Thus, for any $s\in\mathbb{N}$, we can write $\mathbb{E}_{i_{s}}\left[d_{i_{s+1}}\mu_{i_{s+1}}\right]=\mu\mathbb{E}_{i_{1}}\left[d_{i_{2}}\right]=\mu^{2}$.
Finally, from the Neumann series expansion of equation (9) we have
\begin{align*}
a_{i} & =c^{-1}\left(1+\kappa d_{i}+\kappa^{2}d_{i}\mu+\kappa^{3}d_{i}\mu^{2}+\kappa^{4}d_{i}\mu^{3}+...\right)\\
 & =c^{-1}\left(1+\frac{\kappa d_{i}}{1-\kappa\mu}\right)\\
 & =c^{-1}\left(1+\frac{\kappa d_{i}}{1-\kappa\bar{d}}\right)
\end{align*}

Now, suppose $a_{i}=c^{-1}\left(1+\beta d_{i}\right)$, where $\beta=\frac{\kappa}{1-\kappa\bar{d}}$
for all $i$. Then, in vector notation we can write $\mathbf{a}=c^{-1}\left(\mathbf{1}+\beta\boldsymbol{d}\right).$
Comparing this with equation (9) it follows 
\[
c^{-1}\left(\mathbf{1}+\beta\boldsymbol{d}\right)=c^{-1}\cdot\mathbf{1}+\kappa\boldsymbol{M}_{d}\boldsymbol{P}\mathbf{\mathbf{a}}\Rightarrow\beta c^{-1} d_{i}=\kappa d_{i}\left(\boldsymbol{P}\mathbf{\mathbf{a}}\right)_{i}\Rightarrow\frac{\beta c^{-1}}{\kappa}=\left(\boldsymbol{P}\mathbf{\mathbf{a}}\right)_{i}
\]
Moreover
\[
\boldsymbol{P}\mathbf{\mathbf{a}}=c^{-1}\cdot\boldsymbol{P}\left(\mathbf{1}+\beta\boldsymbol{d}\right)=c^{-1}\cdot\boldsymbol{1}+\beta c^{-1}\boldsymbol{P}\boldsymbol{d}
\]
hence, $\left(\boldsymbol{P}\mathbf{\mathbf{a}}\right)_{i}=c^{-1}+\beta c^{-1} \mu_{i}$
and thus. $\tfrac{\beta c^{-1}}{\kappa}=c^{-1}+\beta c^{-1} \mu_{i}$. This implies
\[
\beta\mu_{i}=\frac{\beta }{\kappa}-1=\frac{\kappa\bar{d}}{1-\kappa\bar{d}}=\beta\bar{d}
\]
so that $\mu_{i}=\bar{d}$ for all $i$. It follows that $\mu_{i}=\mu_{j}=\mu$
for all $i\neq j$.

\subsection*{Proof of Proposition \ref{prop:mv}}

We begin with the following lemma.
\begin{lem}
The expectation and variance of the equilibrium actions of any agent
is given by
\begin{align*}
\mathbb{E}_{\pi}(a) & =c^{-1}\left(1+\kappa\bar d+\sum_{s\ge 2}\kappa^s\mathbb{E}_\pi[g^{(s)}]\right)\\
Var_{\pi}(a) & =c^{-2}\sum_{s,\,l\geq1}\kappa^{s+l}\left(\mathbb{E}_{\pi}\left[g^{(s)}g^{(l)}\right]-\mathbb{E}_{\pi}[g^{(s)}]\mathbb{E}_{\pi}[g^{(l)}]\right)
\end{align*}
where
\begin{align*}
\mathbb{E}_{\pi}\left[g^{(s)}\right] & =\bar{d}^{s}+\sum_{\substack{b\in\{0,1\}^{s}\\
|b|>1
}
}\sigma^{\vert b\vert}\bar{d}^{s-\vert b\vert}r^{b}\\
\mathbb{E}_{\pi}\left[g^{(s)}g^{(l)}\right] & =\sum_{b\in\{0,1\}^{s}}\sum_{c\in\{0,1\}^{l}}\sigma^{|b|+|c|}\bar{d}^{s+l-|b|-|c|}r^{b,\,c}
\end{align*}
\label{lem:Prop5lem}
\end{lem}

The proof of the lemma is contained in the online appendix.

Recall that for any binary tuple $b=(b_{0},...,b_{k-1})\in\{0,1\}^{k}$ we
can write 
\[
r^{b}=\mathbb{E}_{\pi_{i_{0}}P_{i_{0}i_{1}}...P_{i_{k-2}i_{k-1}}}\left[\prod_{t=0}^{k-1}x\left(X_{t}\right)^{b_{t}}\right].
\]
If the tuple $b$ has ones at times $0\leq\tau_{0}<\tau_{1}<...<\tau_{m}\leq k-1$,
define its span by $span(b):=\tau_{m}-\tau_{0}$.

Let $1=\rho_{1}\geq\rho_{2}\geq..\geq\rho_{n-1}$ denote the spectrum
of $P$ and set $\rho_{*}=max_{m\geq2}|\rho_{m}\vert$. Moreover,
denote by $<v,w>_{\pi}=\sum_{i}\pi_{i}v_{i}w_{i}$, and $\left\Vert v\right\Vert _{2,\pi}^{2}=<v,v>_{\pi}$.
Note that since $P1=1$, then for any mean zero vector $v$, i.e.,
$<v,1>_{\pi}=0$, its decomposition into the eigenbasis of $P$ gives
\[
\left\Vert P^{k}v\right\Vert _{2,\pi}\leq\rho_{*}^{k}\left\Vert v\right\Vert _{2,\pi}
\]
where and $P^{k}$ is the transition matrix raised to the $k^{th}$
power.

Consider first the special case where $b=(1,0,...,0,1)$ with ones
at time $0$ and $k$. Then we have 
\begin{align*}
|r^{(1,0,...,0,1)}| & =|\sum_{i_{0},\,i_{k-1}}\pi_{i_{0}}P_{i_{0}i_{k-1}}^{(k)}d_{i_{0}}d_{i_{k-1}}|=|<x,P^{k}x>_{\pi}|\leq\rho_{*}^{k}\left\Vert x\right\Vert _{2,\pi}=\rho_{*}^{k}
\end{align*}
where we have used the fact that $\left\Vert x\right\Vert _{2,\pi}^{2}=\sum_{i}\pi_{i}\left(\frac{d_{i}-\bar{d}}{\sigma}\right)^{2}=1.$
Next, without loss of generality suppose $b$ has ones times $0=\tau_{0}<\tau_{1}<...<\tau_{m}$,
so that $\vert b\vert=m+1$ and $span(b)=\tau_{m}-\tau_{0}$ (if $\tau_{0}\neq0$
then use reversibility of $P$ to translate the start of the walk
forward). Then, letting $\omega=\left\Vert x\right\Vert _{\infty}$
and $U=\prod_{j=1}^{m-1}x\left(X_{\tau_{j}}\right)$, we can then
write 
\begin{align*}
|r^{b}| & =\left|\mathbb{E}_{\pi_{i_{0}}P_{i_{0}i_{1}}...P_{i_{k-2}i_{k-1}}}\left[x\left(X_{\tau_{0}}\right)Ux\left(X_{\tau_{m}}\right)\right]\right|\\
 & \leq\omega^{|b|-2}\left|\mathbb{E}_{\pi_{i_{0}}P_{i_{0}i_{1}}...P_{i_{k-2}i_{k-1}}}\left[x\left(X_{\tau_{0}}\right)x\left(X_{\tau_{m}}\right)\right]\right|\\
 & \leq\omega^{|b|-2}\rho^{span(b)}\\
 & \leq\omega^{|b|-2}\rho^{|b|-1}
\end{align*}
where we have used the fact that $\rho_{*}\leq1$ and $span(b)\geq|b|-1$.

{\textbf{Mean Bounds}}

To bound the mean, from Lemma \ref{lem:Prop5lem} it follows that
\begin{align*}
\left|\mathbb{E}_{\pi}(a)-c^{-1}\left(1+\frac{\kappa\bar{d}}{1-\kappa\bar{d}}\right)\right| & =\left| c^{-1}\sum_{s\geq2}^{\infty}\kappa^{s}\sum_{\substack{b\in\{0,1\}^{s}\\
|b|>1
}
}\sigma^{\vert b\vert}\bar{d}^{s-\vert b\vert}r^{b}\right|\\
 & \leq c^{-1}\sum_{s\geq2}^{\infty}|\kappa|^{s}\sum_{m\geq2}^{s}{s \choose m}\sigma^{m}\bar{d}^{s-m}\omega^{m-2}\rho_{*}^{m-1}\\
 & =\frac{c^{-1}}{\omega^{2}\rho_{*}}\sum_{k\geq2}|\kappa|^{s}\left(\xi^{s}-\bar{d}^{s}-s(\sigma\omega\rho_{*})\bar{d}^{s-1}\right)\\
 & =\frac{\left|\kappa\right|^{2}c^{-1}}{\omega^{2}\rho_{*}}\left(\frac{\xi^{2}}{1-|\kappa|\xi}-\frac{\bar{d}^{2}}{1-|\kappa|\bar{d}}-\frac{2\bar{d}-\left|\kappa\right|\bar{d}^{2}}{(1-|\kappa|\bar{d})^{2}}\left(\xi-\bar{d}\right)\right)
\end{align*}
where $\xi=\bar{d}+\sigma\omega\rho_{*}$. 

We consider the function $h\left(.\right):\mathbb{R}\rightarrow\mathbb{R}$
such that 
\[
h\left(t\right)=\frac{t^{2}}{1-\left|\kappa\right|t}
\]
and use the Taylor's theorem with mean value remainder form on $\left[\bar{d},\xi\right]$
to get that

\[
\frac{\xi^{2}}{1-|\kappa|\xi}-\frac{\bar{d}^{2}}{1-|\kappa|\bar{d}}-\frac{2\bar{d}-\left|\kappa\right|\bar{d}^{2}}{(1-|\kappa|\bar{d})^{2}}\left(\xi-\bar{d}\right)\leq\frac{\left(\xi-\bar{d}\right)^{2}}{\left(1-\left|\kappa\right|\xi\right)^{3}}=\frac{\sigma^{2}\omega^{2}\rho_{*}^{2}}{\left(1-\left|\kappa\right|\xi\right)^{3}}
\]
Finally, 
\[
\left|\mathbb{E}_{\pi}(a)-c^{-1}\left(1+\frac{\kappa\bar{d}}{1-\kappa\bar{d}}\right)\right|\leq\frac{\kappa^{2}\sigma^{2}c^{-1}}{\left(1-\left|\kappa\right|\xi\right)^{3}}\rho_{*}
\]

{\textbf{Variance Bound}}

To bound the variance, recall that
for two binary patterns \textbf{$b\in\{0,1\}^{k}$ }and $c\in\{0,1\}^{l}$
with $X_{0}=Y_{0}\sim\pi$ we have 
\[
r^{b,\,c}:=\mathbb{E}_{\pi_{i_{0}}P_{i_{0}i_{1}}...P_{i_{k-2}i_{k-1}}P_{i_{0}j_{1}}\ldots P_{j_{l-2}j_{l-1}}}\left[\prod_{t=0}^{k-1}x\left(X_{t}\right)^{b_{t}}\prod_{s=0}^{l-1}x\left(Y_{s}\right)^{b_{s}}\right]
\]
As with the one-arm path moment, suppose $b$ has ones at $0=\tau_{0}<...<\tau_{m}$
and $c$ at times $0=t_{0}<...<t_{n}$ and define the products $U_{X}=\prod_{j=1}^{m-1}x\left(X_{\tau_{j}}\right)$
and $U_{Y}=\prod_{s=1}^{n-1}x\left(Y_{t_{s}}\right)$. Then, using
the same arguments as in the one-arm case it follows that:

\begin{align*}
|r^{b,\,c}| & :=\left|\mathbb{E}_{\pi_{i_{0}}P_{i_{0}i_{1}}...P_{i_{k-2}i_{k-1}}}\left[x\left(X_{\tau_{0}}\right)x\left(Y_{t_{0}}\right)U_{X}U_{Y}x\left(X_{\tau_{m}}\right)x\left(Y_{t_{n}}\right)\right]\right|\\
 & \leq\omega^{|b|+|c|-4}\rho_{*}^{|b|+|c|}\\
 & \leq\omega^{|b|+|c|-2}\rho_{*}^{|b|+|c|-2}
\end{align*}

Note that 
\[
Var_{\pi}\left(a^{H}\right)=c^{-2}\left(\frac{\kappa\sigma}{1-\kappa\bar{d}}\right)^{2}=c^{-2}\sum_{s,l\geq1}\kappa^{s+l}\sigma^{2}\bar{d}^{s+l-2}
\]

Thus
\[
\left|Var_{\pi}\left(a\right)-Var_{\pi}\left(a^{H}\right)\right|\leq c^{-2}\sum_{s,\,l\geq1}\left|\kappa\right|^{s+l}\left|Cov\left(g^{(s)},g^{(l)}\right)-\sigma^{2}\bar{d}^{s+l-2}\right|
\]

where $Cov\left(g^{(s)},g^{(l)}\right)=\mathbb{E}_{\pi}\left[g^{(s)}g^{(l)}\right]-\mathbb{E}_{\pi}[g^{(s)}]\mathbb{E}_{\pi}[g^{(l)}]$.
Denote by $\Delta_{s}=\mathbb{E}_{\pi}[g^{(s)}]-\bar{d}^{s}$. Then
we can write 
\[
Cov\left(g^{(s)},g^{(l)}\right)-\sigma^{2}\bar{d}^{s+l-2}=\left(\mathbb{E}_{\pi}\left[g^{(s)}g^{(l)}\right]-\bar{d}^{s+l}-\sigma^{2}\bar{d}^{s+l-2}\right)-\bar{d}^{s}\Delta_{l}-\bar{d}^{l}\Delta_{s}-\Delta_{s}\Delta_{l}
\]
Thus, 
\[
\left|Var_{\pi}\left(a\right)-Var_{\pi}\left(a^{H}\right)\right|\leq c^{-2}\sum_{s,\,l\geq1}\left|\kappa\right|^{s+l}\left|\eta_{s,l}\right|+2\left(\sum_{s\geq1}\left|\kappa\right|^{s}\bar{d}^{s}\right)\left(\sum_{l\geq1}\left|\kappa\right|^{l}\left|\Delta_{l}\right|\right)+\left(\sum_{s\geq1}\left|\kappa\right|^{s}\left|\Delta_{s}\right|\right)^{2}
\]
where $\eta_{s,l}=\mathbb{E}_{\pi}\left[g^{(s)}g^{(l)}\right]-\bar{d}^{s+l}-\sigma^{2}\bar{d}^{s+l-2}$. 
\begin{claim}
It can be shown that 
\[
\sum_{s,\,l\geq1}\left|\kappa\right|^{s+l}\left|\eta_{s,l}\right|\leq\frac{3\left|\kappa\right|^{2}\sigma^{2}}{\left(1-\left|\kappa\right|\xi\right)^{6}}\rho_{*}
\]
\label{cl:eta}
\end{claim}

The proof of the claim is contained in the online appendix.

Now deriving a bound for $\Delta_{s}=\sum_{\substack{b\in\{0,1\}^{s}\\
|b|>1
}
}\sigma^{\vert b\vert}\bar{d}^{s-\vert b\vert}r^{b}$, from the proof of mean bound we had that 
\begin{equation}
\sum_{s\geq1}\left|\kappa\right|^{s}\left|\Delta_{s}\right|\leq\frac{1}{\omega^{2}\rho_{*}}\left(\frac{\left|\kappa\right|\xi}{1-\left|\kappa\right|\xi}-\frac{\left|\kappa\right|\bar{d}}{1-\left|\kappa\right|\bar{d}}-\frac{\left|\kappa\right|}{\left(1-\left|\kappa\right|\bar{d}\right)^{2}}\left(\xi-\bar{d}\right)\right)\leq\frac{\left|\kappa\right|^{2}\sigma^{2}\rho_{*}}{\left(1-\left|\kappa\right|\xi\right)^{3}}\label{eq:delta}
\end{equation}
where in the last inequality we use Taylor's theorem as previously.

Also we have, 
\begin{equation}
\sum_{s\geq1}\left|\kappa\right|^{s}\bar{d}^{s}=\frac{\left|\kappa\right|\bar{d}}{1-\left|\kappa\right|\bar{d}}\leq\frac{\left|\kappa\right|\xi}{1-\left|\kappa\right|\xi}\label{eq:dbar}
\end{equation}
Thus, combining eqn (\ref{eq:delta}) and (\ref{eq:dbar}) we get,
\begin{equation}
\left(\sum_{s\geq1}\left|\kappa\right|^{s}\bar{d}^{s}\right)\left(\sum_{l\geq1}\left|\kappa\right|^{l}\left|\Delta_{l}\right|\right)\leq\frac{\left|\kappa\right|\xi}{1-\left|\kappa\right|\xi}\cdot\frac{\left|\kappa\right|^{2}\sigma^{2}\rho_{*}}{\left(1-\left|\kappa\right|\xi\right)^{3}}\leq\frac{\left|\kappa\right|^{2}\sigma^{2}\rho_{*}}{\left(1-\left|\kappa\right|\xi\right)^{6}}\label{eq:second part}
\end{equation}
where we have used the fact that $\left|\kappa\right|\xi\leq1$.
And, we have
\begin{equation}
\left(\sum_{s\geq1}\left|\kappa\right|^{s}\left|\Delta_{s}\right|\right)^{2}\leq\left(\frac{\left|\kappa\right|^{2}\sigma^{2}\rho_{*}}{\left(1-\left|\kappa\right|\xi\right)^{3}}\right)^{2}\leq\frac{\left|\kappa\right|^{2}\sigma^{2}\rho_{*}}{\left(1-\left|\kappa\right|\xi\right)^{6}}\label{eq:third part}
\end{equation}
where we have used the fact that $\left|\kappa\right|\sigma\leq\left|\kappa\right|d_{max}\leq1$. 

Combining Claim \ref{cl:eta}, and inequalities in (\ref{eq:second part})
and (\ref{eq:third part}) we have 
\[
\left|Var_{\pi}\left(a\right)-Var_{\pi}\left(a^{H}\right)\right|\leq\frac{6\left|\kappa\right|^{2}\sigma^{2}c^{-2}}{\left(1-\left|\kappa\right|\xi\right)^{6}}\rho_{*}
\]

\subsection*{Proof of Lemma \ref{lem:Prop5lem}}
We have that
\[
a(\boldsymbol{P})=c^{-1}\sum_{s\ge0}\kappa^{s}g^{(s)},\quad\text{where}\quad g^{(0)}= \boldsymbol{1},\quad g^{(1)}=\boldsymbol{d},\quad g^{(s+1)}=M_{d}\boldsymbol{P}g^{(s)}.
\]
Hence, $\mathbb{E}_\pi \left[\boldsymbol{g}^{\left(1\right)}\right]=\mathbb{E}_\pi \left[\boldsymbol{d}\right]=\bar{d}$.
And,
\[
\E_\pi\left[a\right]=c^{-1}\left(\E_\pi\left[\1\right]+\sum_{s\geq 1}\kappa^s\E_\pi[g^{(s)}]\right)
=c^{-1}\left(1+\kappa\bar d+\sum_{s\ge 2}\kappa^s\E_\pi[g^{(s)}]\right).
\]

For the variance, we can expand
\begin{align*}
\Var_\pi(a)
&=\Var_\pi\!\left(c^{-1}\sum_{s\ge 1}\kappa^s g^{(s)}\right)
=c^{-2}\sum_{s,l\ge 1}\kappa^{s+l}\Cov_\pi(g^{(s)},g^{(l)})\\
&=c^{-2}\sum_{s,l\ge 1}\kappa^{s+l}\Big(\E_\pi[g^{(s)}g^{(l)}]-\E_\pi[g^{(s)}]\E_\pi[g^{(l)}]\Big),
\end{align*}
which is the stated variance expansion.

\medskip
Let $(X_t)_{t\ge 0}$ be the Markov chain with transition kernel $\boldsymbol{P}$.
For each $i$ and $s\ge 1$, set
\[
\Gamma_i^{(s)}:=\E\!\left[\prod_{t=0}^{s-1} d_{X_t}\ \Big|\ X_0=i\right].
\]
Then $\Gamma^{(1)}=d=g^{(1)}$, and by the Markov property,
\[
\Gamma_i^{(s)}=d_i\sum_j P_{ij}\Gamma_j^{(s-1)}=\big(M_d\boldsymbol{P}\,\Gamma^{(s-1)}\big)_i.
\]
Since $g^{(s)}=M_d\boldsymbol{P}\,g^{(s-1)}$ by definition, induction gives $\Gamma^{(s)}=g^{(s)}$.

If we now start at stationarity $X_0\sim\pi$, then
\[
\E_\pi[g^{(s)}]=\sum_i\pi_i g_i^{(s)}=\E\!\left[\prod_{t=0}^{s-1}d_{X_t}\right].
\]

\medskip
We can write $d_{X_t}=\bar d+\sigma x(X_t)$ where $x(i)=(d_i-\bar d)/\sigma$.
Then
\[
\prod_{t=0}^{s-1}(\bar d+\sigma x(X_t))
=\sum_{b\in\{0,1\}^s}\sigma^{|b|}\bar d^{s-|b|}\prod_{t=0}^{s-1}x(X_t)^{b_t}.
\]
Taking expectations and using the definition
\[
r^{b}:=\E\!\left[\prod_{t=0}^{s-1}x(X_t)^{b_t}\right]
\]
gives
\[
\E_\pi[g^{(s)}]=\sum_{b\in\{0,1\}^s}\sigma^{|b|}\bar d^{s-|b|}r^b.
\]
The term $|b|=0$ equals $\bar d^s$. If $|b|=1$, say $b_\tau=1$, then
$r^b=\E[x(X_\tau)]=\E_\pi[x]=0$.
Hence
\[
\E_\pi[g^{(s)}]=\bar d^s+\sum_{\substack{b\in\{0,1\}^s\\|b|>1}}
\sigma^{|b|}\bar d^{s-|b|}r^b.
\]

\medskip

Let $I\sim\pi$. Conditional on $I$, generate two independent chains $(X_t)_{t\ge 0}$ and $(Y_w)_{w\ge 0}$
with kernel $\boldsymbol{P}$, with common root $X_0=Y_0=I$. Then
\[
\E_\pi[g^{(s)}g^{(l)}]
=\E\!\left[\prod_{t=0}^{s-1}d_{X_t}\prod_{w=0}^{l-1}d_{Y_w}\right].
\]
Expanding each factor as $\bar d+\sigma x$ and collecting terms gives
\[
\E_\pi[g^{(s)}g^{(l)}]
=\sum_{b\in\{0,1\}^s}\sum_{c\in\{0,1\}^l}
\sigma^{|b|+|c|}\bar d^{s+l-|b|-|c|}\,r^{b,c},
\]
where
\[
r^{b,c}:=\E\!\left[\prod_{t=0}^{s-1}x(X_t)^{b_t}\prod_{w=0}^{l-1}x(Y_w)^{c_w}\right].
\]

\subsection*{Proof of Claim \ref{cl:eta}}

Note that $\mathbb{E}_{\pi}\left[g^{(k)}g^{(l)}\right]$ evaluated
at $b,c$ such that $\left|b\right|=\left|c\right|=0$, gives $\bar{d}^{k+l}$. 

Denote by $e_{0}^{k}=\left(1,0,\ldots,0\right)\in\left\{ 0,1\right\} ^{k}$
and $e_{0}^{l}=\left(1,0,\ldots,0\right)\in\left\{ 0,1\right\} ^{l}$.
Then for $\left(b,c\right)=\left(e_{0}^{k},e_{0}^{l}\right)$ for
some $k,l\geq1$, we have that $r^{b,c}=1$ and $\mathbb{E}_{\pi}\left[g^{(k)}g^{(l)}\right]$
evaluated at $\left(b,c\right)=\left(e_{0}^{k},e_{0}^{l}\right)$,
gives $\sigma^{2}\bar{d}^{k+l-2}$. 

We can write 
\[
\left|\eta_{k,l}\right|\leq\left|\sum_{\substack{\substack{\left(b,c\right)\neq\left(e_{0}^{k},e_{0}^{l}\right)\\
\left|b\right|=\left|c\right|=1
}
}
}\sigma^{|b|+|c|}\bar{d}^{k+l-|b|-|c|}r^{b,\,c}\right|+\left|\sum_{\left|b\right|+\left|c\right|\geq3}\sigma^{|b|+|c|}\bar{d}^{k+l-|b|-|c|}r^{b,\,c}\right|
\]

Now, consider $b\in\left\{ 0,1\right\} ^{k}$ and $c\in\left\{ 0,1\right\} ^{l}$
such that $\left|b\right|=\left|c\right|=1$ but $\left(b,c\right)\neq\left(e_{0}^{k},e_{0}^{l}\right)$.
Suppose there exists $s\in\left\{ 0,\ldots,k-1\right\} $ and $t\in\left\{ 0,\ldots,l-1\right\} $
with $\left(s,t\right)\neq\left(0,0\right)$ such that 
\[
r^{b,\,c}=\sum_{i_{0}}\pi_{i_{0}}\left(\sum_{i}P_{i_{0}i}^{\left(s\right)}x\left(i\right)\right)\left(\sum_{j}P_{i_{0}j}^{\left(t\right)}x\left(j\right)\right)=\sum_{i_{0},i,j}\pi_{i_{0}}P_{i_{0}i}^{\left(s\right)}P_{i_{0}j}^{\left(t\right)}x\left(i\right)x\left(j\right)
\]
Since, $P$ is reversible, $P^{\left(s\right)}$ is reversible as
well and we can write, 
\[
\sum_{i_{0},i,j}\pi_{i_{0}}P_{i_{0}i}^{\left(s\right)}P_{i_{0}j}^{\left(t\right)}x\left(i\right)x\left(j\right)=\sum_{i,j}\pi_{i}x\left(i\right)x\left(j\right)\sum_{i_{0}}P_{ii_{0}}^{\left(s\right)}P_{i_{0}j}^{\left(t\right)}
\]
where the last equality can be simplified more to be written as 
\[
\sum_{i,j}\pi_{i}x\left(i\right)x\left(j\right)\sum_{i_{0}}P_{ii_{0}}^{\left(s\right)}P_{i_{0}j}^{\left(t\right)}=\sum_{i}\pi_{i}x\left(i\right)\sum_{j}P_{ij}^{\left(s+t\right)}x\left(j\right)=\sum_{i}\pi_{i}x\left(i\right)\left(P_{ij}^{\left(s+t\right)}x\right)\left(i\right)
\]
And hence we can write $r^{b,c}=\left\langle x,P^{s+t}x\right\rangle _{\pi}$and
as shown previously 
\[
\left|r^{b,c}\right|\leq\rho_{*}^{s+t}\leq\rho_{*}
\]
where the last equality is due to the fact that $s+t\geq1$. Thus,
\[
\left|\sum_{\substack{\left(b,c\right)\neq\left(e_{0}^{k},e_{0}^{l}\right)\\
\left|b\right|=\left|c\right|=1
}
}\sigma^{|b|+|c|}\bar{d}^{k+l-|b|-|c|}r^{b,\,c}\right|\leq\sum_{\substack{\left(b,c\right)\neq\left(e_{0}^{k},e_{0}^{l}\right)\\
\left|b\right|=\left|c\right|=1
}
}\sigma^{2}\bar{d}^{k+l-2}\rho_{*}\leq\sigma^{2}kl\bar{d}^{k+l-2}\rho_{*}
\]
and hence,
\begin{equation}
\sum_{k,l\geq1}\left|\kappa\right|^{k+l}\sigma^{2}kl\bar{d}^{k+l-2}\rho_{*}=\sigma^{2}\rho_{*}\left(\frac{\left|\kappa\right|}{\left(1-\left|\kappa\right|\bar{d}\right)^{2}}\right)^{2}\leq\frac{\kappa^{2}\sigma^{2}}{\left(1-\left|\kappa\right|\xi\right)^{6}}\rho_{*}\label{eq:1st part}
\end{equation}
Evaluating the second part of $\eta_{kl}$ we get 
\[
\left|\sum_{\left|b\right|+\left|c\right|\geq3}\sigma^{|b|+|c|}\bar{d}^{k+l-|b|-|c|}r^{b,\,c}\right|\leq\sigma^{2}\sum_{\left|b\right|+\left|c\right|\geq3}\left(\sigma\omega\rho_{*}\right)^{|b|+|c|-2}\bar{d}^{k+l-|b|-|c|}
\]
Using binomial theorem and then subtracting terms at $m=0$ and $n=0,$we
get
\[
\sum_{m=1}^{k}\sum_{n=1}^{l}\binom{k}{m}\binom{l}{n}\left(\sigma\omega\rho_{*}\right)^{m+n-2}\bar{d}^{k+l-m-n}=\frac{\left(\xi^{k}-\bar{d}^{k}\right)\left(\xi^{l}-\bar{d}^{l}\right)}{\left(\xi-\bar{d}\right)^{2}}
\]
Thus, we have
\begin{align*}
    \sum_{\left|b\right|+\left|c\right|\geq3}\left(\sigma\omega\rho_{*}\right)^{|b|+|c|-2}\bar{d}^{k+l-|b|-|c|}&=\sum_{m+n\geq3}\binom{k}{m}\binom{l}{n}\left(\sigma\omega\rho_{*}\right)^{m+n-2}\bar{d}^{k+l-m-n}\\
    &=\frac{\left(\xi^{k}-\bar{d}^{k}\right)\left(\xi^{l}-\bar{d}^{l}\right)}{\left(\xi-\bar{d}\right)^{2}}-kl\bar{d}^{k+l-2}
\end{align*}

And, we can write
\begin{align*}
\sum_{k,l\geq1}\sum_{\left|b\right|+\left|c\right|\geq3}\left|\kappa\right|^{k+l}\sigma^{|b|+|c|}\bar{d}^{k+l-|b|-|c|}\left|r^{b,\,c}\right| & \leq\sigma^{2}\sum_{k,\,l\geq1}\left|\kappa\right|^{k+l}\left(\frac{\left(\xi^{k}-\bar{d}^{k}\right)\left(\xi^{l}-\bar{d}^{l}\right)}{\left(\xi-\bar{d}\right)^{2}}-kl\bar{d}^{k+l-2}\right)\\
 & =\left|\kappa\right|^{2}\sigma^{2}\left(\frac{1}{\left(1-\left|\kappa\right|\xi\right)^{2}\left(1-\left|\kappa\right|\bar{d}\right)^{2}}-\frac{1}{\left(1-\left|\kappa\right|\bar{d}\right)^{4}}\right)\\
 & =\left|\kappa\right|^{3}\sigma^{2}\left(\xi-\bar{d}\right)\frac{2-\left|\kappa\right|\left(\bar{d}+\xi\right)}{\left(1-\left|\kappa\right|\xi\right)^{2}\left(1-\left|\kappa\right|\bar{d}\right)^{4}}\\
 & \leq\frac{2\left|\kappa\right|^{3}\sigma^{2}\left(\xi-\bar{d}\right)}{\left(1-\left|\kappa\right|\xi\right)^{6}}
\end{align*}
Since $\left|\kappa\right|\sigma\omega\leq\left|\kappa\right|d_{max}<1$,
we have that $\left|\kappa\right|\left(\xi-\bar{d}\right)=\left|\kappa\right|\sigma\omega\rho_{*}<\rho_{*}$
and thus can write. 
\begin{equation}
\sum_{k,l\geq1}\sum_{\left|b\right|+\left|c\right|\geq3}\left|\kappa\right|^{k+l}\sigma^{|b|+|c|}\bar{d}^{k+l-|b|-|c|}\left|r^{b,\,c}\right|\leq\frac{2\left|\kappa\right|^{2}\sigma^{2}}{\left(1-\left|\kappa\right|\xi\right)^{6}}\rho_{*}\label{eq:2nd part}
\end{equation}

Combining (\ref{eq:1st part}) and (\ref{eq:2nd part}) we get,
\begin{equation}
\sum_{s,\,l\geq1}\left|\kappa\right|^{s+l}\left|\eta_{s,l}\right|\leq\frac{3\left|\kappa\right|^{2}\sigma^{2}}{\left(1-\left|\kappa\right|\xi\right)^{6}}\rho_{*}\label{eq:eta}
\end{equation}

\subsection*{Derivation of Equation (\ref{eq:config})}

Let $\mathcal{D}=\left\{ d,d^{\prime}\right\} $ and the probability
kernel on $\mathcal{D}$ is 
\[
\boldsymbol{P}=\left(\begin{array}{cc}
p & 1-p\\
1-q & q
\end{array}\right)
\]
with $p,q>0$ and $p\neq q$. Without loss of generality, assume that $d<d^{\prime}$. Then, $\rho_{*}=\left|p+q-1\right|$ and the stationary distribution
$\pi^{T}=\left[\pi_{d},\pi_{d^{\prime}}\right]$ is given by
\begin{equation*}
    \pi_{d} =\frac{1-q}{2-q-p}\qquad\text{ and }\qquad \pi_{d^{\prime}} =\frac{1-p}{2-q-p}
\end{equation*}

Let $\rho=p+q-1$. Thus, $\rho_{*}=\left|\rho\right|$. Equilibrium
actions are then given by 
\begin{align*}
a_{d} & =c^{-1}\delta^{-1}\left(\left(1-\kappa qd^{\prime}\right)+\kappa d\left(1-p\right)\right)\\
a_{d^{\prime}} & = c^{-1}\delta^{-1} \left(\kappa d^{\prime}\left(1-q\right)+\left(1-\kappa dp\right)\right)
\end{align*}
where, $\delta=1-\kappa dp-\kappa d^{\prime}q+\kappa^{2}dd^{\prime}\rho$.

\textbf{Claim: }$\delta>0$.
\begin{proof}
    We can rewrite $\delta$ as
\begin{equation*}
    \delta = \left(1-\kappa d p\right)\left(1-\kappa d^{\prime} q\right) - \kappa^2d d^{\prime} \left(1-p\right)\left(1-q\right)
\end{equation*}
Note that,
\[
1-\kappa dp > \kappa d \left(1-p\right)\qquad\text{ and }\qquad 1-\kappa d^{\prime}q > \kappa d^\prime \left(1-q\right)
\]
Thus multiplying these two inequalities we get, 
\begin{equation*}
    \left(1-\kappa dp\right)\left(1-\kappa d^{\prime}q\right) > \kappa^2 d d^\prime \left(1-p\right) \left(1-q\right)\Rightarrow \delta >0
\end{equation*}
\end{proof}

\
The expected action and variance are given by,
\begin{align*}
    \mathbb{E}_\pi \left[a\right] &= c^{-1}\delta^{-1}\left(1-\kappa \rho \left(d+d^\prime-\bar{d}\right)\right)\\
    Var_{\pi}\left[a\right] &= c^{-2}\delta^{-2}\kappa^2\sigma^2
\end{align*}
where $\sigma^{2}=Var_{\pi}\left[d\right]=\pi_{d}\pi_{d^{\prime}}\left(d-d^{\prime}\right)^{2}$.

Now define the homogeneous kernel, 
\[
\boldsymbol{P}^{H}=\left(\begin{array}{cc}
\pi_{d} & \pi_{d^{\prime}}\\
\pi_{d} & \pi_{d^{\prime}}
\end{array}\right)
\]
so that
\[
\mathbb{E}_{\pi}\left[a^{H}\right]=c^{-1}\left(1-\kappa\bar{d}\right)^{-1}
\]
\[
Var_{\pi}\left[a^{H}\right]=c^{-2}\kappa^{2}\sigma^2\left(1-\kappa\bar{d}\right)^{-2}
\]
We assume that there is high assortativity, i.e. $p+q>1$, so that $\rho>0$ and hence $\rho=\rho_*=p+q-1$. 

Denote $\beta=\left(\bar{d}+\kappa dd^{\prime}-d-d^{\prime}\right)$ and $x=1-\kappa\bar{d}$. Also note that using the identity $dp+d^\prime q=\rho_* \left(d+d^\prime\right)+\left(1-\rho_*\right)\bar{d}$, we  can write $\delta = x + \kappa \rho_*\beta $. Then the difference in expected actions can be written as
\[
\mathbb{E}_{\pi}\left[a\right]-\mathbb{E}_{\pi}\left[a^{H}\right]=C_1\left(\rho_*\right)\rho_{*}
\]
with 
\[
C_1\left(\rho_*\right)=\kappa^2\sigma^2\left(c\delta x\right)^{-1}
\]
From the previous claim we know that $\delta>0$ and hence, we have that $C_1\left(\rho_*\right)>0$. To establish that it is monotonically increasing, note that 
\[
C_1^{\prime}\left(\rho_*\right) = -\frac{\kappa^3\sigma^2\beta}{cx\delta^2}
\]

\textbf{Claim: } $\beta<0$
\begin{proof}
From the way $\beta$ is defined
\begin{align*}
    \beta &= \kappa dd^{\prime}-\left(d+d^{\prime}-\bar{d}\right)\\
    &= \kappa dd^{\prime} - d\pi_{d^{\prime}} - d^\prime\pi_d\\
    &= \kappa dd^{\prime} - d\left(1-\pi_d\right) - d^\prime\pi_d\\
    &= d \left(\kappa d^{\prime}-1\right) + \left(d-d^{\prime}\right)\pi_d\\
    &< 0
\end{align*}
\end{proof}
Thus, we have $\delta>0, x>0$ and $\beta<0$, implying $C_1^{\prime}\left(\rho_*\right) >0$.

Next, we can write the differences in variances as,
\[
Var_{\pi}\left[a\right]-Var_{\pi}\left[a^{H}\right] = \frac{\kappa^{2}\sigma^{2}}{c^{2}}\left(\frac{1}{\delta^{2}}-\frac{1}{x^{2}}\right)
\]
Thus, we have
\[
\frac{1}{\delta^{2}}-\frac{1}{x^{2}} =-\kappa\beta\delta^{-2} x^{-2}\left(x+\delta\right)\rho_*
\]

Let, 
\[
C_2\left(\rho_*\right)=-\kappa^3\sigma^2\beta \left(x+\delta\right)\left(c\delta x\right)^{-2}
\]

Then, note that from the first claim we have $\delta = x + \kappa\rho_* \beta >0$ and hence, $2x + \kappa\rho_* \beta > x > 0$. Moreover, from the second claim we have $\beta<0$ which establishes that $C_2\left(\rho_*\right)>0$. 

To establish that monotonically increasing, note that
\[
C_2^{\prime}\left(\rho_*\right) = \kappa^4 \sigma^2\beta^2 \left(\delta+2x\right)\left(cx\right)^{-2}\delta^{-3}
\]
Thus we have $\delta+2x>0$ and $\delta>0$, implying $C_2^{\prime}\left(\rho_*\right) >0$. Hence we have,
\[
Var_{\pi}\left[a\right]-Var_{\pi}\left[a^{H}\right] = C_2\left(\rho_*\right)\rho_*
\]
with $C_2\left(\rho_*\right)>0$ and  $C_2^\prime\left(\rho_*\right)>0$.

\section*{Appendix C. Iterative Belief Centrality Under Biases}


In the main body of the paper, we focused on environments in which agents are fully rational. They form interim beliefs by Bayesian updating from local observations, and their equilibrium actions are computed on the basis of those rational beliefs. While we have already shown that local network sampling can rationally lead to biased perceptions of the network, in many settings, individuals are subject to additional observational and cognitive biases. For instance, people have been shown to overweight salient ties, misremember connections, and often treat multiple signals of the same underlying tie as independent confirmation even when they are redundant. 

To account for such biases, rather than abandoning our framework, we use our theory as a counterfactual prediction tool. Specifically, we ask how equilibrium behavior would change if agents computed Iterative Belief Centrality using a biased representation of the network.

Suppose we have first order information under isomorphic priors. Let $\boldsymbol{P}$ be the true conditional degree kernel, with behavioral operator $\boldsymbol{M}_{d}\boldsymbol{P}$.%
\footnote{In an applied setting, we may think of the rows of $\boldsymbol{P}$ as representing the true conditional degree distributions on a known network which can be computed by counting the relative frequencies of degree tuples within it.}
We model biases by assuming that instead of $\boldsymbol{P}$, individuals behave as though there is some counterfactual kernel $\widehat{\boldsymbol{P}}$ that captures the biased way they perceive the network. This produces an effective operator $\boldsymbol{M}_{d}\widehat{\boldsymbol{P}}$ and effective actions $a(\widehat{\boldsymbol{P}})=(I-\lambda \boldsymbol{M}_{d}\widehat{\boldsymbol{P}})^{-1}\mathbf{1}$. Comparison to the unbiased benchmark $a(\boldsymbol{P})=\left(I-\lambda \boldsymbol{M}_{d}\boldsymbol{P}\right)^{-1}\mathbf{1}$ then reveals the behavioral consequences of different biases. In this sense, each bias corresponds to a \emph{distortion of the belief hierarchy}. Agents are still iterating beliefs, but they do so through a misspecified kernel.

We consider a general class of normalized counterfactual kernels of the form
\begin{equation}
\widehat{P}_{ij}=\frac{P_{ij}w_{i}(j)}{\sum_{k}P_{ik}w_{i}(k)}.
\label{eq:biasP}
\end{equation}
Below, we describe how several well-documented observational and cognitive biases correspond to counterfactual kernels belonging to this class.

\textbf{Correlation Neglect.} Correlation neglect refers to people's tendency to ignore correlations across signals, treating them as independent evidence \citep{enke2019correlation}. A particularly important form is common source neglect, documented in political belief formation and repeated exposure to misinformation \citep{pennycook2018prior}. To model this, we consider a power-transformed kernel:
\[
\widehat{P}_{ij}=\frac{P_{ij}^{\gamma}}{\sum_{k}P_{ik}^{\gamma}}.
\]
When $\gamma<1$, large entries shrink and small entries grow. Agents therefore treat unlikely neighbor-degrees as more plausible than they actually are. This matches the idea of correlation neglect since the rows of $\widehat{P}$ become flatter relative to $P$ and thus closer to uniform. On the other hand, when $\gamma>1$, large entries become larger and small entries shrink toward zero. In this case, agents overweight the most likely neighbor types and underweight low-probability ones. This can be interpreted as a form of ``correlation hyper-awareness'' whereby repeated exposure to the same underlying signal is mistakenly treated as multiple independent confirmations.

\textbf{Homophily.} The tendency to associate with similar others is one of the best-documented phenomena in social networks, and people can hold biased perceptions about how segregated groups are \citep{lee2019homophily}. In our framework, this can be represented by a convex combination of the identity kernel and the true kernel:
\[
\widehat{P}_{ij}=h \mathbb{I}(i=j)+(1-h)P_{ij},\qquad h\in[0,1).
\]
The higher $h$, the stronger the perceived within-type concentration is.

\textbf{Limited Attention.} Cognitive limits mean individuals cannot process all available information equally. Online experiments show that people disproportionately attend to a subset of their neighbors or to especially salient contacts \citep{bakshy2012role, weng2012competition}. This creates a limited-attention bias in which some neighbors are overweighted while others are ignored. Formally,
\[
\widehat{P}_{ij}=\frac{P_{ij}w_{ij}}{\sum_{j}P_{ij}w_{ij}},
\]
where $w_{ij}\in[0,1]$ are attention weights. When a degree-$i$ node forms beliefs about the degrees of its neighbors, it effectively conditions on the subset of neighbors it notices. Attention is not evenly spread: individuals focus more on some alters (e.g., those who post more, are more salient, or stand out in memory) and neglect others. If a neighbor's chance of being noticed is proportional to $w_{ij}$, then the experienced distribution of neighbor degrees is no longer the baseline $P_{ij}$ but the reweighted one. If attention tilts toward high-degree neighbors, the agent overestimates neighborhood connectivity; if attention tilts toward low-degree or close ties, the agent underestimates it. In either case, the kernel $\widehat{P}$ encodes the fact that beliefs are shaped by what is visible rather than by true frequencies.

Recall that a probability kernel $\boldsymbol{P}$ satisfies Row Based Dominance (RBD) if for any increasing function $\boldsymbol{f}$, the map $i\mapsto(\boldsymbol{P}\boldsymbol{f})_{i}$ is also increasing.

\textbf{Proposition B1.} \textit{Suppose that $\boldsymbol{P}$ satisfies RBD and $\lambda>0$. Then, if $w_{i}(.)$ is increasing for all $i$, $a(\widehat{\boldsymbol{P}})\geq a(\boldsymbol{P})$, $\mathbb{E}_{\pi}(a(\widehat{\boldsymbol{P}}))\geq\mathbb{E}_{\pi}(a(\boldsymbol{P}))$ and $Var_{\pi}(a(\widehat{\boldsymbol{P}}))\geq Var_{\pi}(a(\boldsymbol{P}))$. If $w_{i}(.)$ is decreasing, then inequalities are reversed.}


Each of the biases described above can be understood as overweighting certain neighbor types relative to the truth  (e.g., correlation neglect with $\gamma>1$, homophily with high $h$, or limited attention with weights increasing in degree). Through the lens of Iterative Belief Centrality, these distortions work by changing the belief kernel through which perceived influence propagates. When agents overweight high-degree neighbors, their local samples become more correlated: high-degree agents keep ``seeing'' high-degree agents while low-degree agents keep ``seeing'' low-degree agents. As a result, actions rise on average and disparities widen because perceived influence becomes more concentrated.

The result can also be understood in terms of perceived assortativity. It is easy to show that for any $b\in\{0,1\}^{k}$, $|r^{b}(\widehat{\boldsymbol{P}})|\geq|r^{b}(\boldsymbol{P})|$. Our assortativity measures quantify persistence of correlation along belief steps, and under a biased kernel $\widehat{\boldsymbol{P}}$ every such correlation rises in magnitude. As these belief correlations increase, equilibrium behavior pulls away from the true benchmark $\boldsymbol{P}$. Mean action shifts because everyone's perceived environment becomes more selective, and variance increases because groups with different degrees act on increasingly different belief hierarchies. Therefore, biased network perceptions not only induce greater heterogeneity in beliefs, but also allow that heterogeneity to propagate and persist more strongly through the iterative belief process.

\subsubsection*{Proofs}

\noindent \textbf{Proof of Proposition B1}

Suppose $c=1$.

We start by establishing an auxiliary lemma. We say that kernel $\widehat{P}$
dominates $P$ if for any increasing function $f$, $(\widehat{P}f)_{i}\geq(Pf)_{i}$
for all $i$.

\textbf{Lemma B1. } \textit{If $\widehat{P}$ dominates $P$ and are both row-FOSD,
then $\mathbb{E}_{\pi}(a(\widehat{P}))\geq\mathbb{E}_{\pi}(a(P))$
and $Var_{\pi}(a(\widehat{P}))\geq Var_{\pi}(a(P))$.}


The mean comparison follows directly from the definition of kernel
dominance and the fact that action are monotonic under RBD. 

For the variance, we start by using the series expansion for equilibrium
actions. For any two probability kernels $\widehat{P}$ and $P$
we can write 
\[
Var_{\pi}(a(\widehat{P}))-Var_{\pi}(a(P))=\sum_{k,s\geq0}\lambda^{k+s}\left(Cov_{\pi}\left(g_{\widehat{P}}^{(k)},g_{\widehat{P}}^{(s)}\right)-Cov_{\pi}\left(g_{P}^{(k)},g_{P}^{(s)}\right)\right).
\]
where $a(P)=(a_{i,P})_{i\in N}$ denotes the vector of equilibrium
actions under $P$, and $g_{P}^{(k)}=(g_{i,P}^{(k)})_{i\in N}$ is
the vector whose $i^{th}$ elements is the $k^{th}$ contribution
to player $i's$ equilibrium action under $P$. We need the following
claims:

\textit{Claim 1: If $P$ satisfies RBD, then the map $i\mapsto g_{i,P}^{(k)}$
is increasing in $i$ for all $k\in\mathbb{N}.$}

By row-FOSD we know that $i\mapsto g_{i,P}^{(1)}$ is increasing in
$i$. Suppose it also true that $i\mapsto g_{i.P}^{(k)}$ is also
increasing. Then, by row-FOSD and the fact that $g_{i,P}^{(k)}$ is
also increasing it follows that $g_{P}^{(k+1)}=M_{d}Pg_{P}^{(k)}$
is also increasing in $i$. This establishes claim 1.

\textit{Claim 2: If $\widehat{P}$ dominates $P$ and both satisfy RBD,
then the map $i\mapsto g_{i,\widehat{P}}^{(k)}-g_{i,P}^{(k)}$ is
increasing in $i$ for all $k\in\mathbb{N}.$}

Since $\widehat{P}$ dominates $P$ then \textit{$i\mapsto g_{i,\widehat{P}}^{(1)}-g_{i,P}^{(1)}$
}is increasing in $i$. Suppose it also true that $i\mapsto g_{i,\widehat{P}}^{(k)}-g_{i,P}^{(k)}$
is also increasing. Then,
\begin{align*}
g_{\widehat{P}}^{(k+`1)}-g_{P}^{(k+1)} & =M_{d}\hat{P}g_{\widehat{P}}^{(k)}-M_{d}Pg_{P}^{(k)}\\
 & =M_{d}\widehat{P}\left(g_{\widehat{P}}^{(k)}-g_{P}^{(k)}\right)+M_{d}\left(\widehat{P}-P\right)g_{P}^{(k)}.
\end{align*}
The first term is increasing since $g_{i,\widehat{P}}^{(k)}-g_{i,P}^{(k)}$
is increasing and $\widehat{P}$ satisfies RBD. The second term is also
increasing since $\widehat{P}$ dominates $P$ and $g_{i,P}^{(k)}$
is increasing by RBD. This establishes claim 2.

To prove the lemma, let $\Delta^{(k)}:=g_{i,\widehat{P}}^{(k)}-g_{i,P}^{(k)}$
and note that we write: 
\[
Cov_{\pi}\left(g_{\widehat{P}}^{(k)},g_{\widehat{P}}^{(s)}\right)-Cov_{\pi}\left(g_{P}^{(k)},g_{P}^{(s)}\right)=Cov_{\pi}\left(\Delta^{(k)},g_{\widehat{P}}^{(s)}\right)+Cov_{\pi}\left(\beta_{P}^{(k)},\Delta^{(s)}\right).
\]
From the previous claims, we know that both pairs $\left(\Delta^{(k)},g_{\widehat{P}}^{(s)}\right)$
and $\left(\beta_{P}^{(k)},\Delta^{(s)}\right)$ constitute pairs
of increasing functions in $i.$ Therefore, by Chebyshev's covariance
inequality both $Cov_{\pi}\left(\Delta^{(k)},g_{\widehat{P}}^{(s)}\right)\geq0$
and $Cov_{\pi}\left(\beta_{P}^{(k)},\Delta^{(s)}\right)\geq0$ for
all pairs $k,s$. 

To prove the proposition, it suffices to show that $\widehat{P}$
dominates $P$ for the general class we are considering. Letting $f$
be any increasing function, we have 
\[
(\widehat{P}f)_{i}=(Pf)_{i}+\frac{Cov_{P_{ij}}(w_{i},f)}{(Pw_{i})_{i}}
\]
Since both $f$ and $w_{i}$ are increasing, by Chebyshev's covariance
inequality $Cov_{P_{ij}}(w_{i},f)>0$ establishing the result.

$\blacksquare$

\section*{Appendix D. Additional Examples}


In this Appendix, we present a simple example designed to highlight the role of belief hierarchies in shaping equilibrium behavior. The example provides an intuitive illustration of the process by which agents internalize other agents' types shapes and how these beliefs are employed in equilibrium.

\subsection*{D1. Core-Periphery Networks}

We derive the equilibrium for a class of core-periphery priors and compare it to a corresponding complete information equilibrium in which the network assumes such structure

Let $N_{co},$ $N_{p}\subset N$ with $N_{co}\cap N_{p}=\emptyset$
and $N_{co}\cup N_{p}=N$. Moreover, suppose that $\vert N_{co}\vert\equiv n_{co}\in\{0,...,\,n-2\}\cup\{n\}$ so there are either at least two players in the periphery, or none.\footnote{Note the case with one periphery player is equivalent to having the core be all players.} 
A core-periphery network $\mathbf{g}^{cp}$ associated with 
$N_{co},$ $N_{p}$  satisfies: (i) $g_{ij}^{cp}=1$ if $i\in N_{co}$ for all $j\neq i$,
and (ii) $g_{ij}^{cp}=0$ otherwise. This is a special class of core-periphery networks (those in which the core are all connected to each other, and periphery players are not connected to other periphery players) in which each periphery player is connected to all of the core members. Two instances of core-periphery networks are shown in the following figure.

\begin{figure}[H]
\begin{centering}
\begin{minipage}[t]{0.49\columnwidth}%
\begin{center}
\includegraphics[scale=0.51]{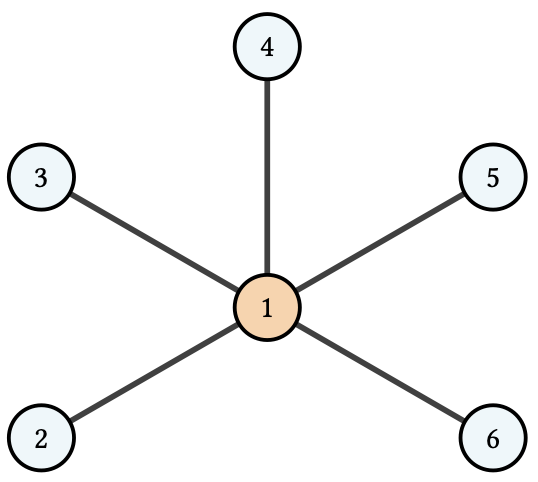} 
\par\end{center}%
\end{minipage}\hfill{}%
\begin{minipage}[t]{0.49\columnwidth}%
\begin{center}
\includegraphics[scale=0.51]{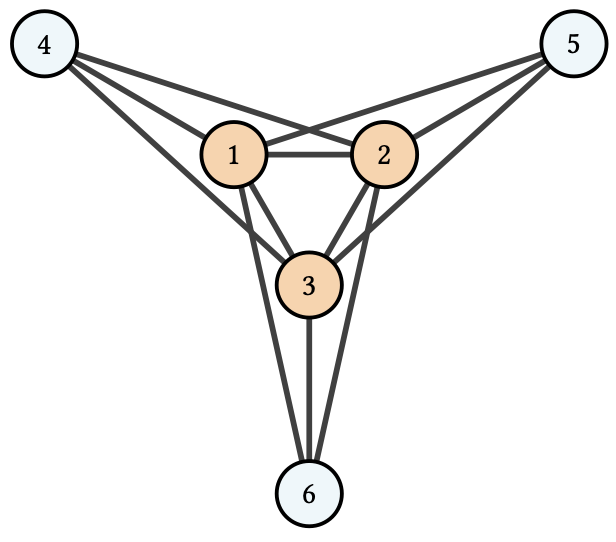} 
\par\end{center}%
\end{minipage}
\par\end{centering}
\caption{Two instances of core-periphery networks on six vertices.}\label{fig:CP}
\end{figure}

Suppose $c=1$. Letting $a_{co}^{c}$ and $a_{p}^{c}$ denote the complete information
Nash equilibrium actions of a core and a peripheral player respectively,
it can be shown that: 
\[
\begin{aligned}a_{co}^{c} & =\frac{1+\lambda n_{p}}{1-\lambda(n_{co}-1)-\lambda^{2}n_{p}n_{co}}\qquad\text{and}\qquad a_{p}^{c}=1+\lambda n_{co}a_{co}^{c}.\end{aligned}
\]
To compare this complete information equilibrium with the incomplete
information equilibrium, we
endow individuals with the ex-ante belief, $P_{cp}\left(.\right)$, that the actual network
has a core-periphery architecture
and their types are their direct connections;
i.e., $\mathcal{A}_{i}(\mathbf{g})=\left\{i\right\}$ for all
 $i$ and $\mathbf{g}$. 
Moreover, 
any such network
is equally likely to be selected by Nature. 
Formally, let $\mathcal{G}_{n}^{CP}\subset \mathcal{G}_{n}^{U}$ be the set of all core periphery networks. Then, priors satisfy $$\hat{p}(\mathbf{g}):=P_{cp}(\mathbf{g})=\frac{1}{\vert \mathcal{G}_{n}^{CP} \vert}$$ if $\mathbf{g}\in \mathcal{G}_{n}^{CP}$ and $P_{cp}(\mathbf{g})=0$ otherwise.

If Nature selects a network according to this distribution, an agent's
type satisfies one of two properties. Either $\sum_{j}g_{ij}^{t_{i}}=n-1$
in which case the agent knows it is in the core, or $\sum_{j}g_{ij}^{t_{i}}=n_{co}<n-1$
where it knows that it is in the periphery. Since players know that 
they are in a
core-periphery network among $n$ players, a peripheral player is able
to infer the architecture of the entire network. However, 
agents in the core do not know which of its
neighbors are in the core versus the periphery. 
This information structure,
together with the fact that each core-periphery networks is equally
likely, leads to the following characterization of equilibrium.

Let $\mathbb{E}_{co}[.]$ represent the expectations of core agents,
and so
\[
\mathbb{E}_{co}\left[n_{p}\right]=(n-1)\frac{\sum_{k=1}^{n-2}\binom{n-2}{k-1}}{2^{n-1}-(n-1)},\quad\mathbb{E}_{co}\left[n_{p}n_{co}\right]=(n-1)\frac{\sum_{k=1}^{n-2}k\binom{n-2}{k-1}}{2^{n-1}-(n-1)},
\]
\[
\mathbb{E}_{co}\left[n_{co}-1\right]=(n-1)\frac{\sum_{k=2}^{n}\binom{n-2}{k-2}-\binom{n-2}{n-3}}{2^{n-1}-(n-1)}.
\]

\textbf{Proposition D1.} \textit{Suppose $P=P_{cp}$ over $n>3$ vertices and that
Nature has chosen a core-periphery network with $n_{co}$ core nodes
and $n_{p}$ peripheral nodes. Let $a_{co}^{*}$ and $a_{p}^{*}$ respectively denote 
the equilibrium action of core and a peripheral agent. Then, 
\[
\begin{aligned}a_{co}^{*} & =\frac{1+\lambda\mathbb{E}_{co}\left[n_{p}\right]}{1-\lambda\mathbb{E}_{co}\left[n_{co}-1\right]-\lambda^{2}\mathbb{E}_{co}\left[n_{p}n_{co}\right]}\qquad\text{and}\qquad a_{p}^{*}=1+\lambda n_{co}a_{co}^{*}.\end{aligned}
\]}


Observe that this equilibrium is similar to the complete
information equilibrium, but with some important distinctions. 

An agent who is
in the core is able to infer the types of walks that it has in
the network,\footnote{For example, walks of length 2 from a core agent can only be of the following types $i_{c}\sim j_{c}\sim k_{c}$, or $i_{c}\sim j_{c}\sim k_{p}$, or $i_{c}\sim j_{p}\sim k_{c}$, or $i_{c}\sim j_{p}\sim k_{p}$, where the subscripts denote whether the agents are core or peripheral. Similarly for other walk lengths.}  but not how many of the walks it has.   It has more walks the larger the core and smaller the periphery.  
Nonetheless,
any two walks of a particular
type provide the same complementarity giving us the characterization
in Proposition D1.

A key implication of Proposition D1 is that even though
peripheral agents know the architecture of the entire network, they
do not exert $a_{p}^{c}$ in equilibrium. This is because they internalize
the fact that core agents cannot infer the 
network, and exert an action that is independent of the number of periphery agents. Consequently, a peripheral agent conditions on the fact that
core agents exert $a_{co}^{*}$ in equilibrium, and exerts an
action that is equal to the actual complementarity it experiences
from the network i.e., $1+\lambda n_{co}a_{co}^{*}$.

An additional implication is that unlike the complete information case where
equilibrium actions of core agents are strictly increasing with the
size of the core, their incomplete information actions do not change
with the network. In contrast, the actions of the peripheral
agents change with the network. The following lemma shows that whenever the size of the core
is less than half the population size, core agents exert actions lesser than that in
the incomplete information equilibrium, and hence so do periphery agents (who are best responding to them). 

\textbf{Lemma D1.} \textit{For any core-periphery network on $n\geq7$ vertices, if the size
of the core satisfies $n_{co}\leq n/2$, then all agents' actions are greater than in the complete information equilibrium.}


Note that the incomplete information equilibrium, given the externalities, can be overall welfare improving for the agents as the complete information equilibrium involves inefficiently low actions. 

Apart from closed-form comparisons between complete and incomplete
information equilibria, this core-periphery example highlights the
interplay between private information and beliefs with strategic behavior
in network games with local complementarities. Even though some individuals
may know the architecture of the entire network, their behavior does
not conform to complete information behavior. Informed individuals
internalize the network uncertainty of others and anticipate that
uniformed player actions do not correspond to complete information
actions.

\subsubsection*{Proofs}

\noindent \textbf{Proof of Proposition D1}

Consider $\left[\mathbf{g}_{k}^{cp}\right]$ to be the equivalence
class consisting of all graphs of core-periphery architecture with
core size being $k$. Employing the notation of Theorem 3 in the main body of the paper, we have $$\left\{ \left[\mathbf{g}_{k}^{cp}\right]:k\in\left\{ 0,\ldots,n-2,n\right\} \right\} \subset\left\{ \left[\mathbf{g}^{1}\right],\ldots,\left[\mathbf{g}^{q}\right]\right\}.$$
Priors
can thus be expressed as 
\[
P\left(\mathbf{g}\right)=\begin{cases}
\frac{1}{\left|\mathcal{G}_{n}^{CP}\right|} & \text{ if }\mathbf{g}\in\left[\mathbf{g}_{k}^{cp}\right]\text{ for some }k\\
0 & \text{ otherwise }
\end{cases}
\]
Hence, the underlying prior satisfies the property, $P\left(\mathbf{g}\right)=P\left(\mathbf{g}^{\prime}\right)$
whenever $\mathbf{g},\mathbf{g}^{\prime}\in\left[\mathbf{g}^{l}\right]$
for some $l$. Thus, from Theorem 3, the actions of
agents can be characterized by their local architecture, so that types can be expressed as $\left\{ P_{k}:k=0,1,2,....,n-2\right\} \cup\left\{ C\right\} $,
where $P_{k}$ is the type of an agent in the periphery with degree
$k$, i.e., the size of the core for that realized network is $k$
and $C$ is the type of an agent who is in the core. 

Let $t_{i}$
denote the type of any agent $i$. If an agent $i$ realizes that
they are in the periphery, i.e. $t_{i}=P_{k}$, they realize that
their direct connections are only the players in the core and hence
the best response for them is given by 
\begin{equation}
a(P_{k})=1+\lambda ka(C)\label{eq:Pk}
\end{equation}
where $a(C)$ is the action taken by a player in the core. The best
response for a Core agent $i$ who is connected to everyone is given
by 
\begin{equation}
\begin{alignedat}{1}a(C)= & 1+\lambda\end{alignedat}
\sum_{j=1}^{n}g_{ij}\left(\sum_{k=1}^{n-2}P\left(t_{j}=P_{k}\mid t_{i}=C\right)a\left(P_{k}\right)+P\left(t_{j}=C\mid t_{i}=C\right)a(C)\right)\label{eq:C1}
\end{equation}
Using Bayes' rule we can compute 
\[
\begin{alignedat}{1}P\left(t_{j}=P_{k}\mid t_{i}=C\right)=\frac{P\left(t_{j}=P_{k},t_{i}=C\right)}{P\left(t_{i}=C\right)} & =\frac{\binom{n-2}{k-1}}{\sum_{k=1}^{n}\binom{n-1}{k-1}-\binom{n-1}{n-2}}\\
P\left(t_{j}=C\mid t_{i}=C\right)=\frac{P\left(t_{j}=C,t_{i}=C\right)}{P\left(t_{i}=C\right)} & =\frac{\sum_{k=2}^{n}\binom{n-2}{k-2}-\binom{n-2}{n-3}}{\sum_{k=1}^{n}\binom{n-1}{k-1}-\binom{n-1}{n-2}}
\end{alignedat}
\]
where $P\left(t_{j}=P_{k},t_{i}=C\right)=\frac{\binom{n-2}{k-1}}{|\mathcal{G}_{n}^{CP}|}$
and $P\left(t_{i}=C\right)=\frac{\sum_{k=1}^{n}\binom{n-1}{k-1}-\binom{n-1}{n-2}}{|\mathcal{G}_{n}^{CP}|}$,
as the underlying distribution is uniform defined over all core-periphery
networks on $n$-nodes. Similar calculations follow for $P\left(t_{j}=C,t_{i}=C\right)$. Plugging these values into (\ref{eq:C1}), the best responses of a
Core agent reduces to: 
\begin{equation}
a(C)=1+\frac{(n-1)\lambda}{\Delta}\left[\sum_{k=1}^{n-2}\binom{n-2}{k-1}a(P_{k})+\left\{ \sum_{k=2}^{n}\binom{n-2}{k-2}-\binom{n-2}{n-3}\right\} a(C)\right]\label{eq:C2}
\end{equation}
where $\Delta=\sum_{k=1}^{n}\binom{n-1}{k-1}-\binom{n-1}{n-2}$. Using (\ref{eq:Pk}) the best response of a core player further reduces to
{\small{}
\[
a(C)=1+\frac{(n-1)\lambda}{\Delta}\left[\sum_{k=1}^{n-2}\binom{n-2}{k-1}+\lambda a(C)\sum_{k=1}^{n-2}k\binom{n-2}{k-1}+\left\{ \sum_{k=2}^{n}\binom{n-2}{k-2}-\binom{n-2}{n-3}\right\} a(C)\right]
\]
}
Therefore, the action exerted by a Core player is then given by

\begin{equation}
a(C)=\frac{1+\lambda X}{1-\lambda Z-\lambda^{2}Y}\label{eq:aC}
\end{equation}
where $X=(n-1)\frac{\sum_{k=1}^{n-2}\binom{n-2}{k-1}}{\Delta}$, $Y=(n-1)\frac{\sum_{k=1}^{n-2}k\binom{n-2}{k-1}}{\Delta}$
and $Z=(n-1)\frac{\sum_{k=2}^{n}\binom{n-2}{k-2}-\binom{n-2}{n-3}}{\Delta}$.
Looking into the terms $X,Y,Z$ we have the following:

(i) Consider the following random variable $X_{j}$ for an agent $i$
who is in the Core (i.e. $i\in C$), where $X_{j}=1$ if $j\in C\text{ for }g_{ij}=1$
and 0, otherwise. Also, $P(X_{j}=1)=\frac{\sum_{k=2}^{n}\binom{n-2}{k-2}-\binom{n-2}{n-3}}{\Delta}$.
Then for the agent $i$ the expected size of the core conditional
on the fact that they are in the core, is given by $\mathbb{E}\left[n_{co}|i\in C\right]=\mathbb{E}_{co}\left[n_{co}\right]=\sum_{j=1}^{n}g_{ij}P(X_{j}=1)+1=Z+1$.
Hence, $\mathbb{E}_{co}\left[n_{co}-1\right]=Z$.

(ii) Consider the following random variable $Y_{j}$ for an agent
$i$ who is in the Core (i.e. $i\in C$), where $Y_{j}=1$ if $j\in P\text{ for }g_{ij}=1$
and $0$, otherwise. Also, $P(Y_{j}=1)=\frac{\sum_{k=1}^{n-1}\binom{n-2}{k-1}-\binom{n-2}{n-2}}{\Delta}$.
Then for the agent $i$ the expected number of agents in the periphery
conditional on the fact that they are in the core, is given by $\mathbb{E}\left[n_{p}|i\in C\right]=\mathbb{E}_{co}\left[n_{p}\right]=\sum_{j=1}^{n}g_{ij}P(Y_{j}=1)=X$.
Hence, $\mathbb{E}_{co}\left[n_{p}\right]=X$.

(iii) For any agent $i\in C$ let $d_{p}$ be the random variable
denoting the degree of their neighbor $j$ if they are in the periphery.
Then the expected degree of agent $i$'s neighbor if they are in the
periphery, conditional on the fact that $i\in C$ is given by, $\mathbb{E}\left[d_{p}|i\in C\right]=\sum_{k=1}^{n-2}k\binom{n-2}{k-1}/\Delta$.
Thus for the agent $i$, the complementary strength that they can
extract from the peripheral nodes through the walks of length 2 is
$\mathbb{E}\left[n_{p}n_{co}|i\in C\right]=\mathbb{E}_{co}\left[n_{p}n_{co}\right]=(n-1)\mathbb{E}\left[d_{p}|i\in C\right]=Y$.

From all the above it follows that the equilibrium action of a core agent
is: 
\[
a_{co}^{*}=a\left(C\right)=\frac{1+\lambda\mathbb{E}_{co}\left[n_{p}\right]}{1-\lambda\mathbb{E}_{co}\left[n_{co}-1\right]-\lambda^{2}\mathbb{E}_{co}\left[n_{p}n_{co}\right]}
\]

$\blacksquare$

\noindent \textbf{Proof of Lemma D1}

Let $n\geq7$.

{\noindent \textbf{Case I:} $n_{co}<n/2$}

Define $X,Y\text{ and }Z$ as in the proof of Proposition 1. Putting the values of $X$ and $Z$, we get 
$$
X+Z =(n-1) =n_{p}+(n_{co}-1)
$$
Then we prove the lemma through the following claims:

{\noindent\textbf{Claim D1.1: }$Yn_{p}-Xn_{p}n_{co}>0$}

We know that $n_{p}\geq0$. Hence, we have to show that $Y-Xn_{co}>0$.
Putting the values of $Y$ and $X$, we can write 
\begin{align*}
Y-Xn_{co} =\frac{(n-1)}{2^{n-1}-(n-1)}\left[n\cdot2^{n-3}-n+1-(2^{n-2}-1)n_{co}\right]
\end{align*}

We see that $Y-Xn_{co}$ is monotonically decreasing in the values
of $n_{co}$. Since $n_{co}<n/2$, if we can show that $Y-Xn_{co}>0$
for $n_{co}=\frac{n}{2}-1$ when $n$ is even and for $n_{co}=\frac{n-1}{2}$
when $n$ is odd, we're done. Consider the case when $n$ is even.
Then for $n_{co}=\frac{n}{2}-1$ we have that 
\begin{align*}
Y-Xn_{co} =\frac{(n-1)}{2^{n-1}-(n-1)}\left[2^{n-2}-\frac{n}{2}\right]
\end{align*}
And as a result, $Y-X(\frac{n}{2}-1)>0$ as $2^{n-2}-\frac{n}{2}>0$
for all $n\geq5$. Similarly for the case when $n$ is odd. This proves
our claim that $Y-Xn_{co}>0$.

{\noindent\textbf{Claim D1.2: }$n_{p}n_{co}-Zn_{p}<Y-X(n_{co}-1)$}

Putting the value $n_{p}=n-n_{co}$ and using $X+Y=n_p+(n_{co}-1)$, we get
\begin{align*}
Y-X(n_{co}-1)-n_{p}n_{co}+Zn_{p} =Y+Zn+X-(2n-n_{co}-1)n_{co}
\end{align*}
Since $n_{co}<n/2$,
we can see that $(2n-n_{co}-1)n_{co}$ increases in $n_{co}$. As
a result, $Y-X(n_{co}-1)-n_{p}n_{co}+Zn_{p}$ decreases in the values
of $n_{co}$. If we can show that $Y-X(n_{co}-1)-n_{p}n_{co}+Zn_{p}>0$
for $n_{co}=\frac{n}{2}-1$ when $n$ is even and for $n_{co}=\frac{n-1}{2}$
when $n$ is odd, we're done. Consider the case when $n$ is even.
Then for $n_{co}=\frac{n}{2}-1$ we arrive at
\begin{align*}
 & Y-X(n_{co}-1)-n_{p}n_{co}+Zn_{p}\\
= & \frac{1}{2^{n-1}-(n-1)}\left[\left(n^{2}-1\right)2^{n}+\left(n^{2}-n\right)\left(2^{n-1}-4\right)-\left(n^{2}-2n\right)\left(n-1+3\cdot2^{n-1}\right)\right]
\end{align*}
is greater than $0$, for all values of $n\geq7$. Where we use the identity,
\begin{align*}
X+Y+Zn = \frac{(n-1)}{2^{n-1}-(n-1)}\left[\left(n+1\right)2^{n-2}+n\left(2^{n-3}-1\right)-n\left(n-2\right)\right]
\end{align*}
Hence $Y-X(n_{co}-1)-n_{p}n_{co}+Zn_{p}>0$ when $n$ is even.Similarly
we can show the same for $n$ being odd. Hence, $n_{p}n_{co}-Zn_{p}<Y-X(n_{co}-1)$.

Thus from the above two claims, we can say that for any $0<\lambda<1/n-1${\footnotesize{}
\[
1+\lambda\left(X+Z\right)+\lambda^{2}\left(Y-X(n_{co}-1)\right)+\lambda^{3}Yn_{p}>1+\lambda\left(n_{p}+n_{co}-1\right)+\lambda^{2}\left(n_{p}n_{co}-Zn_{p}\right)+\lambda^{3}Xn_{p}n_{co}
\]
\[
\Rightarrow1+\lambda X-\lambda\left(n_{co}-1\right)-\lambda^{2}X\left(n_{co}-1\right)-\lambda^{2}n_{p}n_{co}-\lambda^{3}Xn_{p}n_{co}>1+\lambda n_{p}-\lambda Z-\lambda^{2}Zn_{p}-\lambda^{2}Y-\lambda^{3}Yn_{p}
\]
}{\footnotesize\par}

Which results in 
\[
\frac{1+\lambda X}{1-\lambda Z-\lambda^{2}Y}>\frac{1+\lambda n_{p}}{1-\lambda\left(n_{c}-1\right)-\lambda^{2}n_{p}n_{c}}
\]

{\noindent \textbf{Case II:} $n_{co}=n/2$}

{\noindent\textbf{Claim D1.3: $Y+\frac{n}{2}Z-X\left(\frac{n}{2}-1\right)-\frac{n^{2}}{4}+\lambda\frac{n}{2}\left(Y-\frac{n}{2}X\right)>0$}}

Putting the values of $X,Y\text{ and }Z$, we get that 
\begin{align*}
g\left(n,\lambda\right) = Y+\frac{n}{2}Z-X\left(\frac{n}{2}-1\right)-\frac{n^{2}}{4}+\lambda\frac{n}{2}\left(Y-\frac{n}{2}X\right) =\frac{\left(n-2\right)\left(2^{n}-2n\left(1+\lambda\right)\left(n-1\right)\right)}{4\left(2+2^{n}-2n\right)}
\end{align*}
Differentiating with respect to $\lambda$, we get 
\[
\frac{\partial g\left(n,\lambda\right)}{\partial\lambda}=-\frac{n\left(n-1\right)\left(n-2\right)}{2\left(2+2^{n}-2n\right)}<0
\]
Hence, $g\left(n,\lambda\right)$ is decreasing in $\lambda$. As
a result, if we can show that $g\left(n,\lambda\right)>0$ for $\lambda=1/(n-1)$,
then we're done. At $\lambda=1/\left(n-1\right)$ we get 
\[
g\left(n,1/\left(n-1\right)\right)=\frac{\left(n-2\right)\left(2^{n}-2n^{2}\right)}{4\left(2+2^{n}-2n\right)}>0\quad\forall\:n\geq7
\]
Thus, $g\left(n,\lambda\right)>0$ for all $n\geq7\text{ and }\lambda<1/\left(n-1\right)$.
This proves our claim.

Then from the claim we can write that for any $n\geq7\text{ and }\lambda<1/\left(n-1\right)$

{\footnotesize{}
\[
\lambda\left(X+Z\right)+\lambda^{2}\left(Y+\frac{n}{2}Z-X\left(\frac{n}{2}-1\right)-\frac{n^{2}}{4}\right)+\lambda^{3}\frac{n}{2}\left(Y-\frac{n}{2}X\right)>\lambda\frac{n}{2}+\lambda\left(\frac{n}{2}-1\right)
\]
\[
\Rightarrow1-\lambda\left(\frac{n}{2}-1\right)-\lambda^{2}\frac{n^{2}}{4}+\lambda X-\lambda^{2}X\left(\frac{n}{2}-1\right)-\lambda^{3}\frac{n^{2}}{4}X>1-\lambda Z-\lambda^{2}Y+\lambda\frac{n}{2}-\lambda^{2}\frac{n}{2}Z-\lambda^{3}\frac{n}{2}Y
\]
\[
\Rightarrow\frac{1+\lambda X}{1-\lambda Z-\lambda^{2}Y}>\frac{1+\lambda\frac{n}{2}}{1-\lambda\left(\frac{n}{2}-1\right)-\lambda^{2}\frac{n^{2}}{4}}
\]
}Hence, for $n_{co}=n/2$, incomplete information dominates the complete
information equilibrium action.

$\blacksquare$

\clearpage
\bibliographystyle{ecta}
\bibliography{ref}

\end{document}